\documentclass[journal]{IEEEtran}

\usepackage{mathrsfs}  
\usepackage{graphicx}
\usepackage{epstopdf}
\usepackage{subfloat}
\usepackage{float}
\usepackage{amsmath}
\usepackage{CJK}
\usepackage{amssymb}
\usepackage{indentfirst}
\usepackage{mathtools}
\usepackage{subfigure}
\usepackage{stfloats}
\usepackage{mathrsfs}
\usepackage{bm}
\usepackage{blindtext}
\usepackage{algorithm}
\usepackage{algpseudocode}
\usepackage{amsmath}
\usepackage{cite}
\usepackage{colortbl}
\definecolor{mygray}{gray}{.9}
\usepackage[table]{xcolor}
\usepackage{booktabs}
\usepackage{enumerate}

\usepackage{amsmath,amssymb}

\usepackage[Symbol]{upgreek}
\usepackage{booktabs}
\usepackage{multirow}

\usepackage{array}

\ifCLASSINFOpdf

\else

\fi

\newcolumntype{C}[1]{>{\centering\let\newline\\\arraybackslash\hspace{0pt}}m{#1}}
\begin{document}
\title{Model-based Joint Bit Allocation between Geometry and Color for Video-based 3D Point Cloud Compression}

\author{Qi~Liu, Hui~Yuan,~\IEEEmembership{Senior Member,~IEEE,}
        Junhui~Hou,~\IEEEmembership{Member,~IEEE,}\\
        Raouf~Hamzaoui,~\IEEEmembership{Senior Member,~IEEE,} and Honglei~Su
\thanks{This work was supported in part by the National Natural Science Foundation of China under Grants 61571274 and 61871342; in part by the National Key R\&D Program of China under Grants 2018YFC0831003; in part by the Shandong Natural Science Funds for Distinguished Young Scholar under Grant JQ201614, in part by the Shandong Provincial Key Research and Development Plan under Grant 2017CXGC1504; in part by the open project program of state key laboratory of virtual reality technology and systems, Beihang University, under Grant VRLAB2019B03; in part by Shenzhen Science and Technology Research and Development Funds under Grant JCYJ20170818103244664, and in part by the Young Scholars Program of Shandong University (YSPSDU) under Grant 2015WLJH39.}
\thanks{Q. Liu is with the School of Information Science and Engineering, Shandong University, Qingdao 266237, China, also with the Shenzhen Research Institute of Shandong University, Shenzhen 518057, China (Email: sdqi.liu@gmail.com).}
\thanks{H. Yuan is the corresponding author. He is with the Shenzhen Research Institute of Shandong University, Shenzhen 518057, China, also with the School of Control Science and Engineering, Shandong University, Ji'nan 250061, China (Email: huiyuan@sdu.edu.cn).}
\thanks{J. Hou is with the Department of Computer Science, City University of Hong Kong, Hong Kong (Email: jh.hou@cityu.edu.hk).}
\thanks{R. Hamzaoui is with School of Engineering and Sustainable Development, De Montfort University, Leicester, UK (Email: rhamzaoui@dmu.ac.uk).}
\thanks{H. Su is with the School of Electronic Information, Qingdao University, Qingdao 266071, China (Email: suhonglei@qdu.edu.cn).}
}

\markboth{Submitted to IEEE Transactions on Multimedia}%
{Shell \MakeLowercase{\textit{et al.}}: Bare Demo of IEEEtran.cls
for Journals}

\maketitle

\begin{abstract}
Rate distortion optimization plays a very important role in image/video coding. But for 3D point cloud, this problem has not been investigated. In this paper, the rate and distortion characteristics of 3D point cloud are investigated in detail, and a typical and challenging rate distortion optimization problem is solved for 3D point cloud. Specifically, since the quality of the reconstructed 3D point cloud depends on both the geometry and color distortions, we first propose analytical rate and distortion models for the geometry and color information in video-based 3D point cloud compression platform, and then solve the joint bit allocation problem for geometry and color based on the derived models. To maximize the reconstructed quality of 3D point cloud, the bit allocation problem is formulated as a constrained optimization problem and solved by an interior point method. Experimental results show that the rate-distortion performance of the proposed solution is close to that obtained with exhaustive search but at only 0.68\% of its time complexity. Moreover, the proposed rate and distortion models can also be used for the other rate-distortion optimization problems (such as prediction mode decision) and rate control technologies for 3D point cloud coding in the future.
\end{abstract}


\IEEEpeerreviewmaketitle

%
\section{Introduction}
\IEEEPARstart{W}{ith} the rapid development of 3D scanning
techniques~\cite{6296662}, point clouds are now readily available
and popular~\cite{rusu20113d}. There are already many 3D point cloud
applications in the fields of 3D
modeling~\cite{8368133}~\cite{6642077},
automatic driving~\cite{chen2017multi}, 3D
printing~\cite{oropallo2018point}, augmented
reality~\cite{park2008multiple}, etc. Compared with traditional 2D
images and videos, a 3D point cloud describes the 3D scene or object
with the geometry information and the corresponding attributes
(e.g., color, reflectance)~\cite{liu2018model}. To represent a 3D
scene accurately, millions of points must be captured and processed.
This huge data volume poses a severe challenge for efficient storage
and transmission. In the past few years, major progress in both
static and dynamic point cloud compression has been
made~\cite{schwarz2018emerging}. To compress the 3D point cloud
(especially the attribute information) effectively, transforms that
can adapt to the irregular structure of the 3D point cloud were
exploited. These include the shape-adaptive discrete cosine
transform~\cite{cohen2016point}, graph
transforms~\cite{zhang2014point}~\cite{cohen2016attribute}~\cite{8419286},
region-adaptive hierarchical
transforms~\cite{de2016compression,sandri2019integer,m49380,m42640},
Gaussian process
transforms~\cite{chou2016gaussian}~\cite{de2017transform}, and
sparse representation based on virtual adaptive
sampling~\cite{gu20193d}~\cite{hou2017sparse}. Based on the
previously mentioned graph transforms, Shao \emph{et
al.}~\cite{shao2018hybrid} further combined a slice partition scheme
and an intra prediction technique to improve the performance of
attribute compression. Instead of compressing the irregular data
directly, some
researchers~\cite{mekuria2016design}~\cite{tu2016compressing} try to
map the irregular data to a regular representation (e.g., a 2D
plane) to simplify the task. Similar ideas were previously proposed
to compress 3D human motion~\cite{6826514} and 3D facial
expression~\cite{6778752,6470663,5782944,gu2002geometry}. All the
described methods are mainly designed for static point clouds. Since
dynamic point clouds are becoming more and more essential in
practical applications, efficient compression methods for dynamic
point clouds are also required. Because of the inter-frame
redundancy of dynamic point clouds, motion estimation and motion
compensation are the key technologies to effectively compress
dynamic point clouds. Thanou, Chou and
Frossard~\cite{thanou2016graph} focused on motion estimation by
using a spectral graph wavelet descriptor. De Queiroz and
Chou~\cite{de2017motion} proposed a motion compensation approach to
encode dynamic voxelized point clouds. Anis, Chou and
Ortega~\cite{anis2016compression} simplified motion compensation by
representing sets of frames in a consistently evolving
high-resolution subdivisional triangular mesh.

To standardize 3D point cloud compression (PCC) technologies, the
Moving Pictures Expert Group (MPEG) launched a call for proposals in
2017. As a result, three point cloud compression technologies were
developed: surface point cloud compression (S-PCC) with software
platform TMC1~\cite{n17223} for static point cloud data, video-based
point cloud compression (V-PCC) with software platform
TMC2~\cite{n17248} for dynamic content, and LIDAR point cloud
compression (L-PCC) with software platform TMC3~\cite{n17249} for
dynamically acquired point clouds. Recently, L-PCC and S-PCC were
merged under the name geometry-based point cloud compression (G-PCC)
with software platform TMC13~\cite{n18189}.

\begin{figure*}[t!]
  \centering
  \includegraphics[width=1.8\columnwidth]{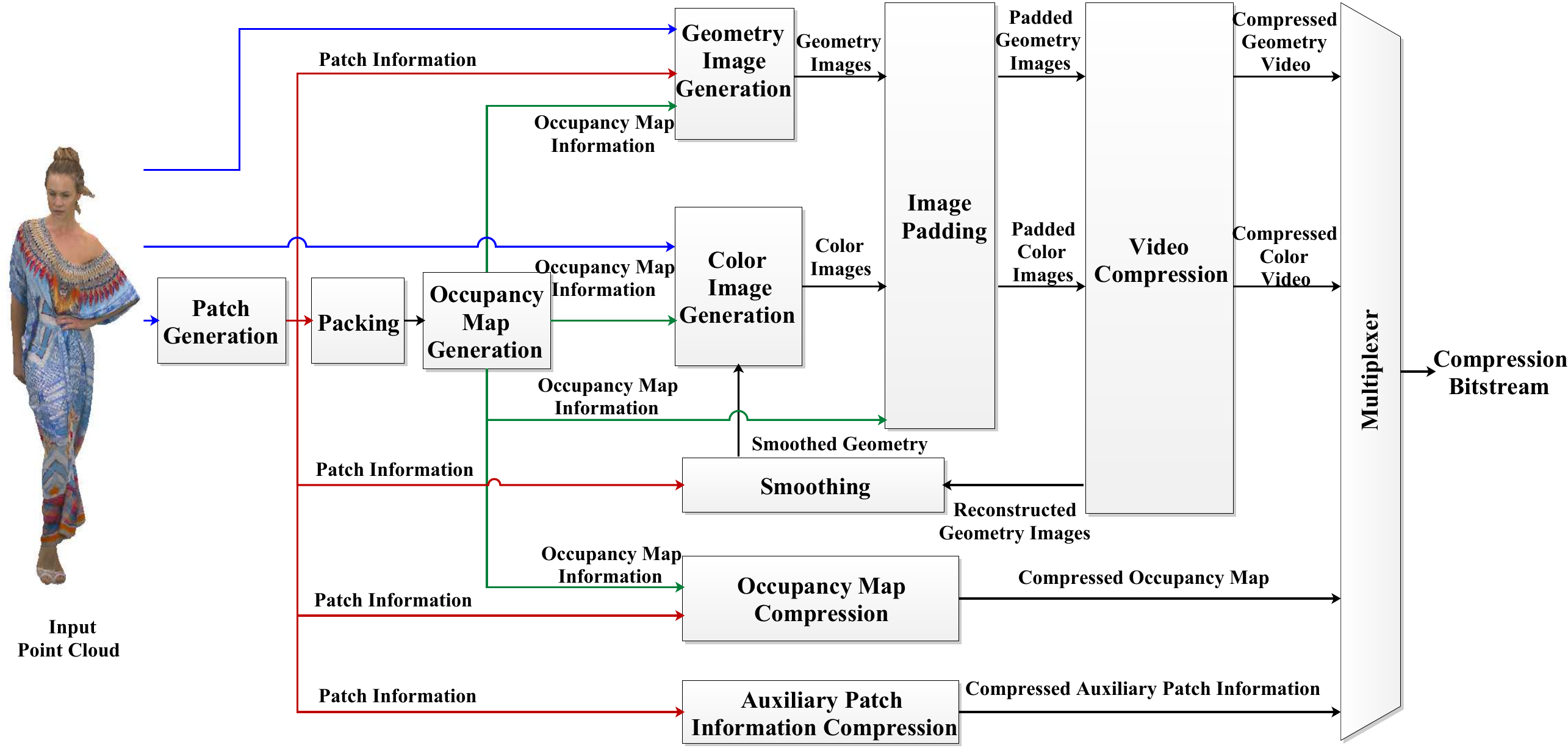}
  \caption{General framework of the V-PCC encoder.}
  \label{tmc2framework}
\end{figure*}

In this paper, we focus on the V-PCC platform (TMC2) due to its
excellent performance for compressing both static and dynamic point
clouds. As shown in Fig.~\ref{tmc2framework}, the main philosophy
behind V-PCC is to leverage state-of-the-art video coders for
PCC~\cite{tmc2v1}.
This is essentially achieved by decomposing each point cloud of a
sequence of 3D point clouds into a set of patches, which are
independently mapped to a 2D grid of uniform blocks. This mapping is
then used to store the geometry and color information as one
geometry image and one color image. The sequences of geometry images
and color images corresponding to the dynamic point cloud are then
compressed separately with a video coder, e.g., H.265/HEVC. Finally,
the geometry and color videos, together with metadata (occupancy map
for the 2D grid, auxiliary patch and block information) are used to
reconstruct the dynamic 3D point cloud. Please refer
to~\cite{haoliutbc} for more details.

The bitstream of a compressed 3D point cloud consists of two parts:
geometry information and color information. For a given platform,
the size of each part is controlled by a quantization parameter,
which can take a large number of values. At the same time,
quantization introduces distortion, which may affect the
reconstruction quality. The aim of this paper is to find the pair of
quantization parameter values (one for the geometry and one for
color) that minimizes the reconstruction error subject to a
constraint on the total number of bits. Since the number of
candidates is very large, finding an optimal solution with
exhaustive search is too time consuming and infeasible in practice
as it requires encoding and decoding the point cloud multiple times.
In this paper, we address this challenge by proposing analytical
models for the rate and distortion and using an interior point
method to efficiently solve the rate-distortion optimization
problem. The contributions in this paper are as follows.
\begin{enumerate}[1)]
\item We propose analytical models that characterize the rate
and distortion of the geometry and color as functions of the V-PCC
quantization steps.
\item We exploit the proposed analytical
models to formulate the bit allocation problem for V-PCC as a
constrained optimization problem.
\item We use an interior point method to efficiently solve the optimization problem.
\end{enumerate}

\textbf{Rate distortion optimization plays a very important role in video/image coding. But for 3D point cloud, rate distortion optimization as well as bit allocation/rate control technologies are not investigated before our research. To the best of our knowledge, this is the first time that the rate and distortion characteristics of 3D point cloud are investigated in detail. The proposed rate and distortion models can not only be used for the joint bit allocation of geometry and color but also used for the rate-distortion optimization and rate control technologies for 3D point cloud coding.} Based on the derived rate and distortion model, in this paper, we focus on the typical and challenging rate distortion optimization problem, i.e., optimal quantization parameters decision for geometry and color, for 3D point cloud coding. Extensive experimental results demonstrate the advantage and effectiveness of our method. In~\cite{liu2018model}, we used a similar approach for the Point Cloud Library (PCL)-based codec~\cite{mekuria2016design}~\cite{PCL}. The PCL-based encoder recursively subdivides the point cloud into eight subsets. This results in an octree data structure, where the position of each voxel is represented by its center whose attribute (color) is set to the average of the attributes of the enclosed points. Then the octree data is encoded by an entropy coder. In order to encode the color attributes, the octree attributes are mapped directly to a structured  image grid using depth first tree traversal and then the image grid is encoded with the JPEG codec.

There are three main differences between this paper and our previous
work~\cite{liu2018model}.

First, since the two codecs are different, both the rate-distortion
analysis and the modeling of the rate and distortion functions are
different.

Second, the objective function for the rate-distortion optimization,
as well as the optimization variables are different. In this paper,
the objective function is the overall distortion, which includes the
distortion of both geometry and color information, and the
optimization variables are the quantization parameters for the
geometry and color information. In~\cite{liu2018model}, only the
color distortion is minimized, and the optimization variables are
the octree level and the JPEG quality factor.

Third, in contrast to our previous paper~\cite{liu2018model}, where the models are obtained using a simple and rough statistical analysis without theoretical justification, we derived the models theoretically in this paper, and then verified the accuracy of the models by experiments.

The remainder of this paper is organized as follows. In Section II,
we formulate the rate-distortion optimization problem for V-PCC and
provide analytical models for the distortion and rate as a function
of the quantization steps. In Section III, we use these models to
solve the rate-distortion optimization problem with an interior
point method. In Section IV, we present experimental results where
we compare the accuracy of our approach to that of exhaustive
search. Section V gives our conclusions and suggests future work.

\section{Rate and Distortion Models Derivation}\label{sec:2}
To efficiently solve the bit allocation problem, we formulate it as
a constrained optimization problem by deriving rate and distortion
models for the 3D point cloud. As the distortion of a 3D point cloud
is determined by the coding distortion of both geometry and color
information, the bit allocation problem can be expressed as
\begin{equation}
\label{eq:rdtotal}
\begin{aligned}
\min_{(\emph{$R_{g},R_{c}$})} & D(D_{g},D_{c}) \\
\mathrm{s.t.} \quad & R_g + R_c \leq \emph{$R_{T}$},
\end{aligned}
\end{equation}
where the distortion $D$ of the reconstructed 3D point cloud is
determined by the color distortion ($D_c$) and the geometry
distortion ($D_g$); $R_g$ and $R_c$ are the geometry and color
bitrate, respectively, and $R_T$ is the target bitrate for both the
geometry and color. It is worth noting that the occupancy map and
auxiliary information also consume bitrate resources. The occupancy
map is a binary array that indicates whether a pixel position is
occupied or not. The auxiliary information is just used to store a
few encoder parameters. In general, both the occupancy map and the
auxiliary information are compressed without loss. In addition,
their bitrate cost is small and fixed for a given 3D point cloud.
Therefore, we do not consider the bitrate of the occupancy map and
auxiliary information in~\eqref{eq:rdtotal}. Because both $R_g$ and
$R_c$ are controlled by the quantization parameters (QPs) of the
color and geometry coders, the problem is to find the values of the
corresponding quantization steps $Q_g$ and $Q_c$ to achieve optimal
quality under a given target bitrate. In the next section, we
develop distortion and rate models with respect to the quantization
steps, so that the bit allocation problem~\eqref{eq:rdtotal} can be
solved analytically.
\subsection{Distortion Model}  
\begin{figure}[t!]
\centering \subfigure[]{ \label{fig2:subfig:a}
\includegraphics[width=0.48\columnwidth]{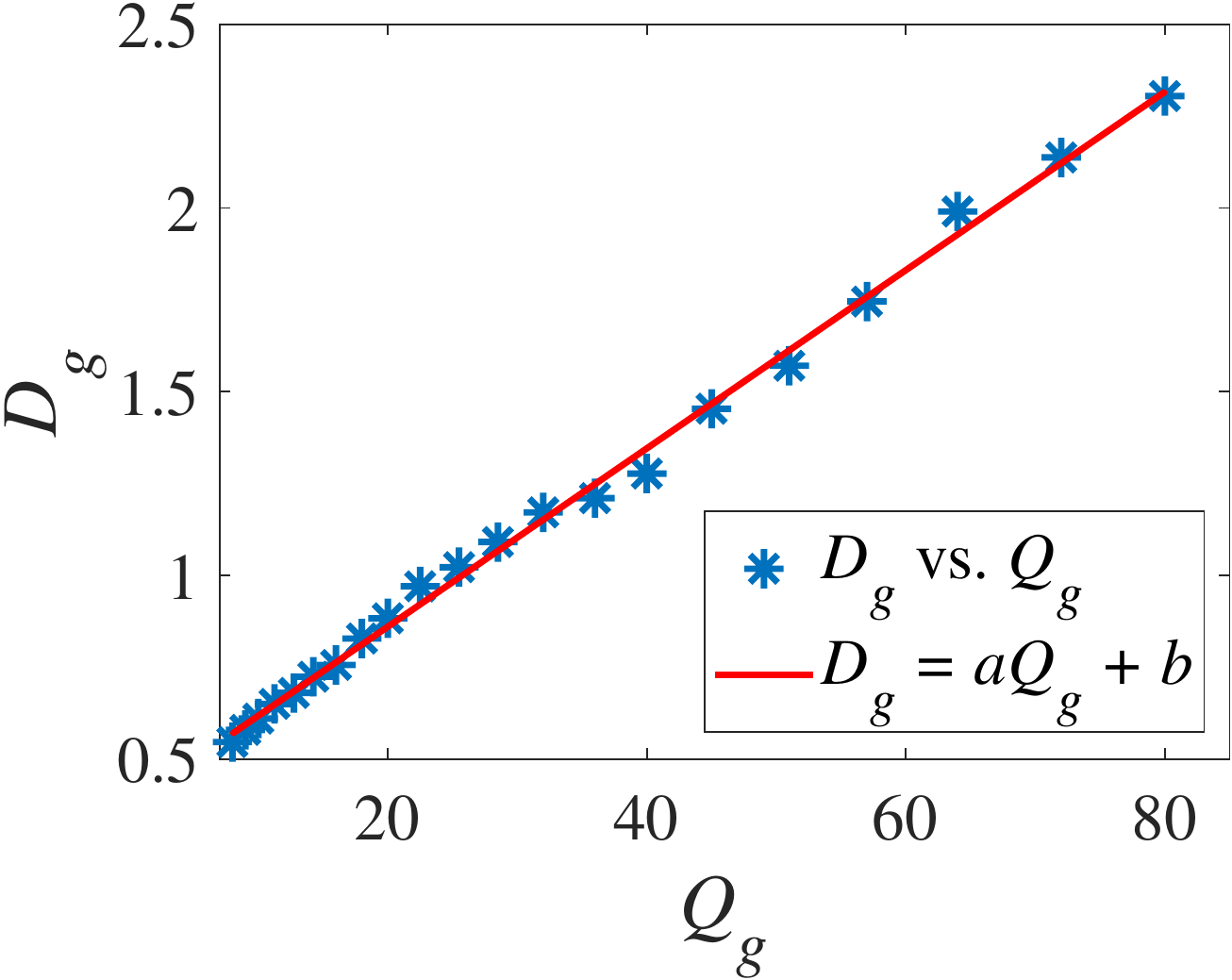}}
\subfigure[]{ \label{fig2:subfig:b}
\includegraphics[width=0.48\columnwidth]{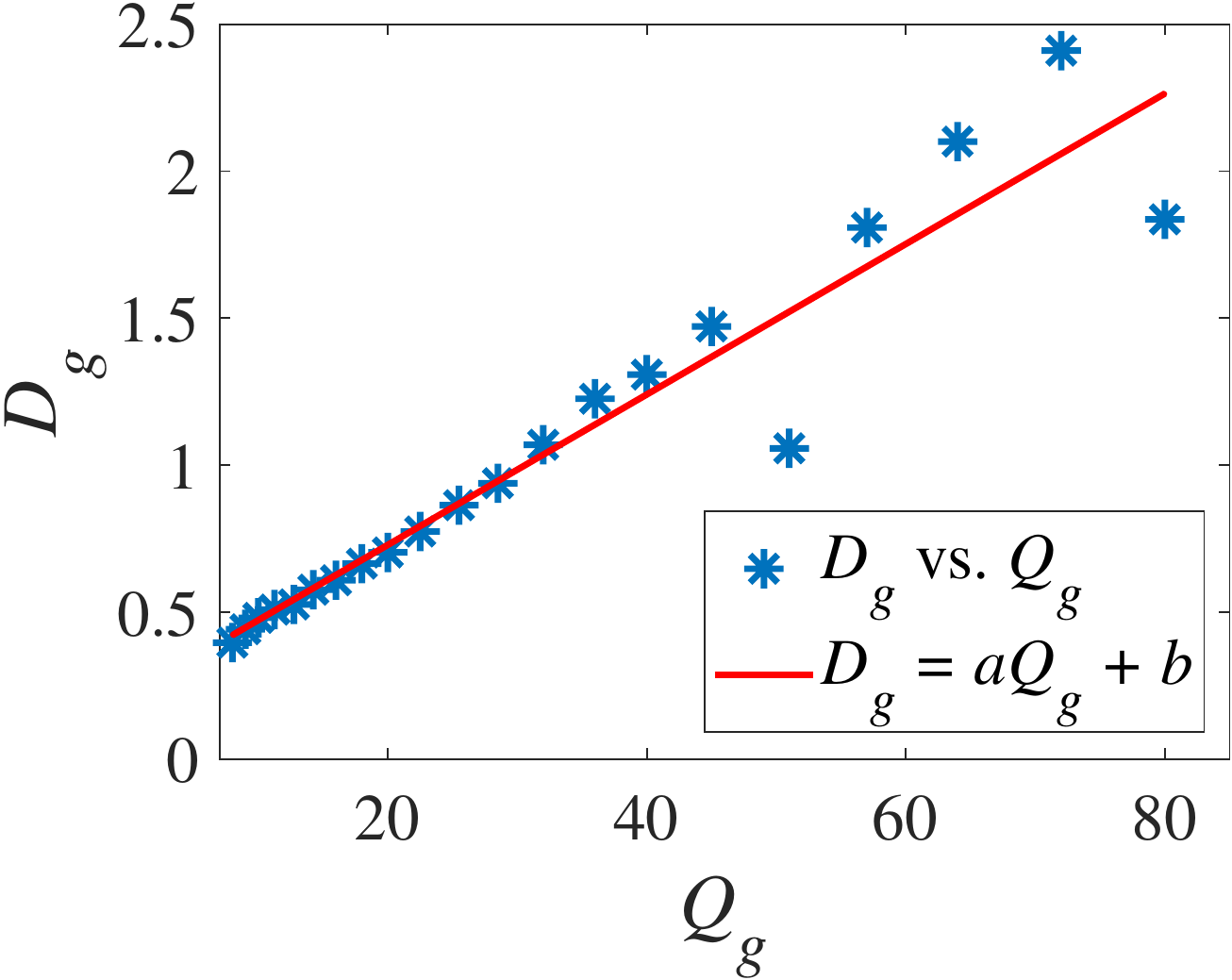}}
\subfigure[]{ \label{fig2:subfig:c}
\includegraphics[width=0.48\columnwidth]{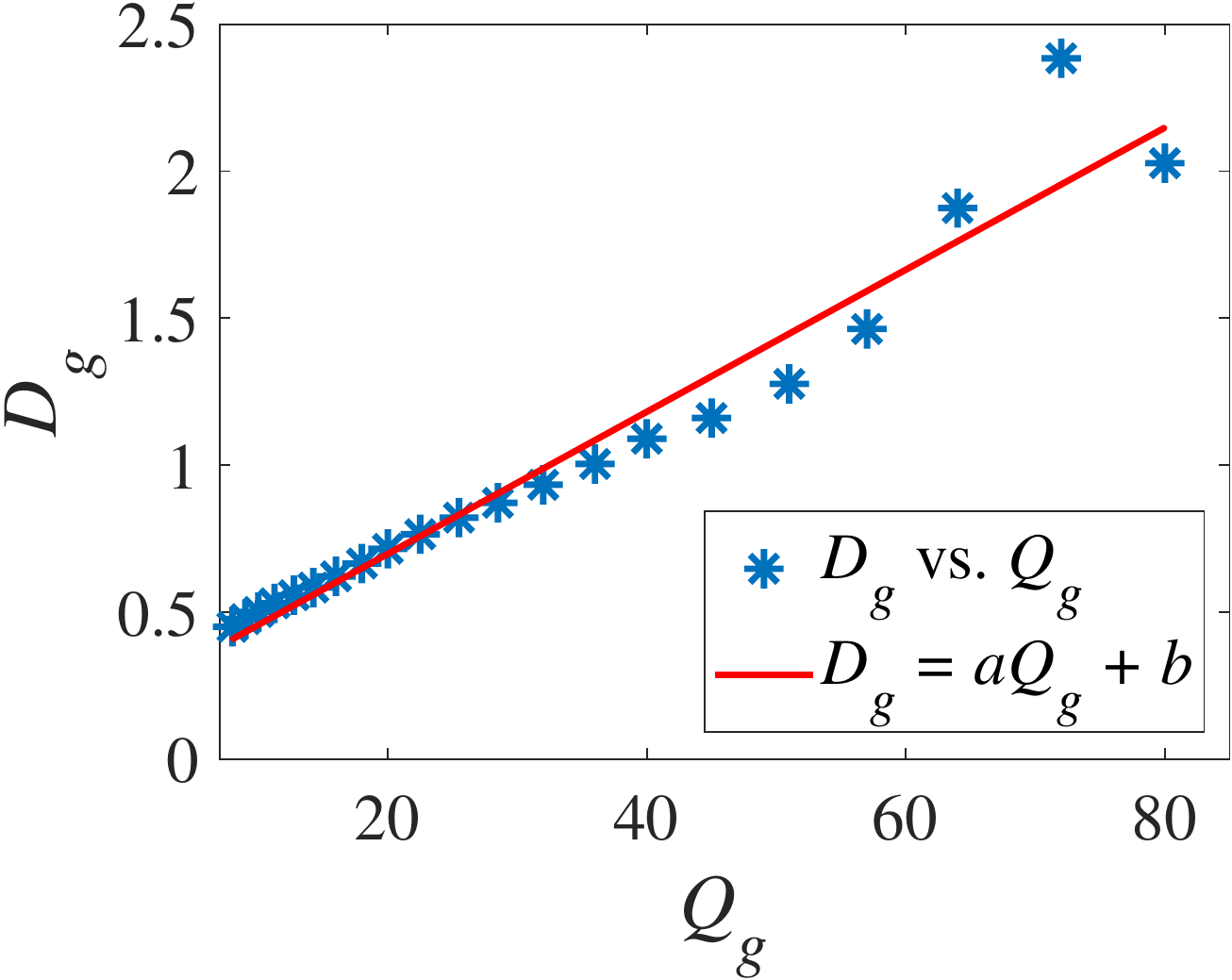}}
\subfigure[]{ \label{fig2:subfig:d}
\includegraphics[width=0.48\columnwidth]{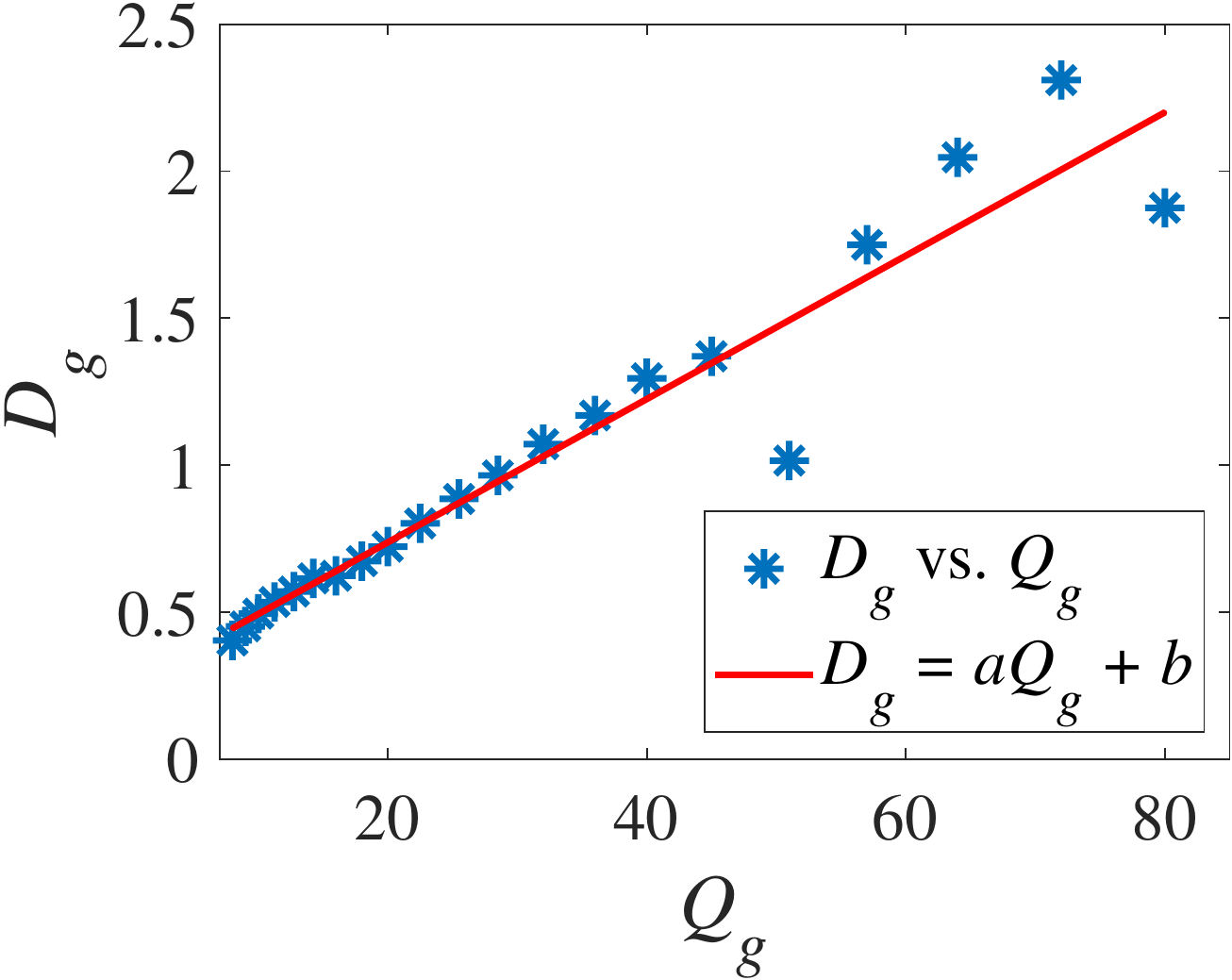}}
\caption{Illustration of the relationship between the geometry
distortion $D_g$ and the geometry quantization step $Q_g$. (a)
\emph{Phil}, (b) \emph{Longdress}, (c) \emph{David}, (d)
\emph{Loot}.} \label{fig2}
\end{figure}

\begin{figure*}[t!]
\centering \subfigure[]{ \label{fig3:subfig:a}
\includegraphics[width=0.485\columnwidth]{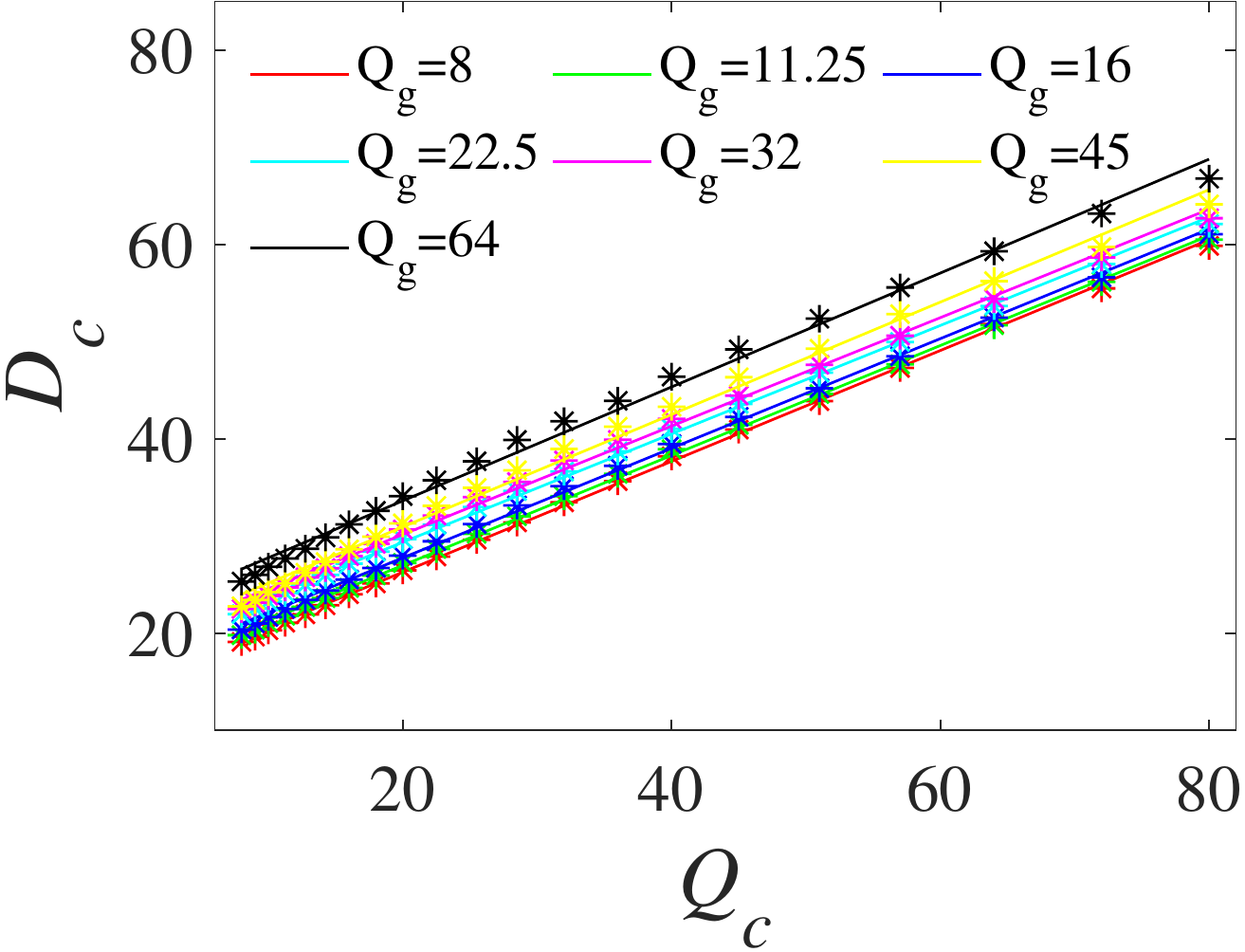}} 
\subfigure[]{ \label{fig3:subfig:b}
\includegraphics[width=0.485\columnwidth]{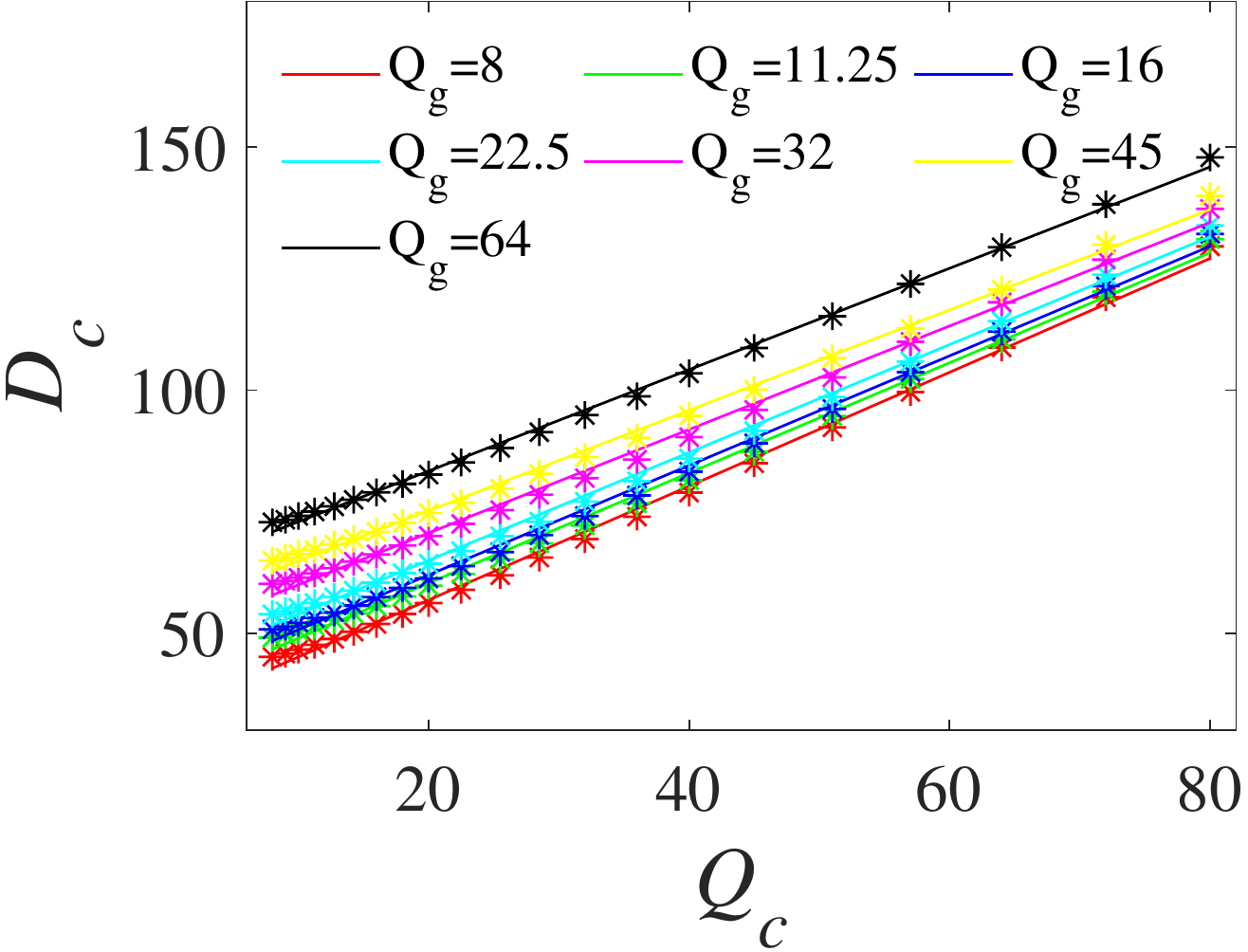}}
\subfigure[]{ \label{fig3:subfig:c}
\includegraphics[width=0.485\columnwidth]{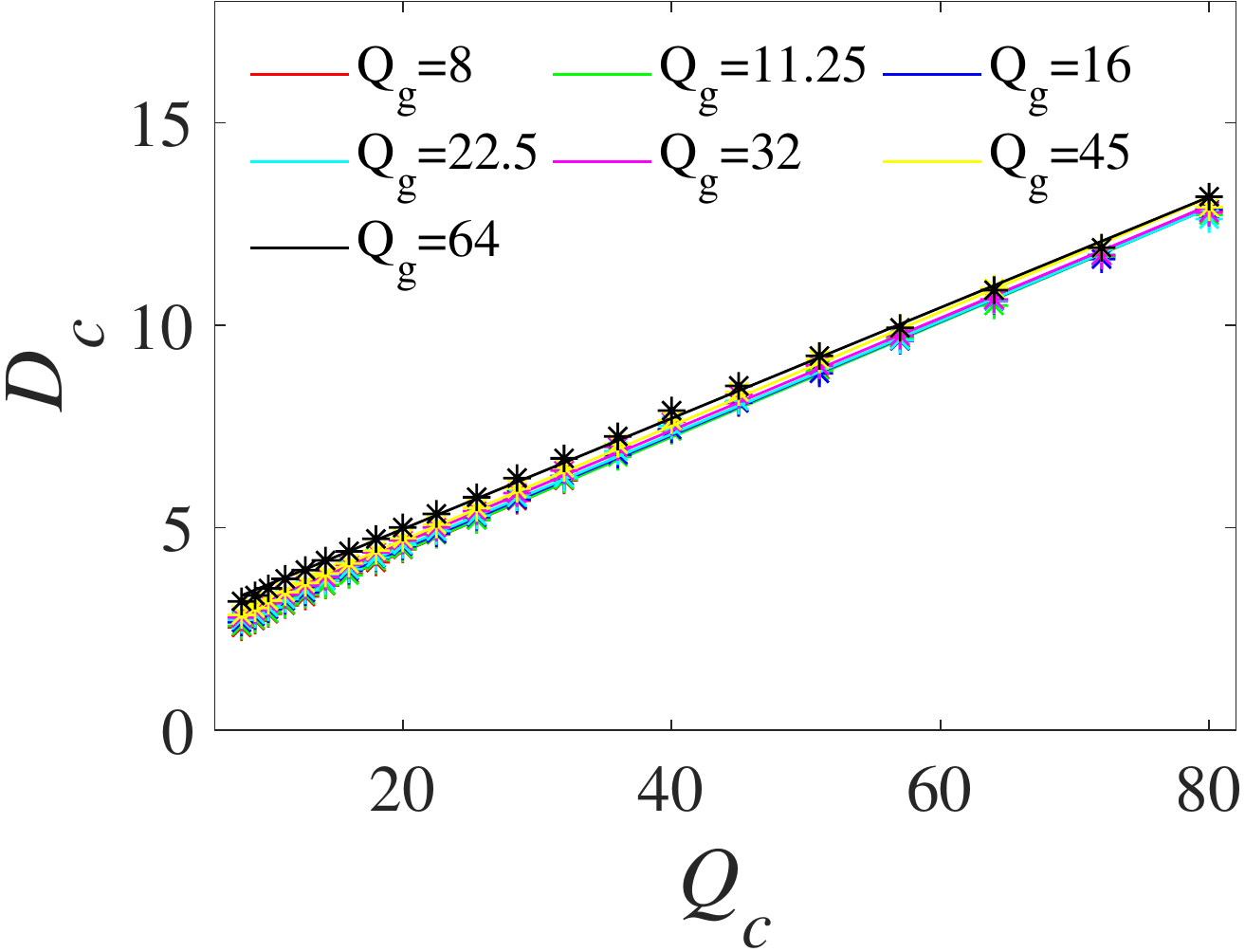}}
\subfigure[]{ \label{fig3:subfig:d}
\includegraphics[width=0.485\columnwidth]{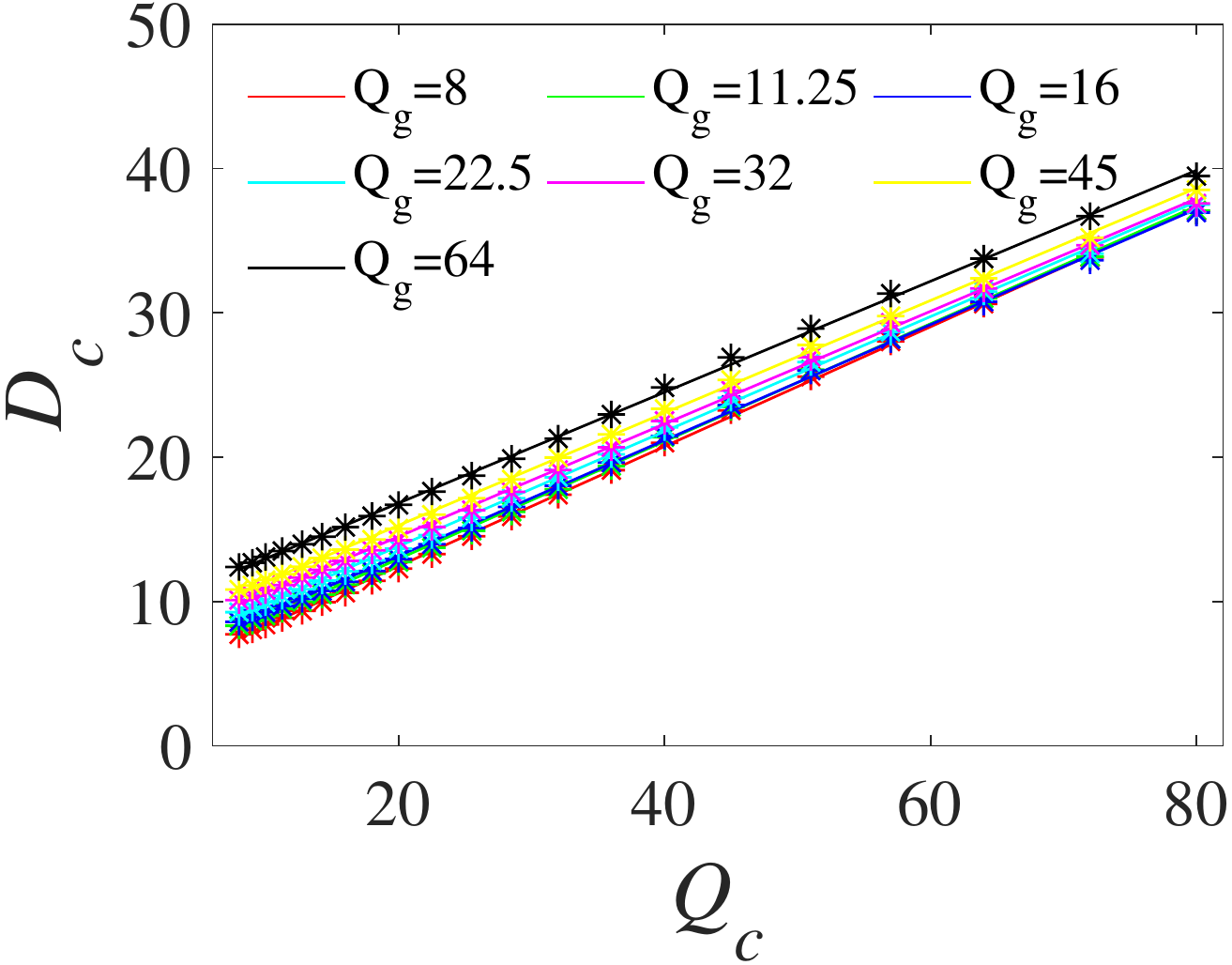}}
\subfigure[]{ \label{fig3:subfig:e}
\includegraphics[width=0.485\columnwidth]{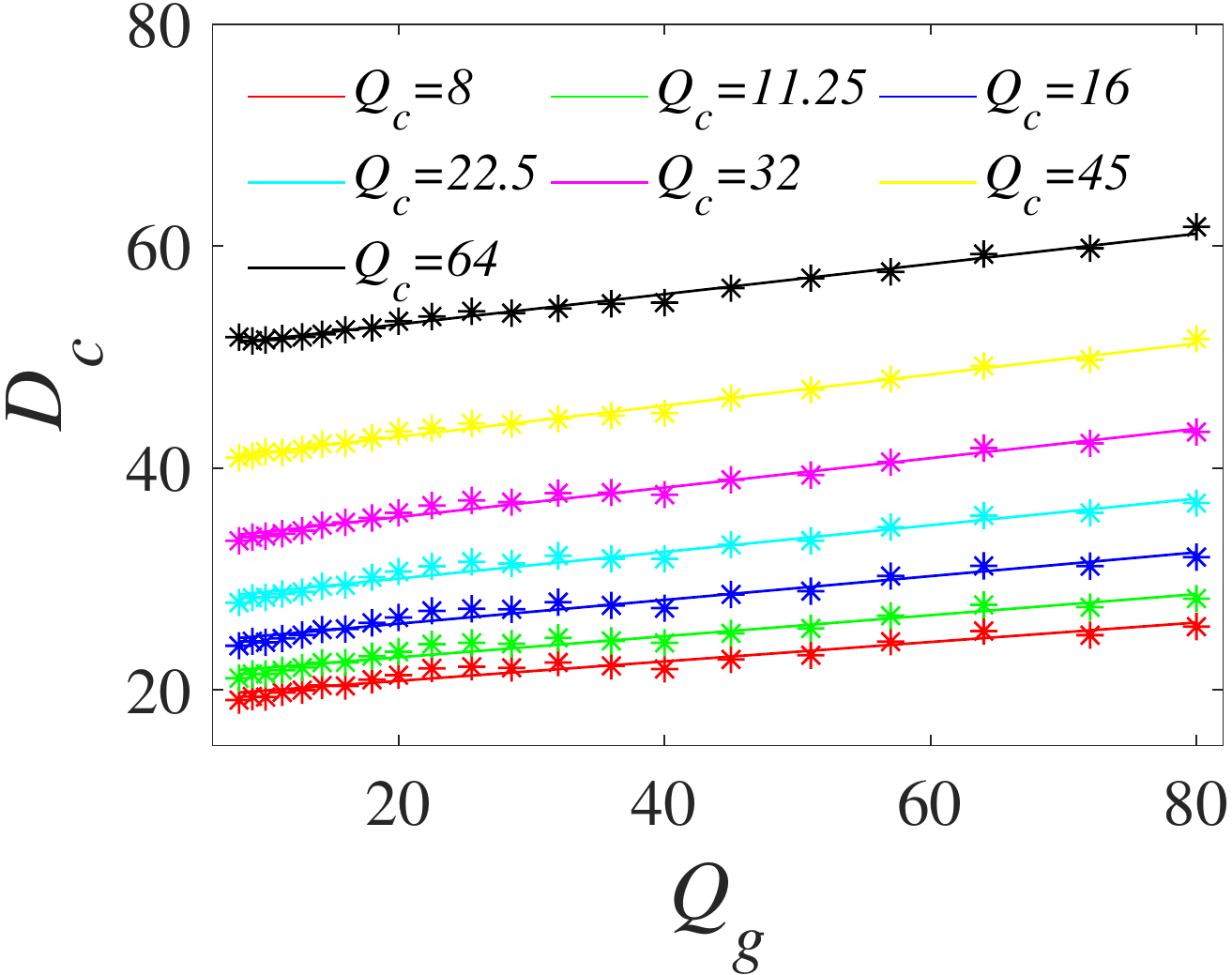}}
\subfigure[]{ \label{fig3:subfig:f}
\includegraphics[width=0.485\columnwidth]{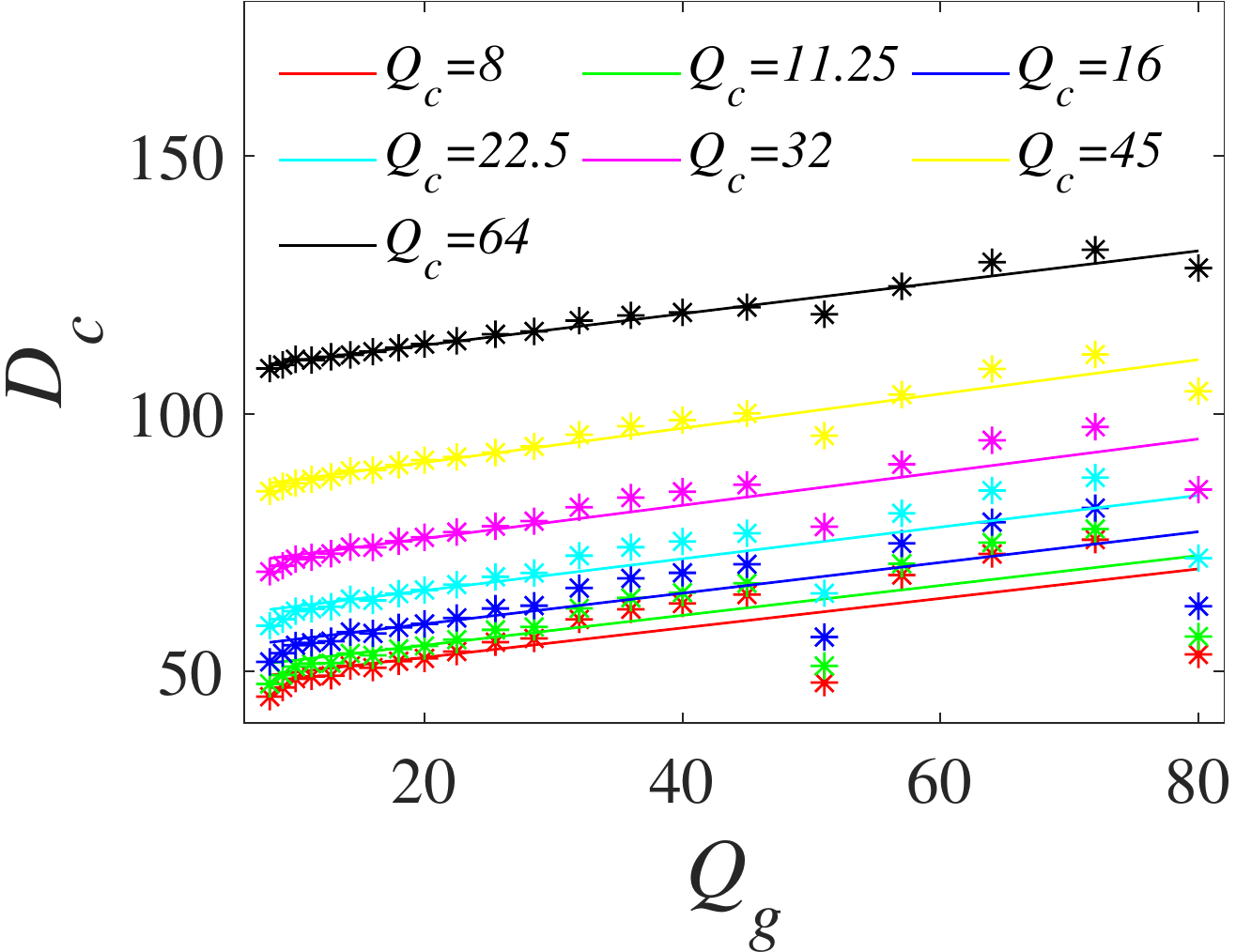}}
\subfigure[]{ \label{fig3:subfig:g}
\includegraphics[width=0.485\columnwidth]{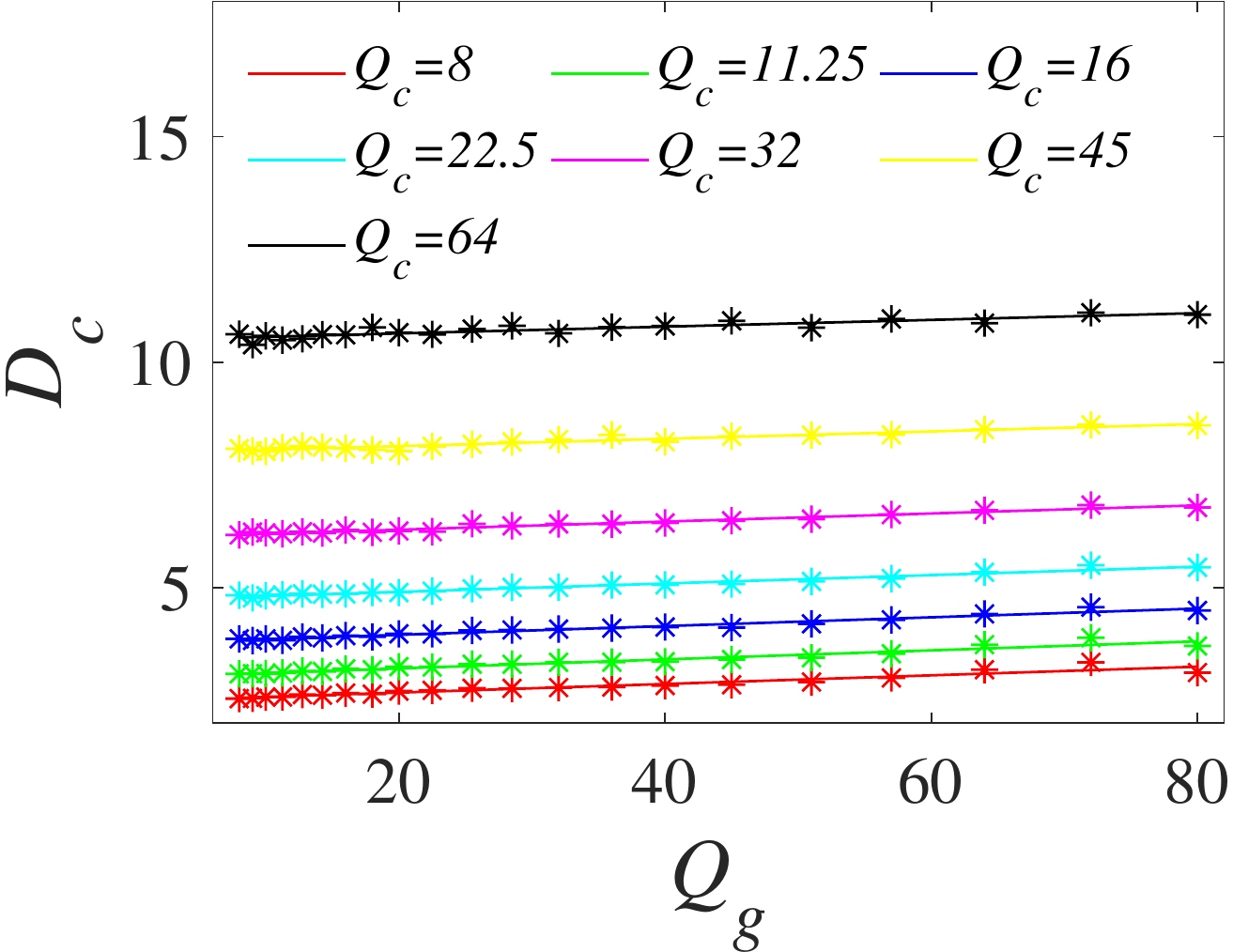}}
\subfigure[]{ \label{fig3:subfig:h}
\includegraphics[width=0.485\columnwidth]{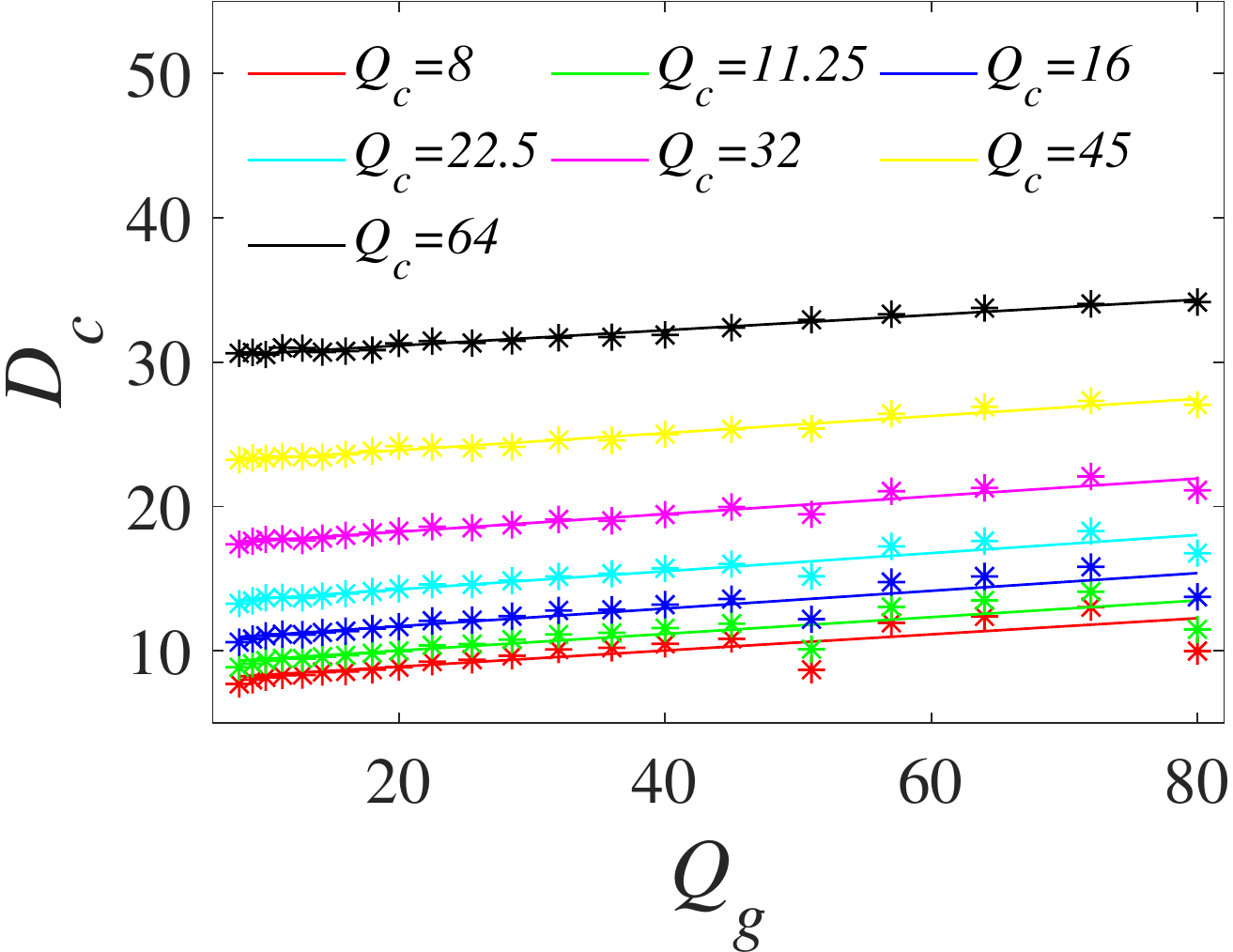}}
\caption{Statistical results for $f_g
(Q_g)=\alpha_{gc}Q_g+\beta_{gc}$ and $f_c (Q_c)=\alpha_{cc} Q_c +
\beta_{cc}$. (a)-(d): relationship between  $D_c$ and $Q_c$ for
\emph{Phil}, \emph{Longdress}, \emph{David}, and \emph{Loot},
(e)-(f): relationship between  $D_c$ and $Q_g$ for \emph{Phil},
\emph{Longdress}, \emph{David}, and \emph{Loot}.} \label{fig3}
\end{figure*}
The mean squared error (MSE) is widely used as the distortion metric for point clouds. We calculate it as a linear combination of the geometry distortion and color distortion. That is,
\begin{equation}
\label{eq:MSE} D=\omega D_g + \left(1-\omega\right) D_c,
\end{equation}
where $D_g$ is the geometry distortion, $D_c$ is the color distortion, and $\omega$ is a weighting factor. In this paper, we use the symmetric point-to-point distortion defined as follows ~\cite{pcerror}. Let \emph{A} and \emph{B} denote the original point cloud and the reconstructed point cloud, respectively. Then $D_g=max(e^{B,A}_{g}, e^{A,B}_{g})$. Here, for two point clouds $A$ and $B$,
\begin{equation}
\label{eq:MSE_geometry} e^{B,A}_{g}=\frac{1}{|B|}\sum_{b_i\in B}\|b_in_A(b_i)\|^2_2,
\end{equation}
where $|B|$ denotes the number of points in $B$, $n_A(b_i)$ is the nearest neighbor of $b_i$ in $A$, and $\|  \|^2_2$ is the Euclidean norm. Similarly, $D_c=max(e^{B,A}_{c}, e^{A,B}_{c})$ where
\begin{equation}
\label{eq:MSE_color} e^{B,A}_{c}=\frac{1}{|B|}\sum_{b_i\in B}|C(b_i)-C(n_A(b_i))|^2,
\end{equation}
Here $C(x)$ is a color attribute of point $x$. For simplicity, we consider only the $Y$ (luminance) component~\cite{mekuria2017performance} in this paper. Because of the structure of the V-PCC encoder, $D_g$ only depends on $Q_g$. To obtain a relationship between $D_g$ and $Q_g$, statistical analysis was conducted by compressing point clouds with different $Q_g$s, as shown in Fig.~\ref{fig2}.

\begin{table}[t!]
\centering \caption{Accuracy of Proposed Geometry Distortion
Model~\eqref{eq:gdgq}} \label{tab:gdmodel}
  \begin{tabular}{c|c|c}
      \toprule
      Point Cloud & SCC & RMSE \\\hline
      \emph{Andrew}  &0.99    &0.04 \\
      \emph{David}   &0.95    &0.13 \\
      \emph{Phil}    &1.00    &0.03 \\
      \emph{Ricardo} &0.99    &0.03 \\
      \emph{Longdress}   &0.91    &0.18 \\
      \emph{Loot}    &0.92    &0.16 \\
      \emph{Queen}   &0.96    &0.10 \\
      \emph{Redandblack} &0.90    &0.18 \\\hline
      Average &0.95    &0.11 \\
      \bottomrule
  \end{tabular}
\end{table}

The results suggest that a linear model
\begin{equation}
\label{eq:gdgq} D_g =\alpha_g   Q_g + \beta_g,
\end{equation}
where $\alpha_g$ and $\beta_g$ are model parameters is appropriate.
This is confirmed by Table~\ref{tab:gdmodel} which gives the squared
correlation coefficient (SCC) and the root mean squared error (RMSE)
between the actual data and the fitted data. We can see that the
SCCs are close to 1, while the maximum RMSE is only 0.18.

Because color compression is based on the reconstructed geometry in the V-PCC encoder, we need to further study the influence of geometry distortion on color distortion. Since $D_c=max(e^{B,A}_{c}, e^{B,A}_{c})$, we will analyze $e^{B,A}_{c}$ and $e^{B,A}_{c}$. Based on ~\eqref{eq:MSE_color}, $e^{B,A}_{c}$ can be rewritten as
\begin{equation}
\label{eq:Dc1}
\begin{aligned}
e^{B,A}_{c} &= \frac{1}{|B|} \sum_{ j=1 }^{|B|}  \big|  \textbf{\emph{C}}_{v_j}-\textbf{\emph{C}}_{v_j^*}  \big| ^2 , \\
\end{aligned}
\end{equation}
where $\textbf{\emph{C}}_{v_j}$ denotes the color of a point $v_j$ in the point cloud $B$, $v_j^*$ is the nearest neighbor of $v_j$ in the point cloud $A$, and $\textbf{\emph{C}}_{v_j^*}$ is the color of $v_j^*$. $e^{B,A}_{c}$ can be decomposed into two functions of $Q_g$ and $Q_c$, i.e.,
\begin{equation}
\label{eq:Dc4f}
\begin{aligned}
e^{B,A}_{c} &=f^{B,A}_g(Q_g)+f^{B,A}_c(Q_c),
\end{aligned}
\end{equation}
where $f^{B,A}_g(Q_g)$ and $f^{B,A}_c(Q_c)$ are the function of $Q_g$ and $Q_c$ respectively. Appendix~\ref{appendix:a} gives the details of the derivation. Similarly, $e^{A,B}_{c}$ can also be decomposed to be the similar functions, i.e., $f^{A,B}_g(Q_g)$ and $f^{A,B}_c(Q_c)$ of $Q_g$ and $Q_c$ respectively, and $D_c$ can be written as
\begin{equation}
\label{eq:Dc5f}
\begin{aligned}
D_c&=max(e^{B,A}_{c}, e^{A,B}_{c}) \\&= max(f^{B,A}_g(Q_g)+f^{B,A}_c(Q_c), f^{A,B}_g(Q_g)+f^{A,B}_c(Q_c)) \\&=f_g(Q_g)+f_c(Q_c).
\end{aligned}
\end{equation}

To obtain an analytical expression for $f_g (Q_g)$ and $f_c (Q_c)$,
statistical experiments were conducted, as shown in Fig.~\ref{fig3}.
To investigate the exact expression of $f_c (Q_c)$, which
characterizes the relationship between $Q_c$ and $D_c$, the
influence of $Q_c$ on the distortion of the comparison color point
cloud was statistically analyzed by setting $Q_g$ to fixed values.
As suggested in Fig.~\ref{fig3:subfig:a}-\ref{fig3:subfig:d}, for a
fixed $Q_g$, a linear model
\begin{equation}
\label{eq:fcqc}
\begin{aligned}
f_c(Q_c)=\alpha_{cc} Q_c + \beta_{cc},
\end{aligned}
\end{equation}
where $\alpha_{cc}$ and $\beta_{cc}$ are model parameters, is
appropriate. Similarly, to derive an analytical expression for $f_g
(Q_g)$, the relationship between $Q_g$ and $D_c$ was statistically
analyzed for a fixed $Q_c$.
Fig.~\ref{fig3:subfig:e}-\ref{fig3:subfig:h} suggests that a linear
model
\begin{equation}
\label{eq:fgqg}
\begin{aligned}
f_g(Q_g)=\alpha_{gc} Q_g + \beta_{gc},
\end{aligned}
\end{equation}
where $\alpha_{gc}$ and $\beta_{gc}$ are model parameters is
suitable.

Consequently, based on~\eqref{eq:Dc4f},~\eqref{eq:fcqc},
and~\eqref{eq:fgqg}, the color distortion can be finally written as
\begin{equation}
\label{eq:dcfinal}
\begin{aligned}
D_c &=\alpha_{gc} Q_g + \beta_{gc} +\alpha_{cc} Q_c + \beta_{cc}\\
&=\alpha_{gc} Q_g +\alpha_{cc} Q_c +\beta_{c},
\end{aligned}
\end{equation}
where $\beta_{c}= \beta_{gc}+\beta_{cc}$. The accuracy
of~\eqref{eq:dcfinal} is verified in Table~\ref{tab:cdmodel}, which
shows that all SCCs are larger than or equal to 0.96, and the
average RMSE is only 1.42.
\begin{table}[t!]
\centering \caption{Accuracy of Proposed Color Distortion
Model~\eqref{eq:dcfinal} } \label{tab:cdmodel}
  \begin{tabular}{c|c|c}
      \toprule
      Point Cloud & SCC & RMSE \\\hline
      \emph{Andrew}  &1.00    &1.16 \\
      \emph{David}   &1.00    &0.11 \\
      \emph{Phil}    &1.00    &0.82 \\
      \emph{Ricardo} &1.00    &0.14 \\
      \emph{Longdress}   &0.97    &4.27 \\
      \emph{Loot}    &1.00    &0.55 \\
      \emph{Queen}   &0.96    &3.46 \\
      \emph{Redandblack} &0.99    &0.81 \\\hline
      Average &0.99    &1.42 \\
      \bottomrule
  \end{tabular}
\end{table}
Finally,~\eqref{eq:MSE} can be rewritten as
\begin{equation}
\label{eq:dfinal}
\begin{aligned}
D &=\omega   D_g + \left(1-\omega\right)   D_c \\
&=\omega (\alpha_g   Q_g + \beta_g)+\left(1-\omega\right)(\alpha_{gc} Q_g +\alpha_{cc} Q_c + \beta_{c})\\
&=a Q_g +b Q_c +c ,
\end{aligned}
\end{equation}
where $a=\omega\alpha_g+\left(1-\omega\right)\alpha_{gc}$,
$b=\left(1-\omega\right) \alpha_{cc}$, and
$c=\omega\beta_g+\left(1-\omega\right) \beta_c$.
Table~\ref{tab:dmodel} shows the SCC and RMSE between the actual $D$
and the one provided by our model. In this table, the SCC and RMSE
were calculated by setting the weighting factor $\omega$ to 0.5. We
can see that the average SCC and RMSE are 0.99 and 0.74,
respectively, which indicates that~\eqref{eq:dfinal} is an accurate
model. Different from the previous work~\cite{liu2018model}, in this
paper, both the color and the geometry distortion are considered as
given in (9).
\begin{table}[t!]
\centering \caption{Accuracy of Proposed Distortion
Model~\eqref{eq:dfinal}} \label{tab:dmodel}
  \begin{tabular}{c|c|c}
      \toprule
      Point Cloud & SCC & RMSE \\\hline
      \emph{Andrew}  &1.00    &0.59 \\
      \emph{David}   &1.00    &0.09 \\
      \emph{Phil}    &1.00    &0.42 \\
      \emph{Ricardo} &1.00    &0.07 \\
      \emph{Longdress}   &0.97    &2.21 \\
      \emph{Loot}    &0.99    &0.34 \\
      \emph{Queen}   &0.96    &1.76 \\
      \emph{Redandblack} &0.99    &0.47 \\\hline
      Average &0.99    &0.74 \\
      \bottomrule
  \end{tabular}
\end{table}
\subsection{Rate Model}  
\begin{figure}[t!]
\centering \subfigure[]{ \label{fig4:subfig:a}
\includegraphics[width=0.48\columnwidth]{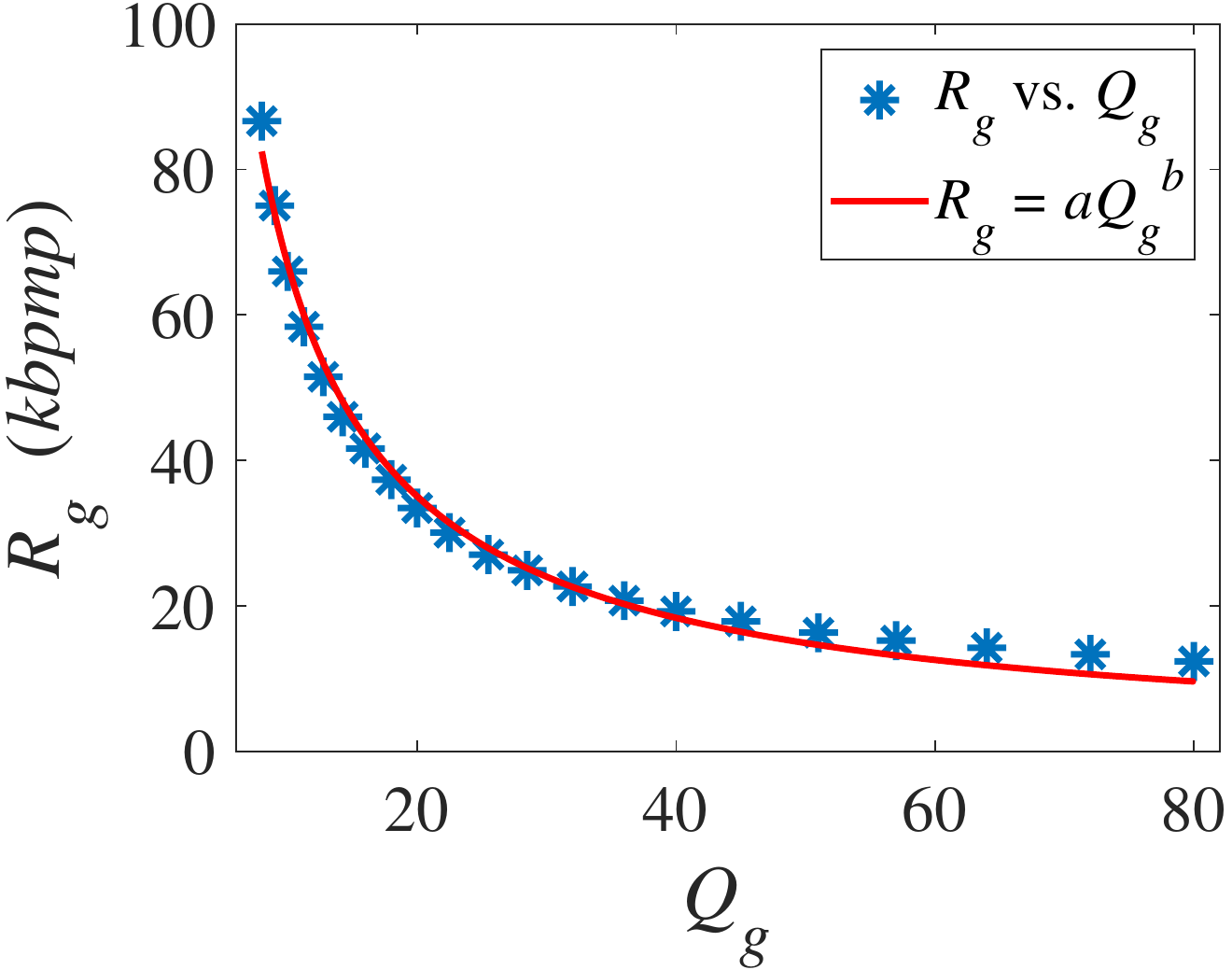}}
\subfigure[]{ \label{fig4:subfig:b}
\includegraphics[width=0.48\columnwidth]{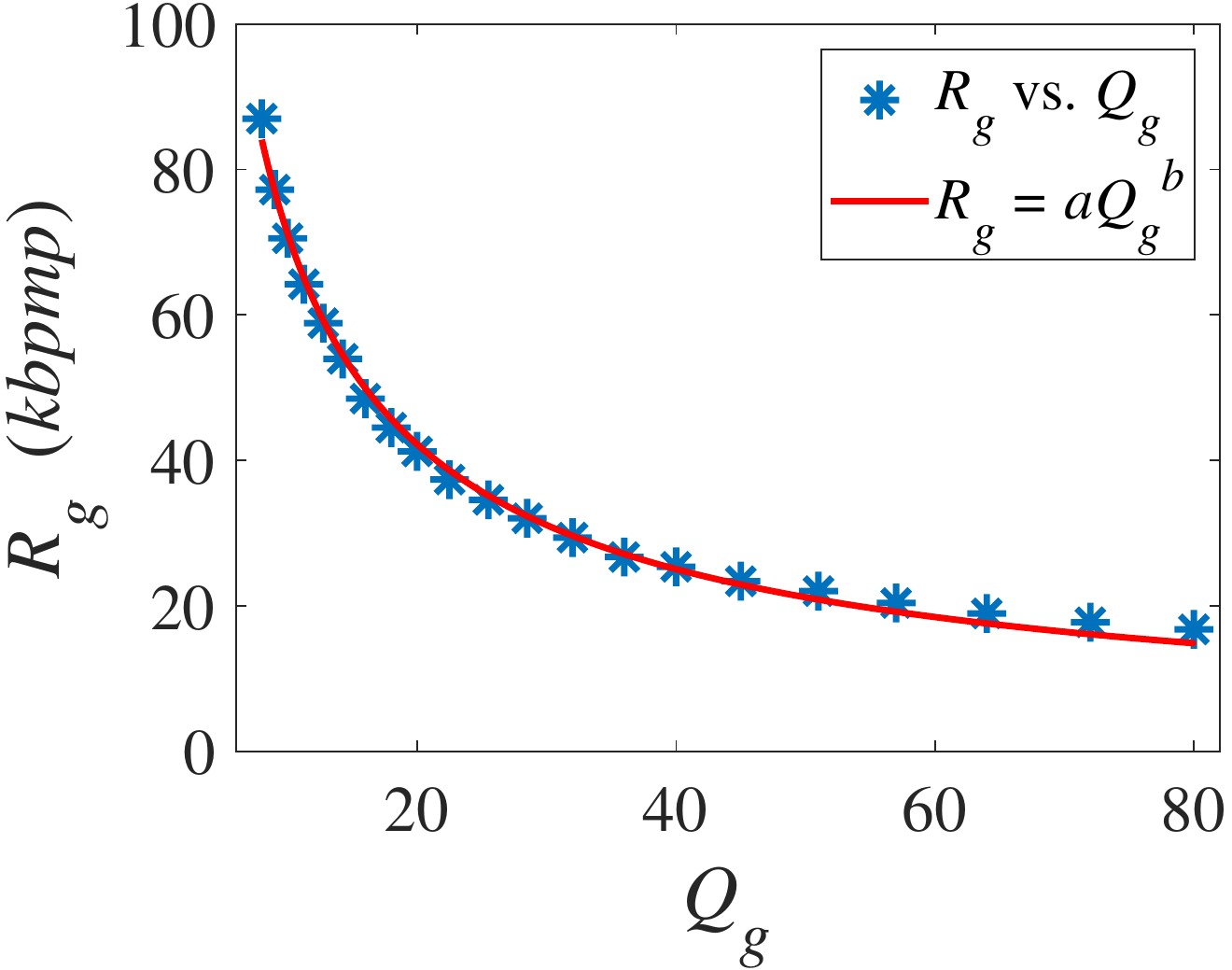}}
\subfigure[]{ \label{fig4:subfig:c}
\includegraphics[width=0.48\columnwidth]{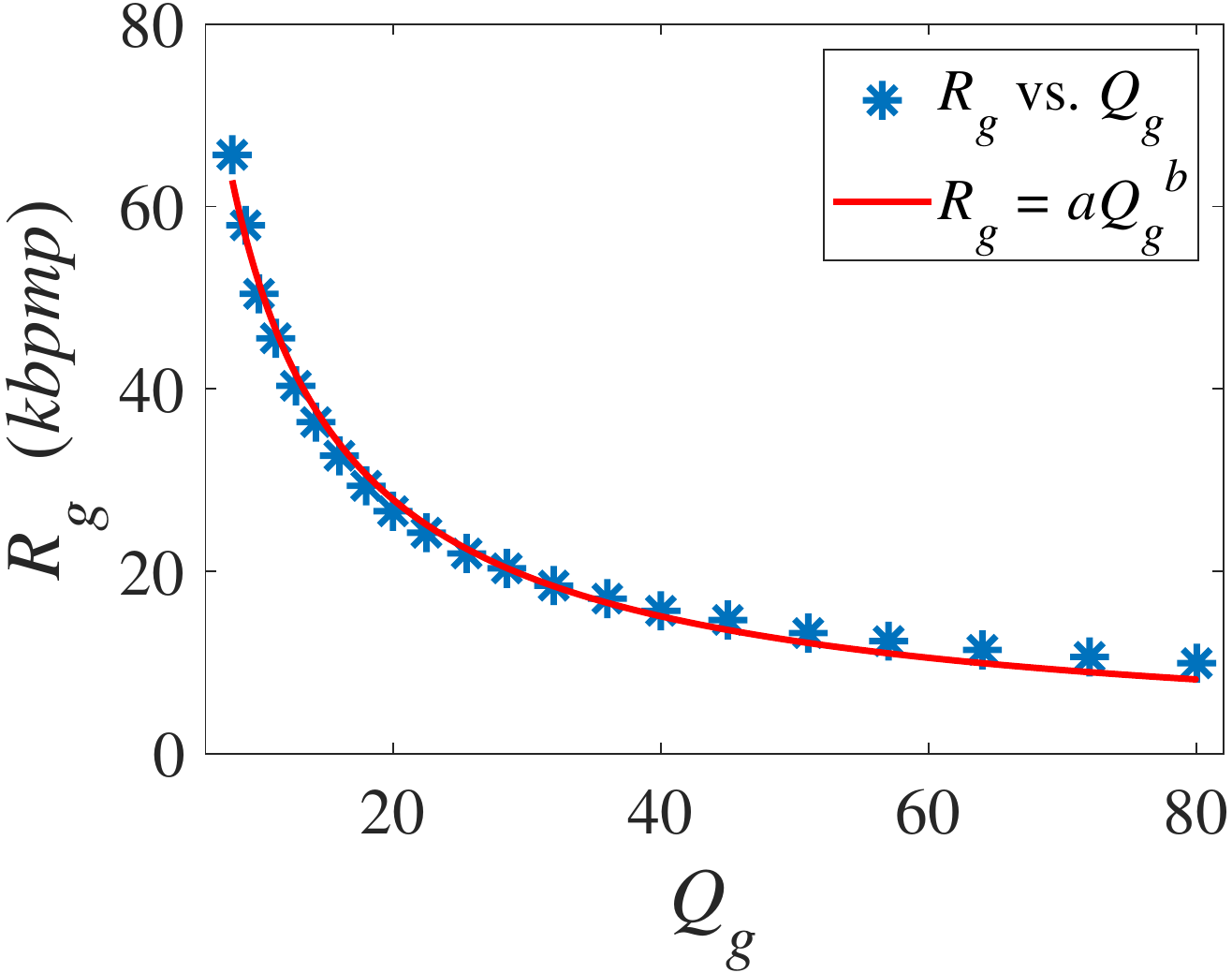}}
\subfigure[]{ \label{fig4:subfig:d}
\includegraphics[width=0.48\columnwidth]{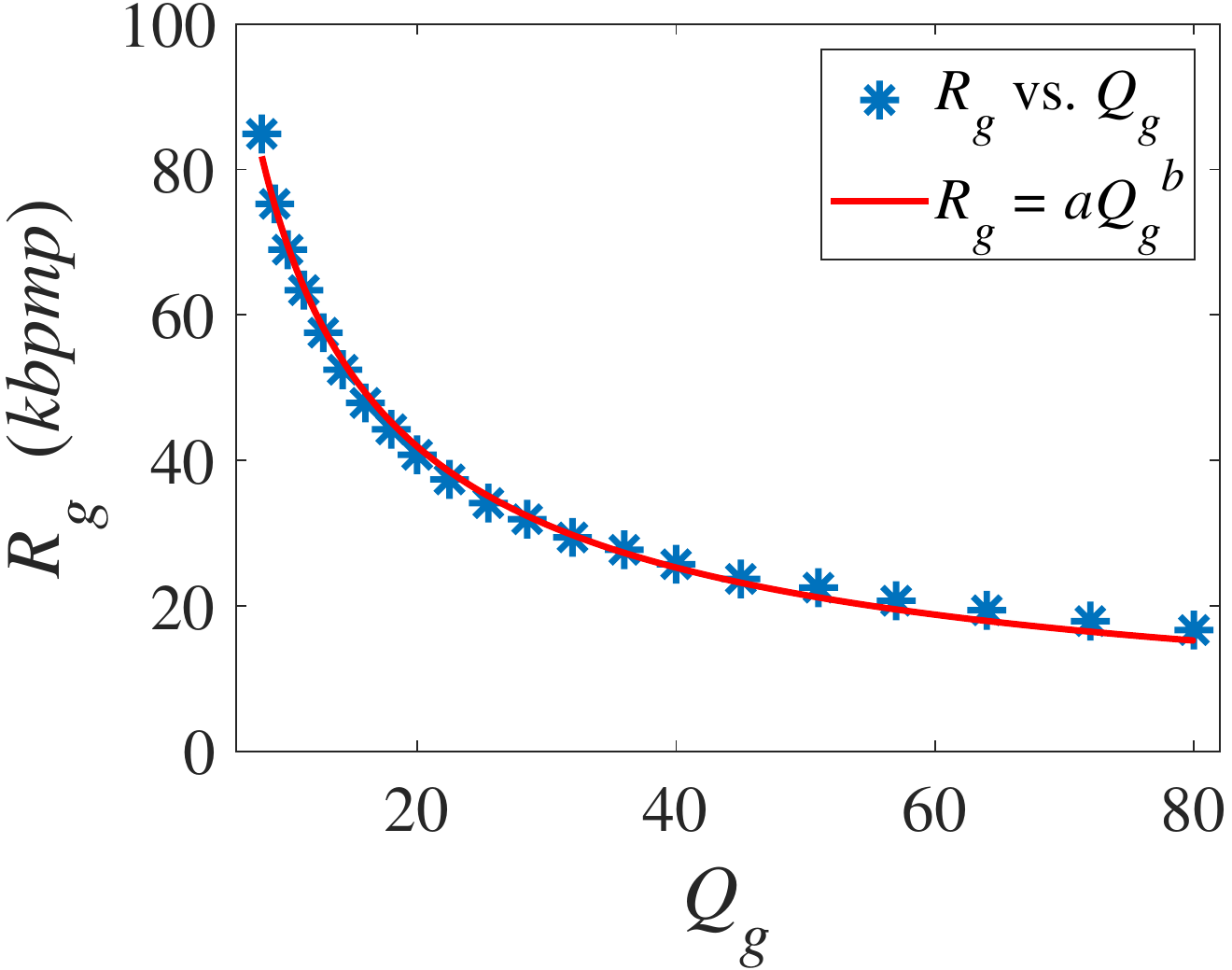}}
\caption{Illustration of the relationship between $R_g$ and $Q_g$.
(a) \emph{Phil}, (b) \emph{Longdress}, (c) \emph{David}, (d)
\emph{Loot}.} \label{fig4}
\end{figure}

The total bitrate $R$ is the sum of the geometry bitrate and color
bitrate, i.e.,
\begin{equation}
\label{eq:rate}
\begin{aligned}
R &=R_g + R_c,
\end{aligned}
\end{equation}
where $R_g$ is the geometry bitrate, which depends only on $Q_g$,
whereas $R_c$ is the color bitrate, which depends on both $Q_g$ and
$Q_c$. For $R_g$, we used the Cauchy-based rate
model~\cite{cen2014efficient}:
\begin{equation}
\label{eq:grgq} R_g=\gamma_g  Q_g^{\theta_g},
\end{equation}
where $\gamma_g$ and $\theta_g$ are model parameters. Because the
bitrate of a 3D point cloud is relatively large, we used kilobits
per million points (\emph{kbpmp}) as the bitrate unit.
Fig.~\ref{fig4} shows the results of statistical experiments to
verify the accuracy of~\eqref{eq:grgq}.
From this figure, we can observe that the $R_g$
model~\eqref{eq:grgq} is appropriate. This is confirmed by
Table~\ref{tab:grmodel}, which shows that the SCCs between the
actual $R_g$ and the fitted values are always larger than or equal
to 0.99.
\begin{figure}[t!]
\centering \subfigure[]{ \label{fig5:subfig:a}
\includegraphics[width=0.48\columnwidth]{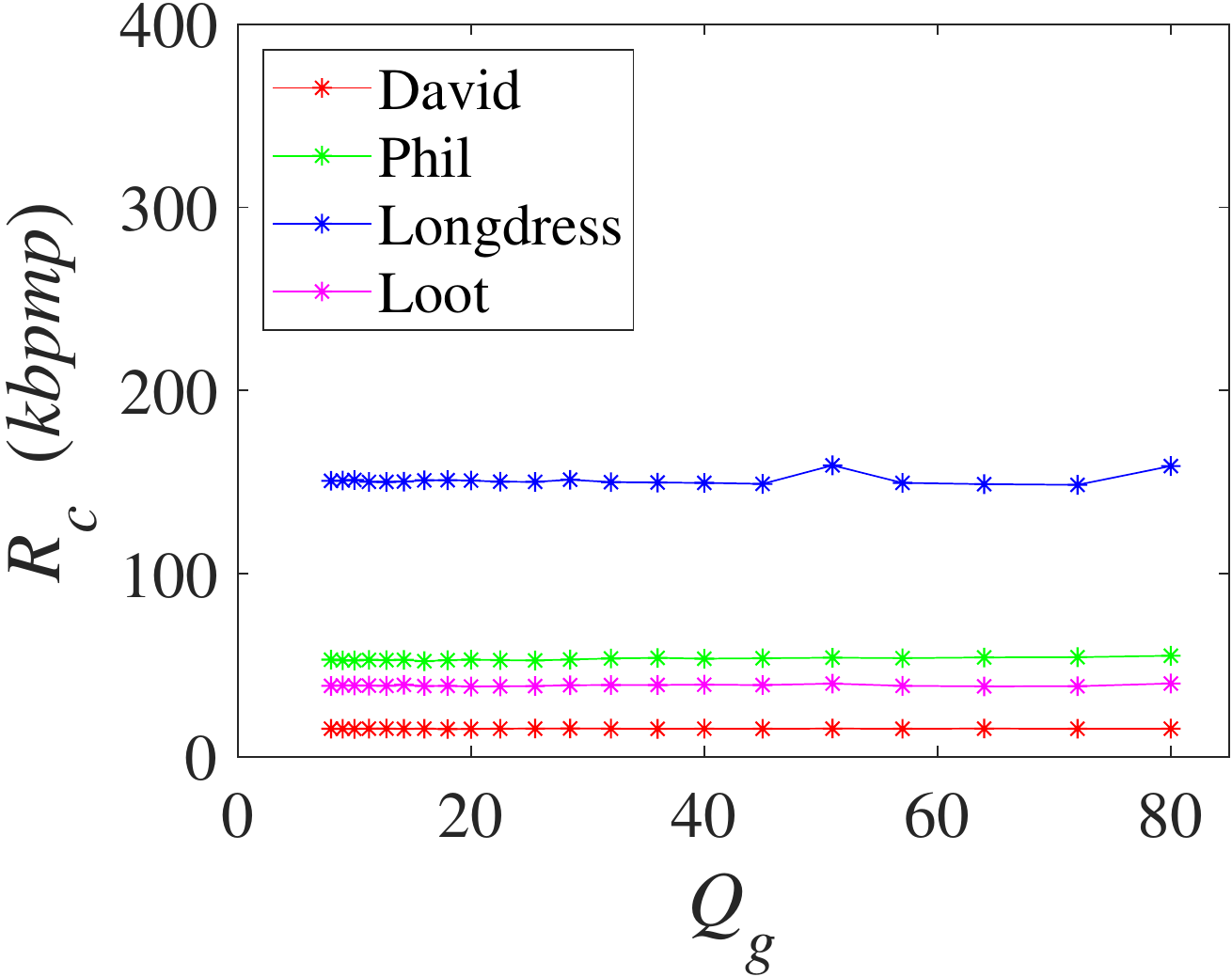}}
\subfigure[]{ \label{fig5:subfig:b}
\includegraphics[width=0.48\columnwidth]{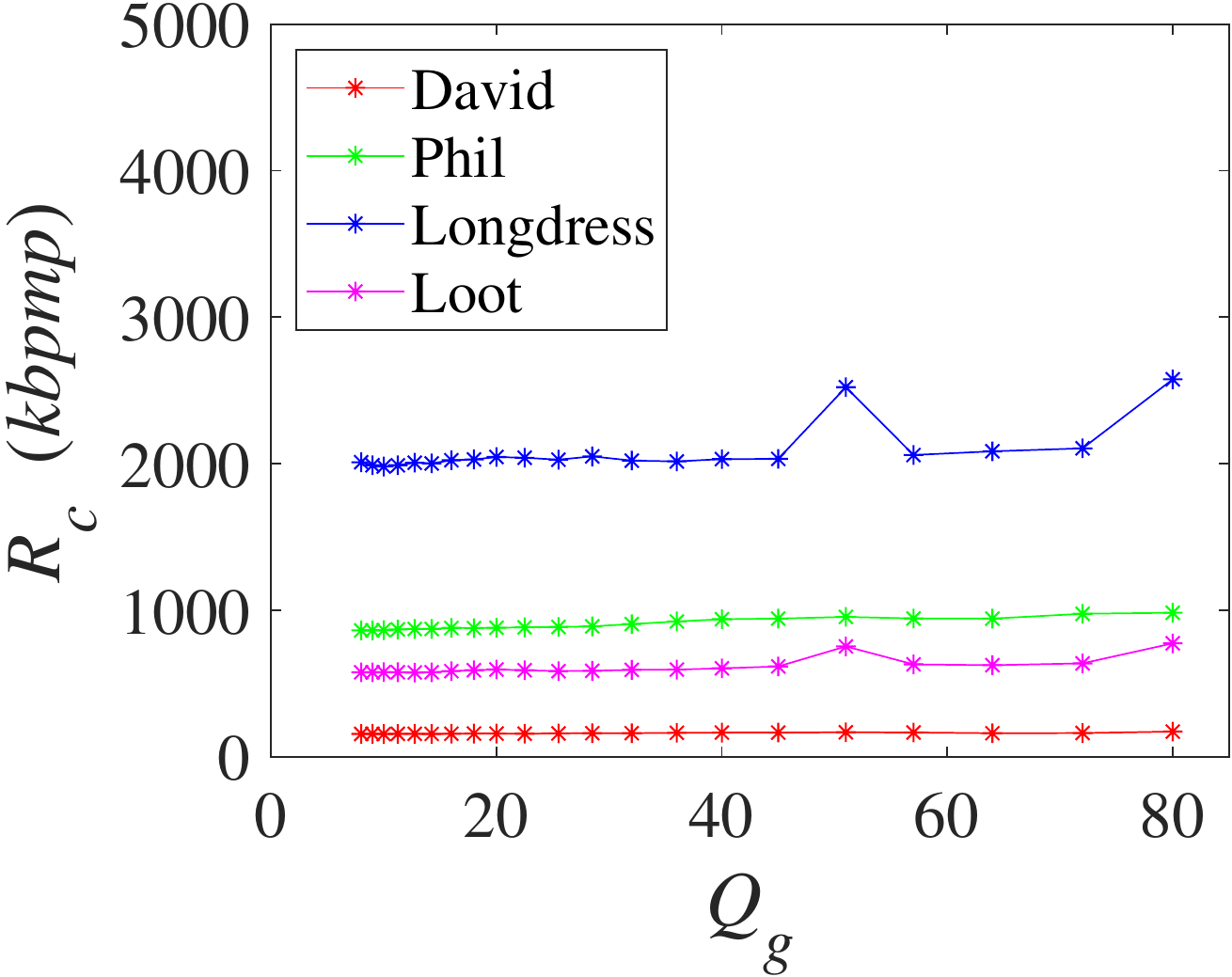}}
\caption{Influence of $Q_g$ on $R_c$. (a) $Q_c$=80,
$Q_g=8,9,\dots,80$; (b) $Q_c$=8, $Q_g=8,9,\dots,80$.} \label{fig5}
\end{figure}
\begin{figure}[t!]
\centering \subfigure[]{ \label{fig6:subfig:a}
\includegraphics[width=0.48\columnwidth]{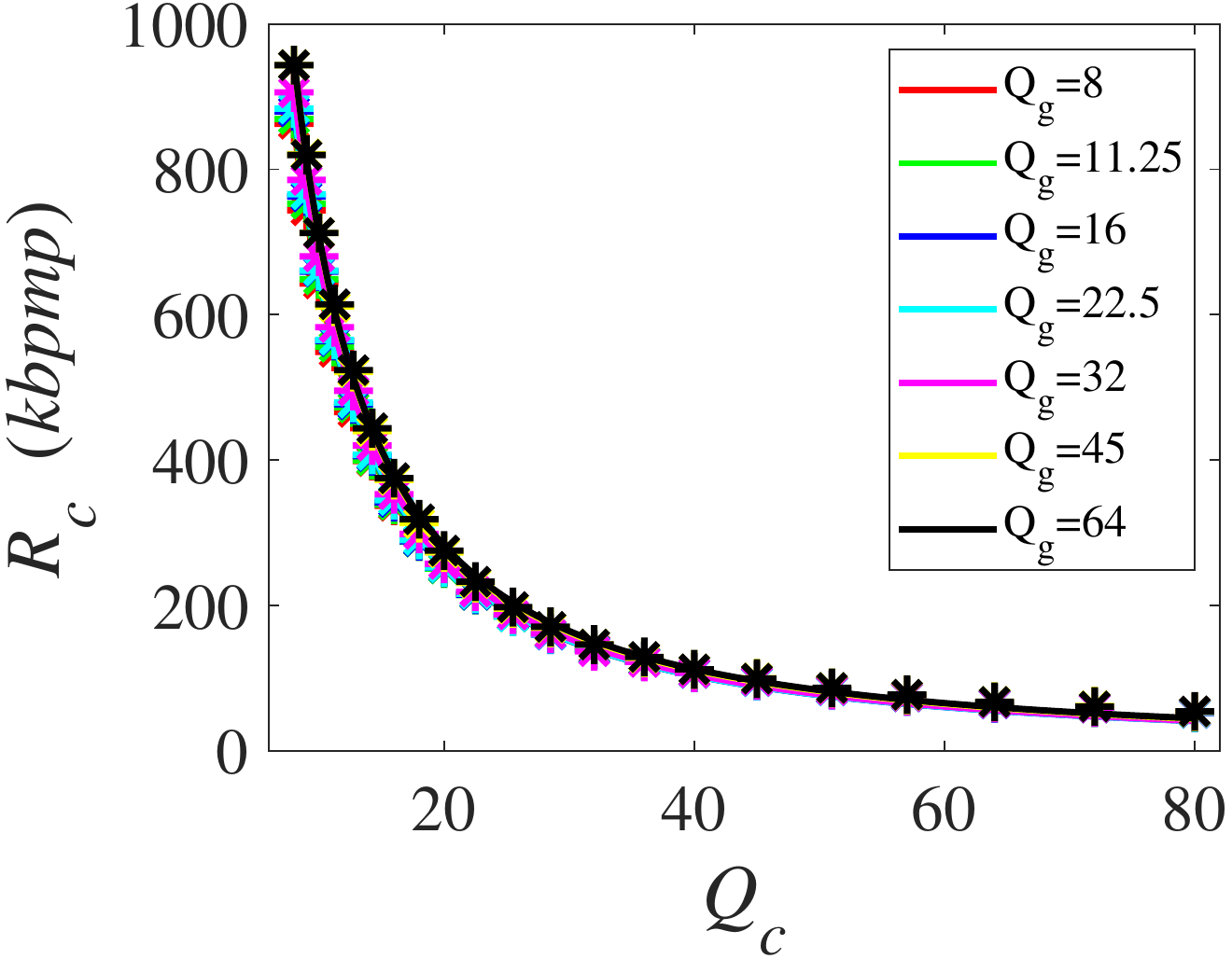}}
\subfigure[]{ \label{fig6:subfig:b}
\includegraphics[width=0.48\columnwidth]{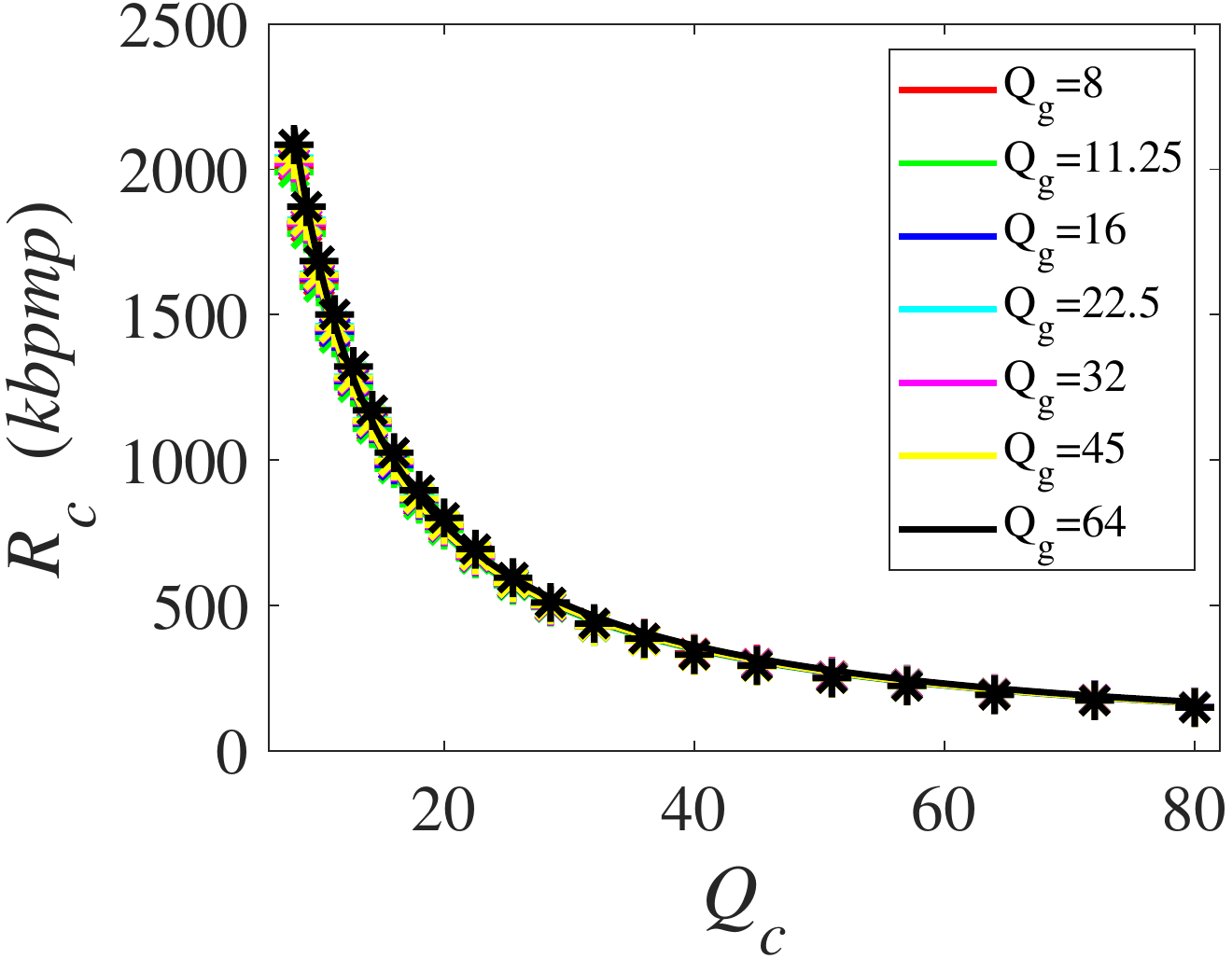}}
\subfigure[]{ \label{fig6:subfig:c}
\includegraphics[width=0.48\columnwidth]{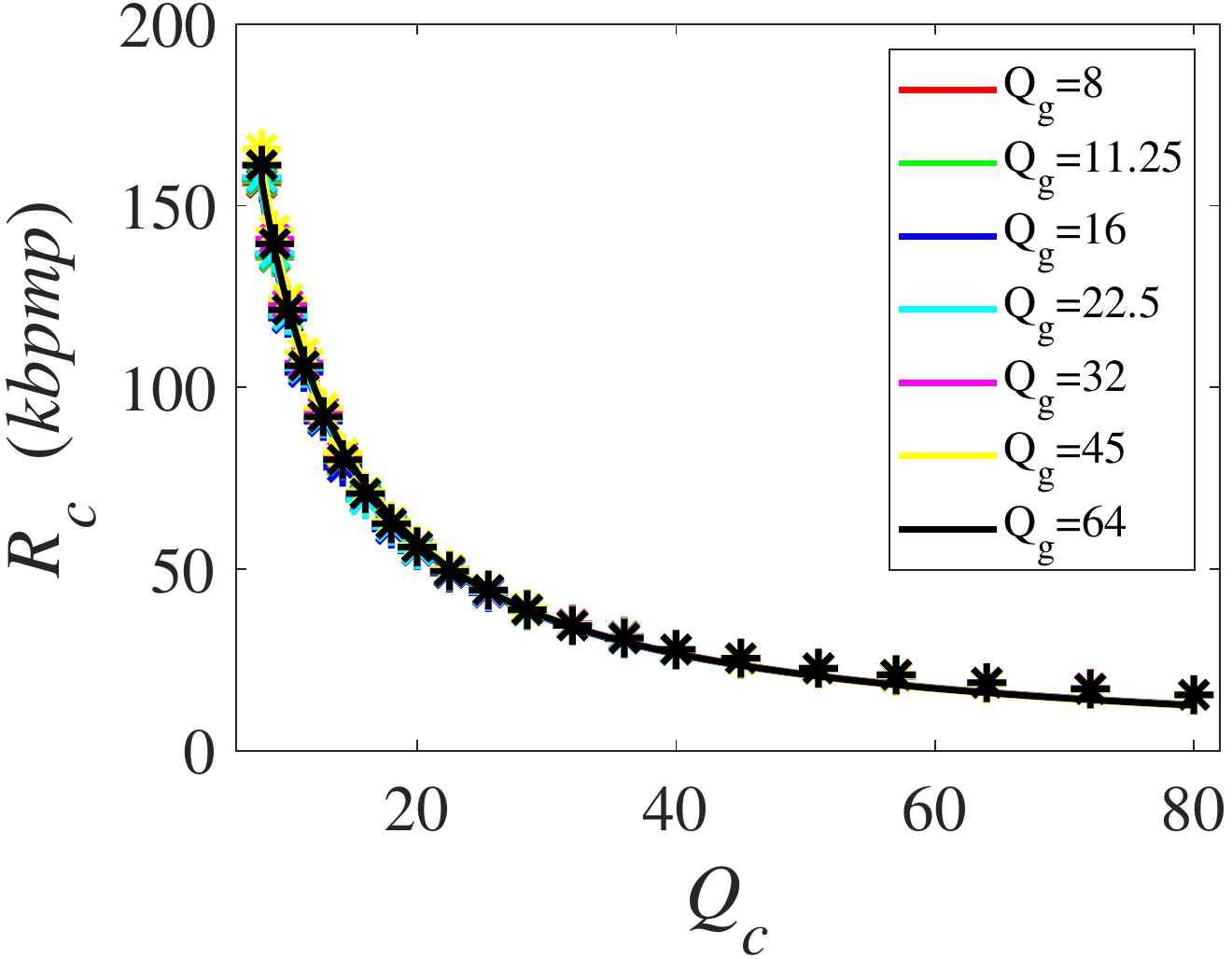}}
\subfigure[]{ \label{fig6:subfig:d}
\includegraphics[width=0.48\columnwidth]{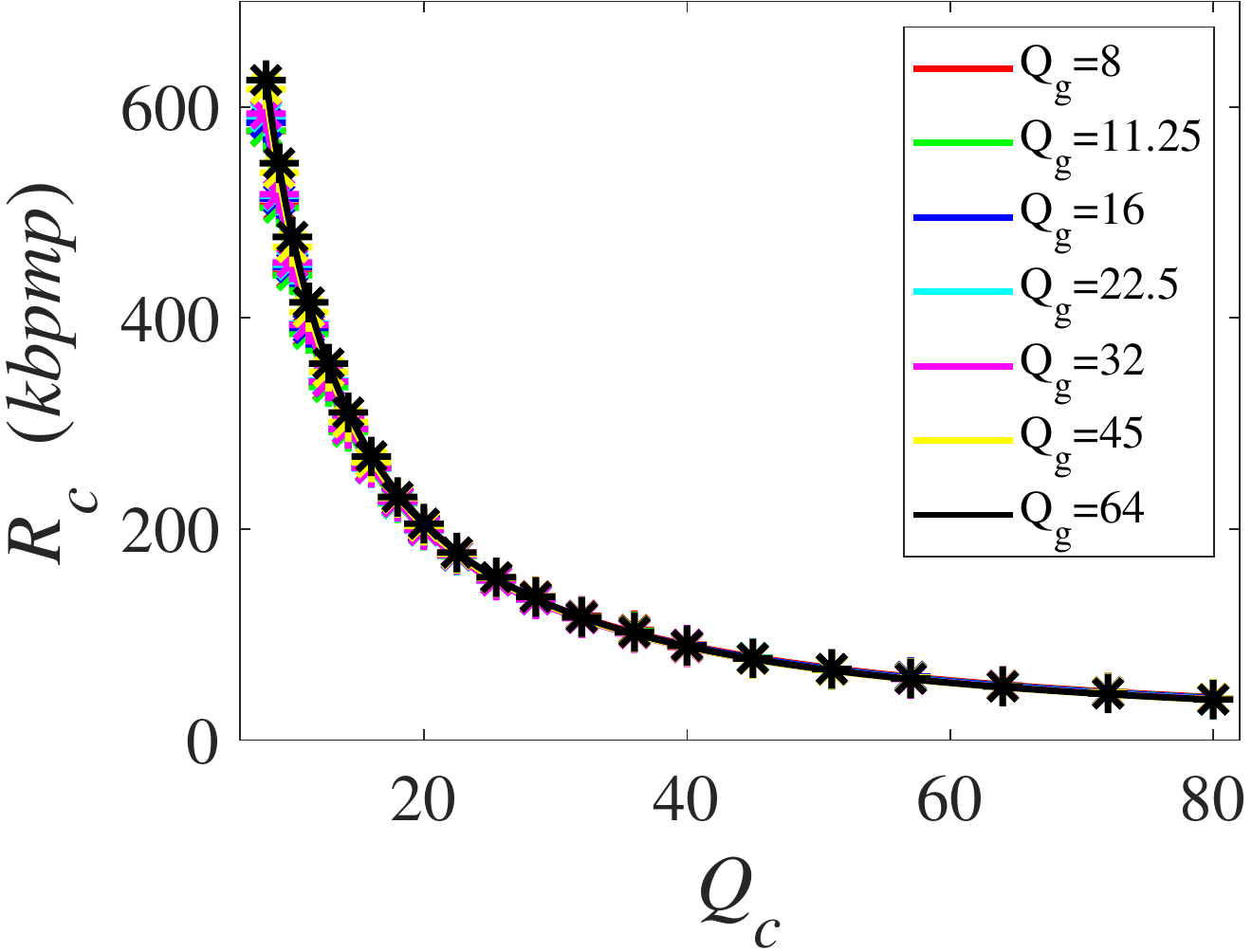}}
\caption{Illustration of the relationship between $R_c$ and $Q_c$
($R_c=\gamma_c Q_c^{\theta_c}$). (a) \emph{Phil}, (b)
\emph{Longdress}, (c) \emph{David}, (d) \emph{Loot}.} \label{fig6}
\end{figure}
\begin{table}[t!]
\newcommand{\tabincell}[2]{\begin{tabular}{@{}#1@{}}#2\end{tabular}}
\centering \caption{Verification of $R_g$ Model~\eqref{eq:grgq}}
\label{tab:grmodel}
  \begin{tabular}{c|c|c|c|c}
      \toprule
      Point Cloud & SCC & \tabincell{l}{RMSE\\(\emph{kbpmp})} &\tabincell{l}{ $\rm Rate_{max}$\\(\emph{kbpmp})} & NRMSE\\\hline
      \emph{Andrew}  &0.99    &1.53 &75.89 &0.02\\
      \emph{David}   &0.99    &1.33 &65.66 &0.02 \\
      \emph{Phil}    &0.99    &1.93 &86.64 &0.02 \\
      \emph{Ricardo} &0.99    &1.40 &67.48 &0.02 \\
      \emph{Longdress}   &1.00    &1.21 &86.99 &0.01\\
      \emph{Loot}    &1.00    &1.23 &84.89 &0.01 \\
      \emph{Queen}   &1.00    &1.11 &69.23 &0.02 \\
      \emph{Redandblack} &1.00    &1.53 &102.74 &0.01 \\
      \bottomrule
  \end{tabular}
\end{table}
Because the bitrate of a 3D point cloud is typically very large, the
RMSE seems to be large. Therefore, we also calculated the normalized
RMSE to illustrate the fitting error effectively. The normalized
RMSE ($NRMSE$) is defined as
\begin{equation}
\label{eq:nrmse}
\begin{aligned}
NRMSE &=\frac{RMSE}{Rate_{\max}},
\end{aligned}
\end{equation}
where $Rate_{\max}$ is the maximum bitrate. Table~\ref{tab:grmodel}
shows that the NRMSE is as low as 0.01, which confirms
that~\eqref{eq:grgq} is accurate.

To study the effect of $Q_g$ on $R_c$, we compressed the color
information with a fixed $Q_c$ and multiple $Q_g$s. The results,
shown in Fig.~\ref{fig5}, indicate that the effect of $Q_g$ on $R_c$
is negligible. This is also confirmed by Fig.~\ref{fig6}, which shows
$R_c$ as a function of $Q_c$ for various $Q_g$s.
\begin{figure*}[t!]
\centering \subfigure[]{ \label{fig7:subfig:a}
\includegraphics[width=0.48\columnwidth, height=3.5cm]{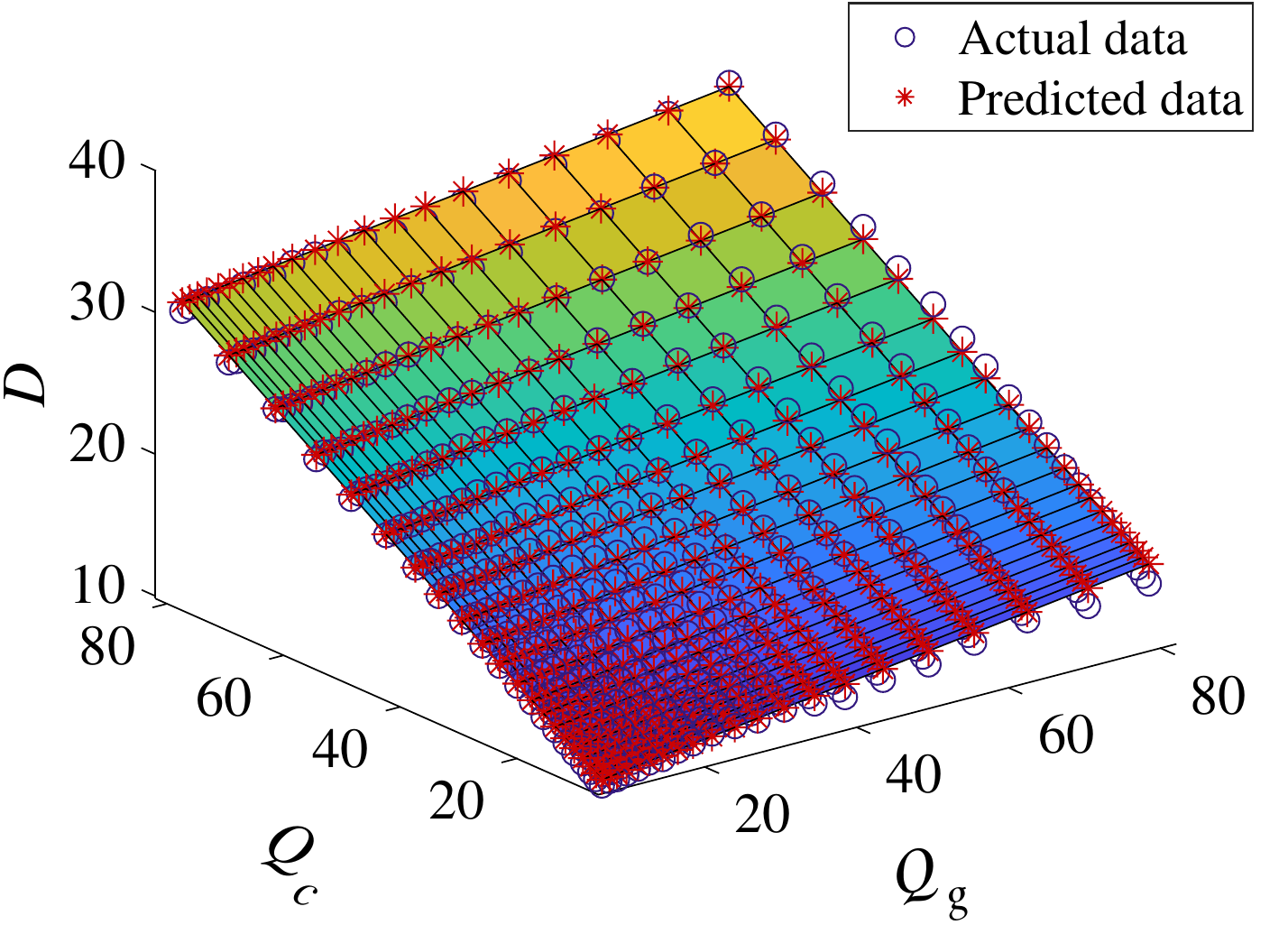}}
\subfigure[]{ \label{fig7:subfig:b}
\includegraphics[width=0.48\columnwidth, height=3.5cm]{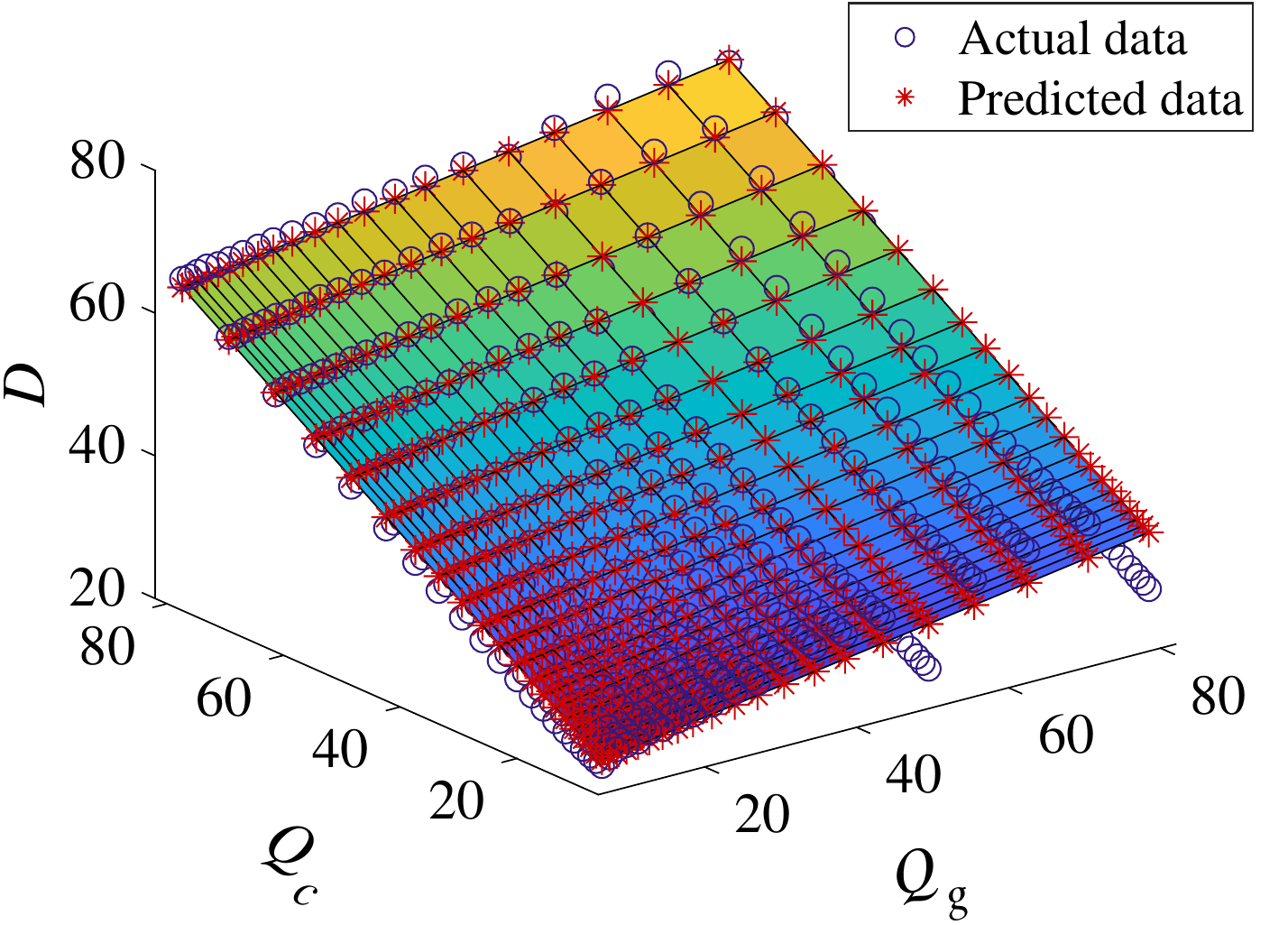}}
\subfigure[]{ \label{fig7:subfig:c}
\includegraphics[width=0.48\columnwidth, height=3.5cm]{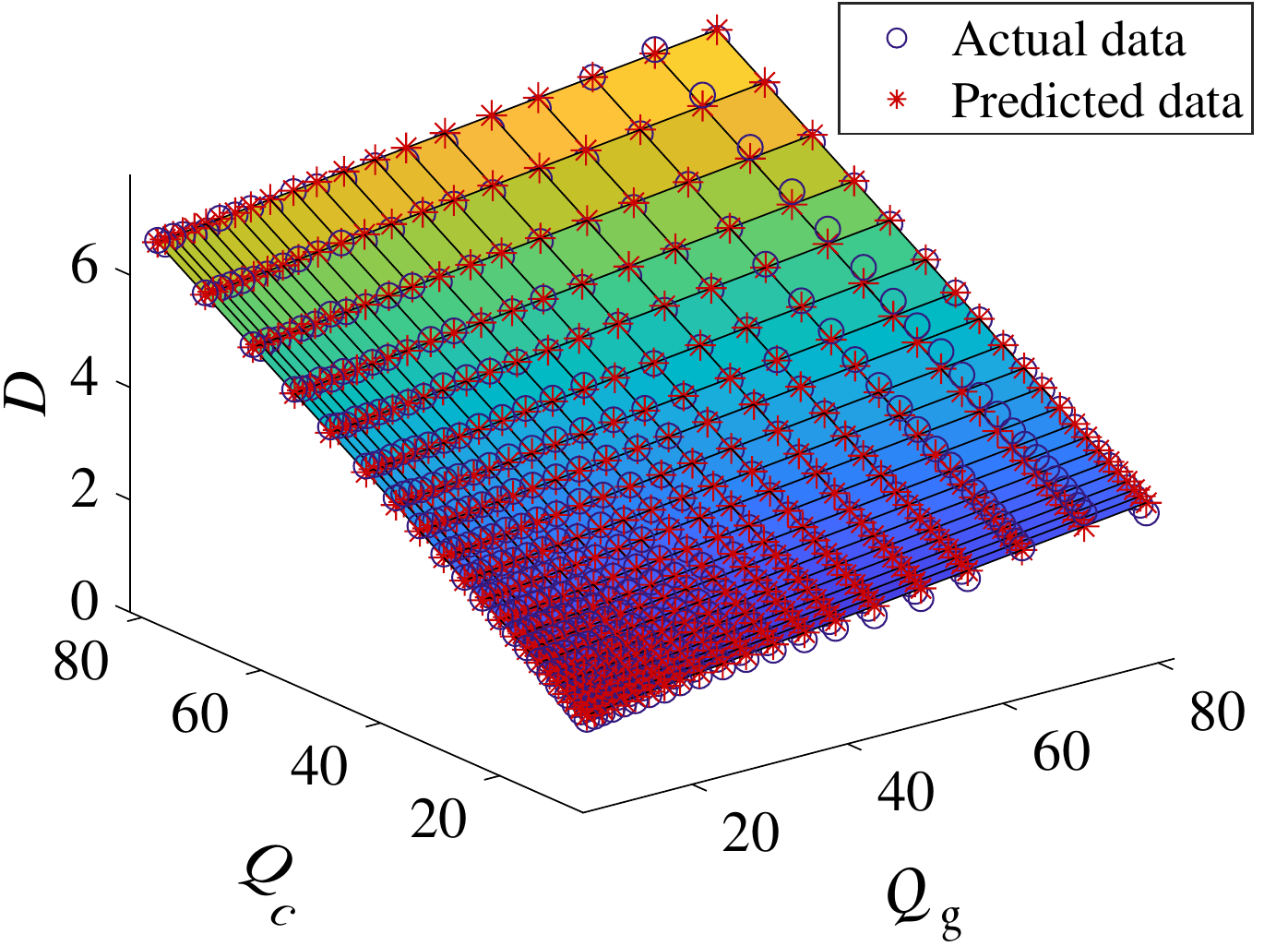}}
\subfigure[]{ \label{fig7:subfig:d}
\includegraphics[width=0.48\columnwidth, height=3.5cm]{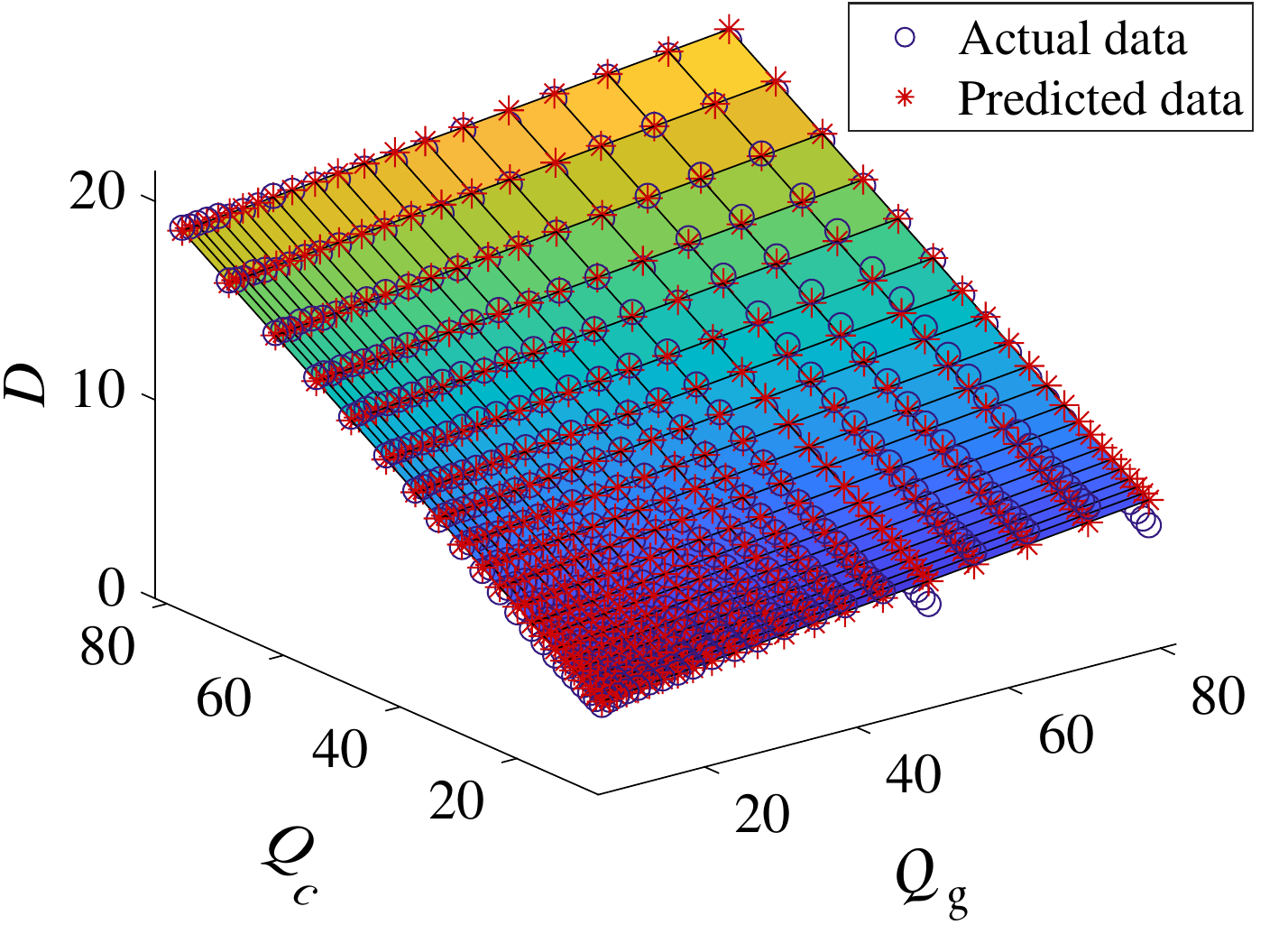}}
\subfigure[]{ \label{fig7:subfig:e}
\includegraphics[width=0.48\columnwidth, height=3.5cm]{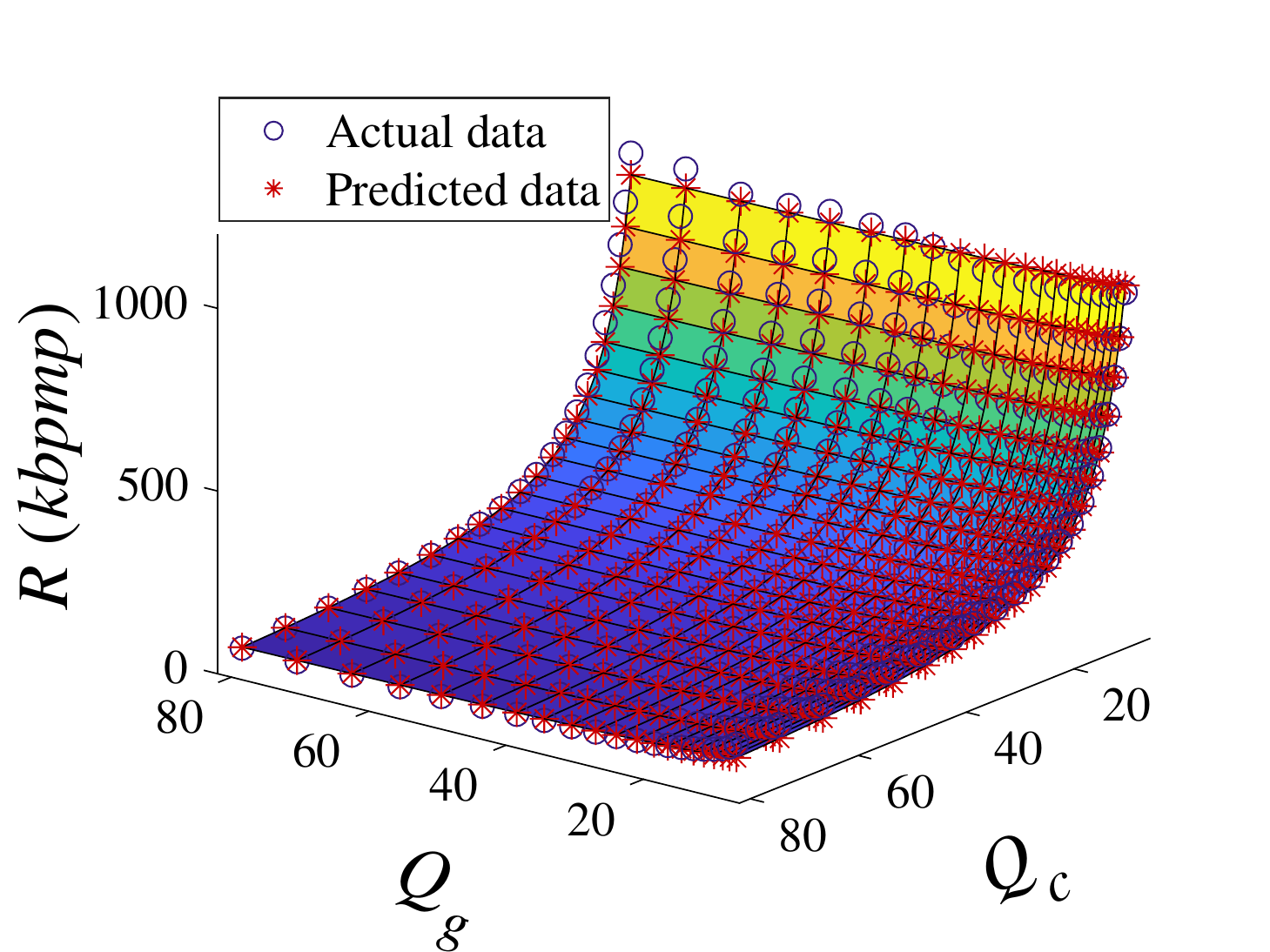}}
\subfigure[]{ \label{fig7:subfig:f}
\includegraphics[width=0.48\columnwidth, height=3.5cm]{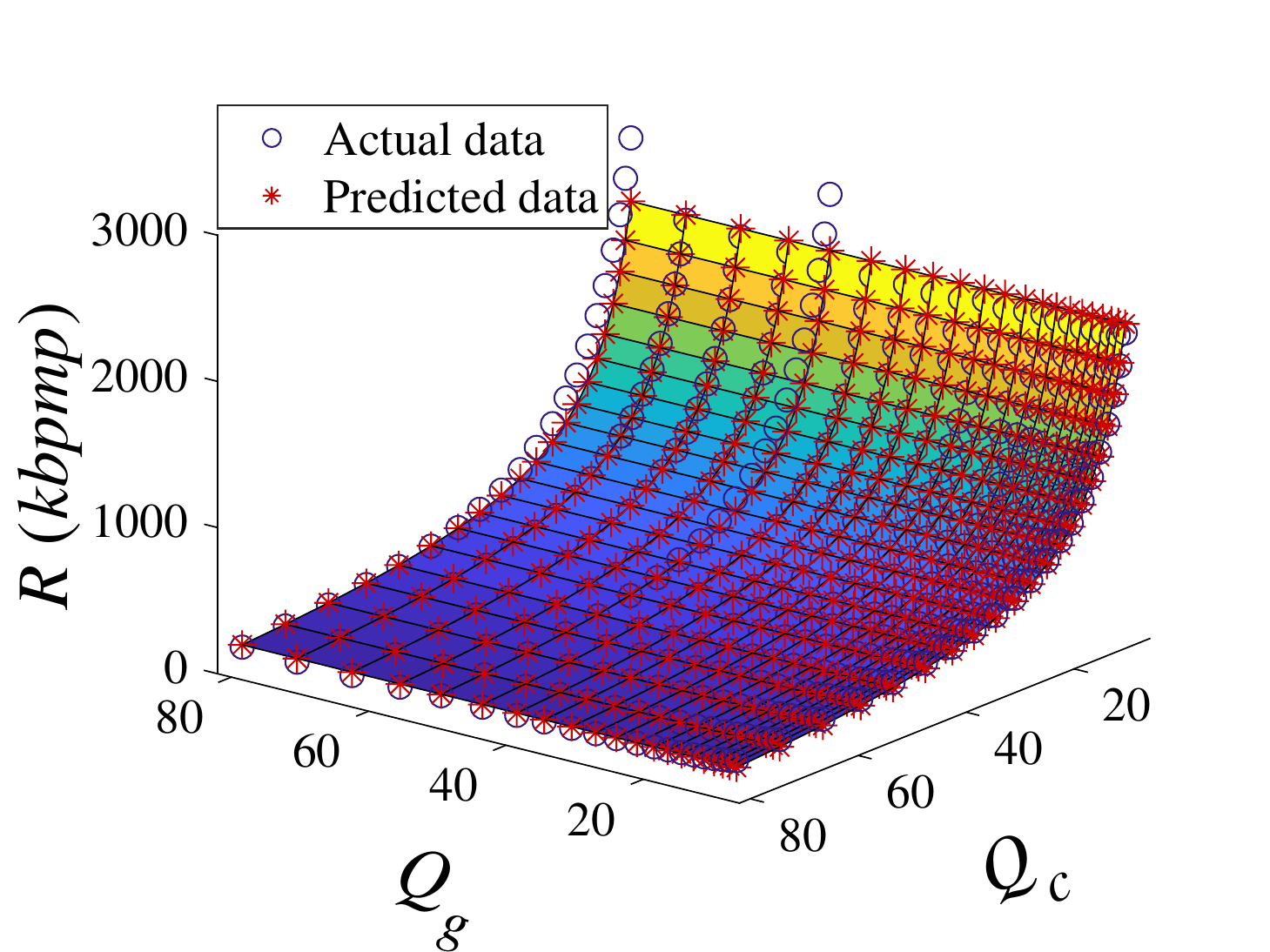}}
\subfigure[]{ \label{fig7:subfig:g}
\includegraphics[width=0.48\columnwidth, height=3.5cm]{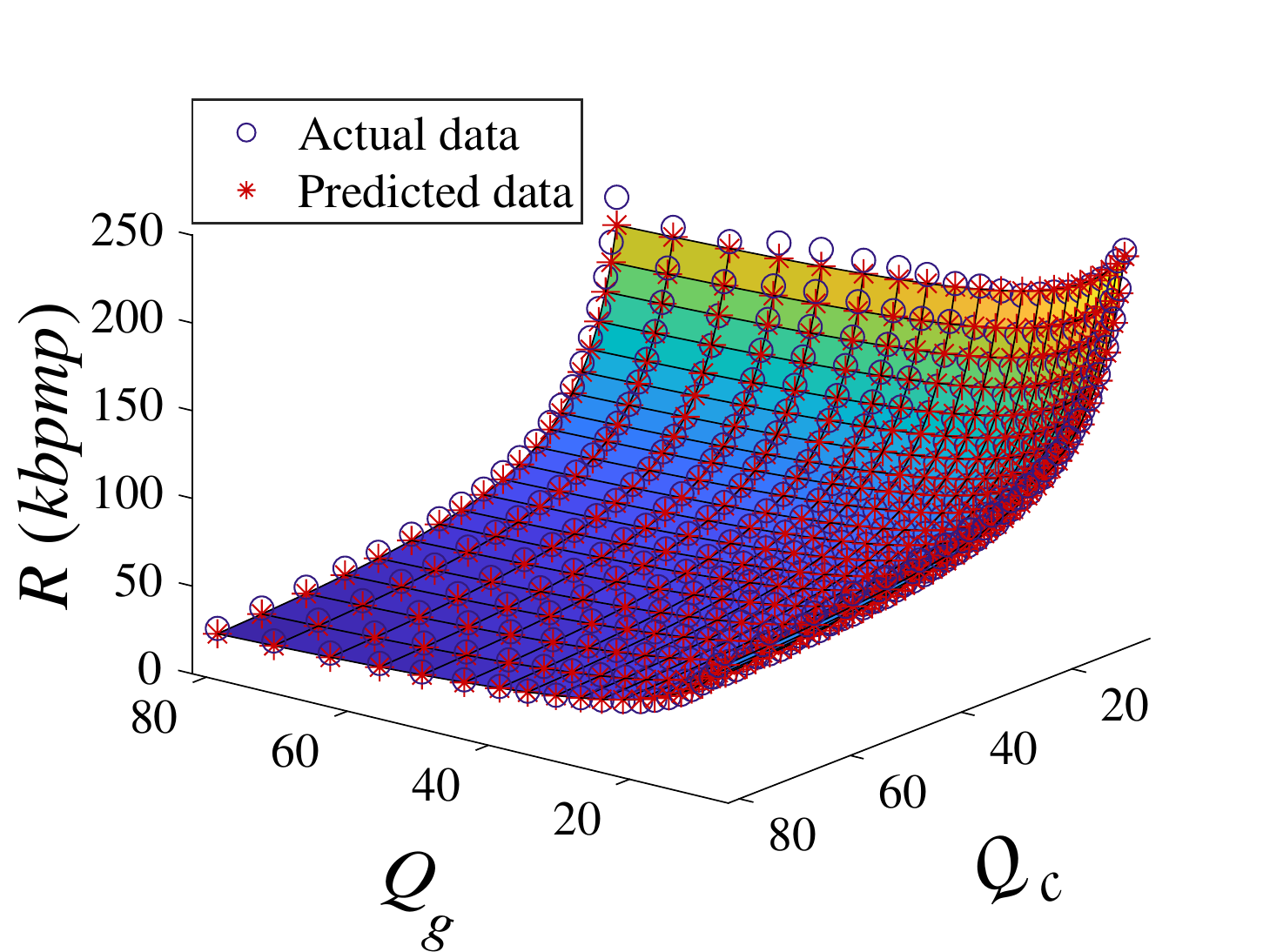}}
\subfigure[]{ \label{fig7:subfig:h}
\includegraphics[width=0.48\columnwidth, height=3.5cm]{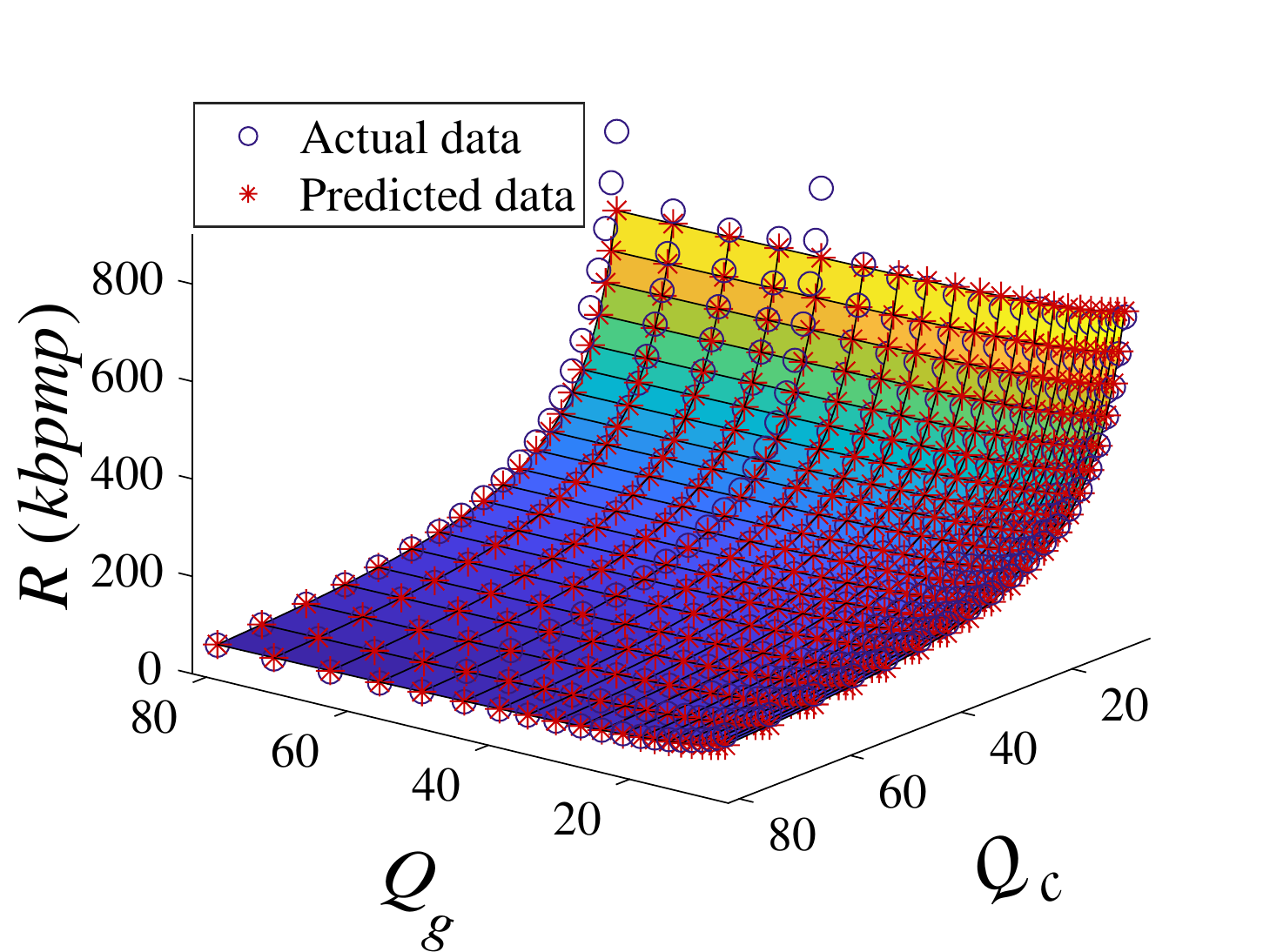}}
\caption{Illustration of the accuracy of the $R$
model~\eqref{eq:ratefinal} and $D$ model~\eqref{eq:dfinal}. (a)-(d):
accuracy of the $D$ model~\eqref{eq:dfinal} for \emph{Phil},
\emph{Longdress}, \emph{David}, and \emph{Loot}, (e)-(f): accuracy
of the $R$ model~\eqref{eq:ratefinal} for \emph{Phil},
\emph{Longdress}, \emph{David}, and \emph{Loot}.} \label{fig7}
\end{figure*}
Thus, for simplicity, we can assume that $R_c$ is only affected by
$Q_c$. Moreover, Fig.~\ref{fig6} suggests that the exponential model
\begin{equation}
\label{eq:crcq} R_c=\gamma_c  Q_c^{\theta_c},
\end{equation}
where $\gamma_c$ and $\theta_c$ are model parameters is appropriate
to describe the relationship between $Q_c$ and $R_c$. This is
confirmed in Table~\ref{tab:crmodel}, which shows that the SCC of
the relationship between $R_c$ and $Q_c$ is always greater than or
equal to 0.98, while the NRMSE is always smaller than 0.03.
\begin{table}[t!]
\newcommand{\tabincell}[2]{\begin{tabular}{@{}#1@{}}#2\end{tabular}}
\centering \caption{Verification of $R_c$ Model~\eqref{eq:crcq}}
\label{tab:crmodel}
  \begin{tabular}{c|c|c|c|c}
      \toprule
      Point Cloud & SCC & \tabincell{l}{RMSE\\(\emph{kbpmp})} &\tabincell{l}{ $\rm Rate_{max}$\\(\emph{kbpmp})} & NRMSE\\\hline
      \emph{Andrew}  &1.00    &21.78   &1192.01     &0.02\\
      \emph{David}   &1.00    &2.59    &172.32  &0.02 \\
      \emph{Phil}    &0.99    &19.10   &983.27  &0.02 \\
      \emph{Ricardo} &0.99    &3.41    &179.67  &0.02 \\
      \emph{Longdress}   &0.98    &76.67   &2576.12     &0.03\\
      \emph{Loot}    &0.98    &20.96   &774.43  &0.03 \\
      \emph{Queen}   &0.99    &15.94   &732.89  &0.02 \\
      \emph{Redandblack} &0.99    &30.21   &1182.12    &0.03 \\
      \bottomrule
  \end{tabular}
\end{table}
Accordingly,~\eqref{eq:rate} can be rewritten as
\begin{equation}
\label{eq:ratefinal}
\begin{aligned}
R &=R_g + R_c\\
&=\gamma_g  Q_g^{\theta_g}+\gamma_c  Q_c^{\theta_c}.
\end{aligned}
\end{equation}
 Different from our previous work~\cite{liu2018model}, the rate of color and geometry are considered separately in this paper, which may lead to more accurate bit allocation result. Table~\ref{tab:rmodel} validates~\eqref{eq:ratefinal} by showing that the SCC was always greater than or equal to 0.99, and the NRMSE was always smaller than or equal to 0.03.
\begin{table}[t!]
\newcommand{\tabincell}[2]{\begin{tabular}{@{}#1@{}}#2\end{tabular}}
\centering \caption{Accuracy of Rate Model~\eqref{eq:ratefinal}}
\label{tab:rmodel}
  \begin{tabular}{c|c|c|c|c}
      \toprule
      Point Cloud & SCC & \tabincell{l}{RMSE\\(\emph{kbpmp})} &\tabincell{l}{ $\rm Rate_{max}$\\(\emph{kbpmp})} & NRMSE\\\hline
      \emph{Andrew}  &1.00    &17.85   &1202.71     &0.01\\
      \emph{David}   &1.00    &2.97    &221.89  &0.01 \\
      \emph{Phil}    &1.00    &17.22   &995.64  &0.02\\
      \emph{Ricardo} &0.99    &3.94    &228.18  &0.02 \\
      \emph{Longdress}  &0.99    &71.07   &2592.90     &0.03\\
      \emph{Loot}    &0.99    &20.36   &791.11  &0.03 \\
      \emph{Queen}   &1.00    &11.60   &745.44  &0.02\\
      \emph{Redandblack}&0.99    &29.30   &1201.44     &0.02\\
      \bottomrule
  \end{tabular}
\end{table}
Finally, Fig.~\ref{fig7} illustrates the accuracy of
models~\eqref{eq:ratefinal} and~\eqref{eq:dfinal} by comparing the
actual values to the values predicted by the models.
\section{Model-based Optimal Bit Allocation Algorithm}
Based on the analysis in Section~\ref{sec:2}, the optimal bit
allocation problem~\eqref{eq:rdtotal} can be converted to the
problem of finding the optimal solution of the constrained
optimization problem
\begin{equation}
\label{eq:rdo2}
\begin{aligned}
\min_{(\emph{$Q_g$},\emph{$Q_c$})} &\quad a Q_g +b Q_c +c \\
\mathrm{s.t.} \quad &\gamma_g  Q_g^{\theta_g}+\gamma_c  Q_c^{\theta_c} \leq \emph{$R_T$}. \\
\end{aligned}
\end{equation}
To solve~\eqref{eq:rdo2}, we first need to determine the model
parameters $a$, $b$, $c$, $\gamma_g $, $\theta_g$, $\gamma_c$, and
$\theta_c$. This is done by encoding the 3D point cloud with three
different pairs of quantization steps ($Q_{g,1}$,$Q_{c,1}$),
($Q_{g,2}$,$Q_{c,2}$), ($Q_{g,3}$,$Q_{c,3}$) and solving the systems
of equations~\eqref{eq:deqs} and~\eqref{eq:reqs}:
\begin{equation}
\label{eq:deqs}
\begin{aligned}
\begin{cases}
D_1&=a  Q_{g,1}+ b   Q_{c,1} +c \\
D_2&=a  Q_{g,2}+ b   Q_{c,2} +c \\
D_3&=a  Q_{g,3}+ b   Q_{c,3} +c
\end{cases}
\end{aligned}
\end{equation}
\begin{equation}
\label{eq:reqs}
\begin{aligned}
\begin{cases}
R_{g,1}&= \gamma_g   Q_{g,1}^{\theta_g}\\
R_{g,2}&= \gamma_g   Q_{g,2}^{\theta_g} \\
R_{c,1}&= \gamma_c   Q_{c,1}^{\theta_c} \\
R_{c,2}&= \gamma_c   Q_{c,2}^{\theta_c}
\end{cases}
\end{aligned},
\end{equation}
where $D_1$, $D_2$, $D_3$ are the corresponding distortions and
$R_{g,1}$, $R_{g,2}$, $R_{c,1}$, $R_{c,2}$ are the corresponding
geometry and color bitrates, respectively. Because both the
objective function and the constraint function in~\eqref{eq:rdo2}
are convex, the optimal quantization steps, $Q_{g,opt}$ and
$Q_{c,opt}$ can be obtained by using an interior point
method~\cite{zhang1998solving}. In this method, the convex
optimization problem is first converted to an unconstrained
optimization problem using a logarithmic barrier
function~\cite{boyd2004convex}:
\begin{equation}
\label{eq:interior}
\begin{aligned}
\min_{(\emph{$Q_g$},\emph{$Q_c$})}(a Q_g +b Q_c +c)-\mu
\log[-(\gamma_g  Q_g^{\theta_g}+\gamma_c  Q_c^{\theta_c}-R_T)]
\end{aligned},
\end{equation}
where $\mu$ is the barrier parameter. The details of the
interior point method are as follows:\\
Input: a barrier parameter $\mu >0$, a decline factor $\eta < 1$ and
\hspace*{1cm}a desired level of accuracy $\epsilon >0$. \\
Output: ($Q_{g,opt}$, $Q_{c,opt}$), an optimal solution
to~\eqref{eq:rdo2}.\\
Initialization: $k=0$, $(Q_{g,opt}, Q_{c,opt})=(Q^{(k)}_g,Q^{(k)}_c)=$\\
\hspace*{1.8 cm} $(80,80)$, $\mu^{(k)}=\mu$. \\
While $\mu^{(k)} \geq \epsilon$ :\\
\hspace*{0.3 cm}Step 1: compute $(Q_{g,opt}, Q_{c,opt})$ as solution to~\eqref{eq:interior} using \\
\hspace*{1.4 cm}Newton's method initialized with $(Q_g,Q_c)=$\\
\hspace*{1.4 cm}$(Q^{(k)}_g,Q^{(k)}_c)$;\\
\hspace*{0.3 cm}Step 2: update $k=k+1$, $(Q^{(k)}_g,Q^{(k)}_c)=(Q_{g,opt}, Q_{c,opt})$,\\
\hspace*{1.4 cm} $\mu^{(k)}=\eta \mu^{(k-1)}$.\\
The output of the interior point method is subsequently rounded to
obtain a solution that belongs to the finite set of discrete
quantization steps used by the V-PCC coder. While rounding makes the
solution practical for coding, it may lead to a slight violation of
the constraint on the target bitrate.

Unlike exhaustive search, the proposed algorithm does not
necessarily find an optimal solution to the original
problem~\eqref{eq:rdtotal}. This is not only due to rounding but
also because our analytical rate and distortion models are only
approximations. However, the experimental results (see
Section~\ref{sec:5}) show that the rate-distortion performance of
the proposed algorithm is very close to that of exhaustive search.

\begin{figure}[t!]
  \centering
  \includegraphics[width=0.8\columnwidth]{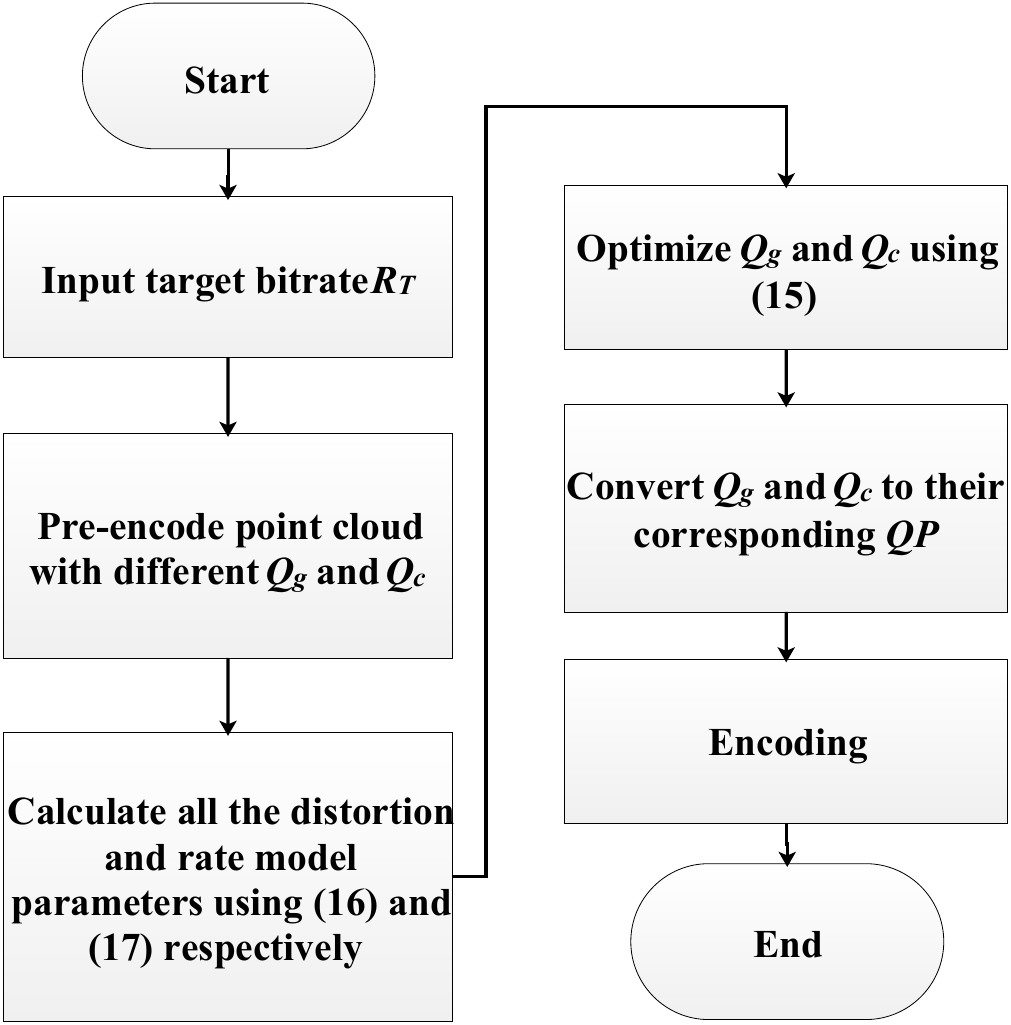}
  \caption{Flow chart of the proposed bit allocation algorithm for geometry and color information.}
  \label{fig8}
\end{figure}

The flowchart of the proposed bit allocation algorithm is shown in
Fig.~\ref{fig8}.

\section{Experimental Results}\label{sec:5} 
\begin{figure*}[t!]
\centering \subfigure[]{ \label{fig9:subfig:a}
\includegraphics[width=0.485\columnwidth,height=3.0cm]{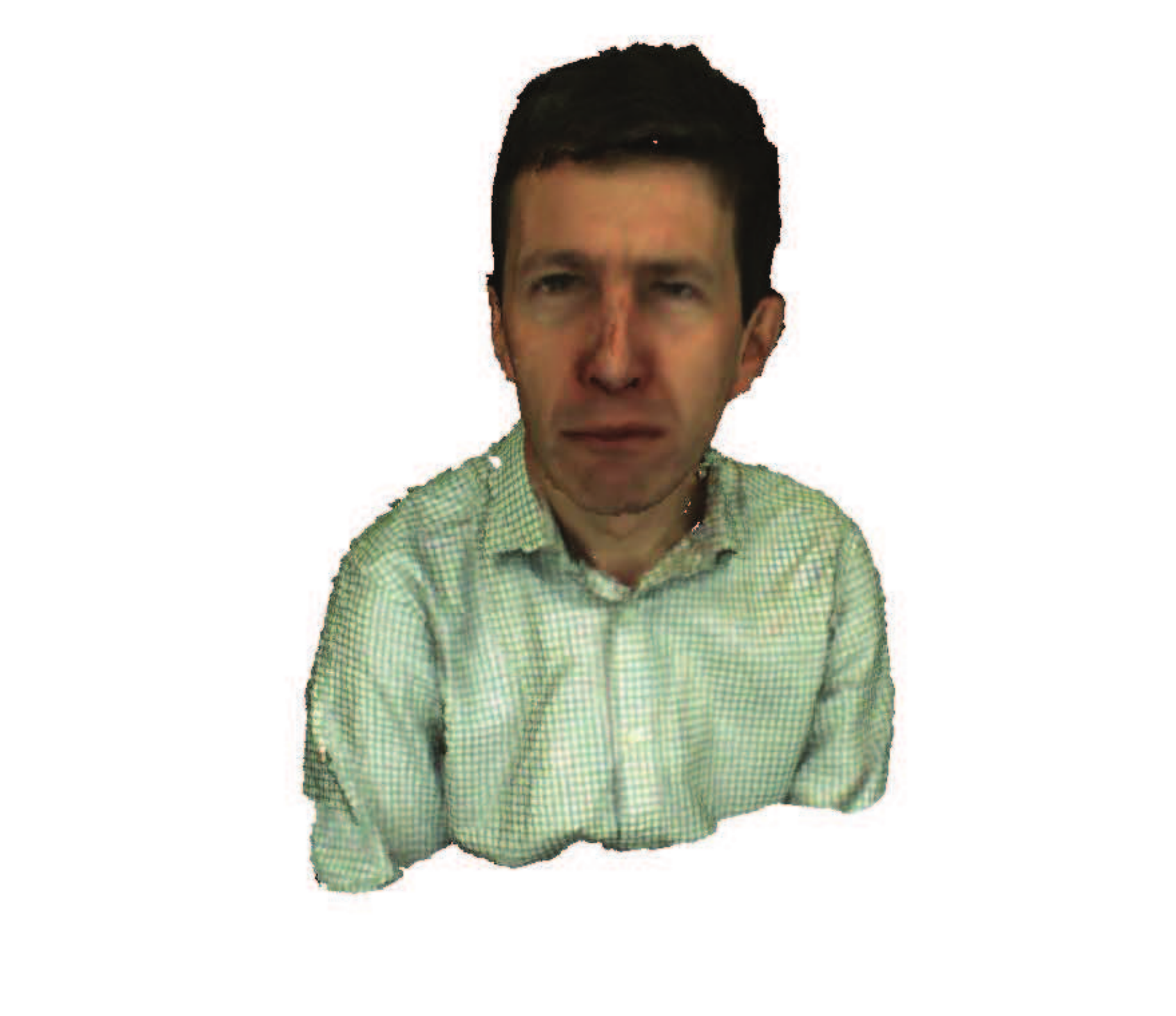}}
\subfigure[]{ \label{fig9:subfig:b}
\includegraphics[width=0.485\columnwidth,height=3.0cm]{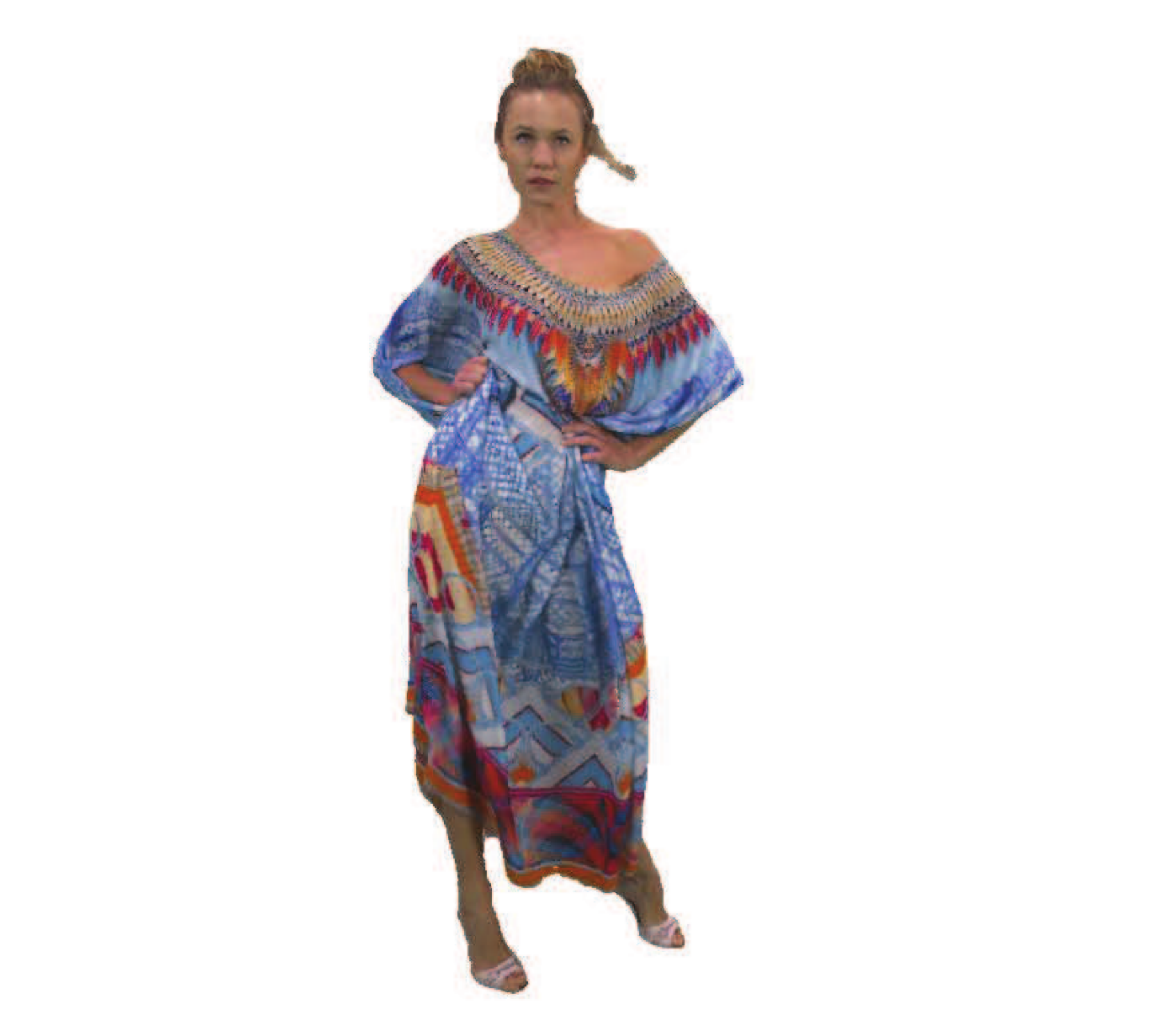}}
\subfigure[]{ \label{fig9:subfig:c}
\includegraphics[width=0.485\columnwidth,height=3.0cm]{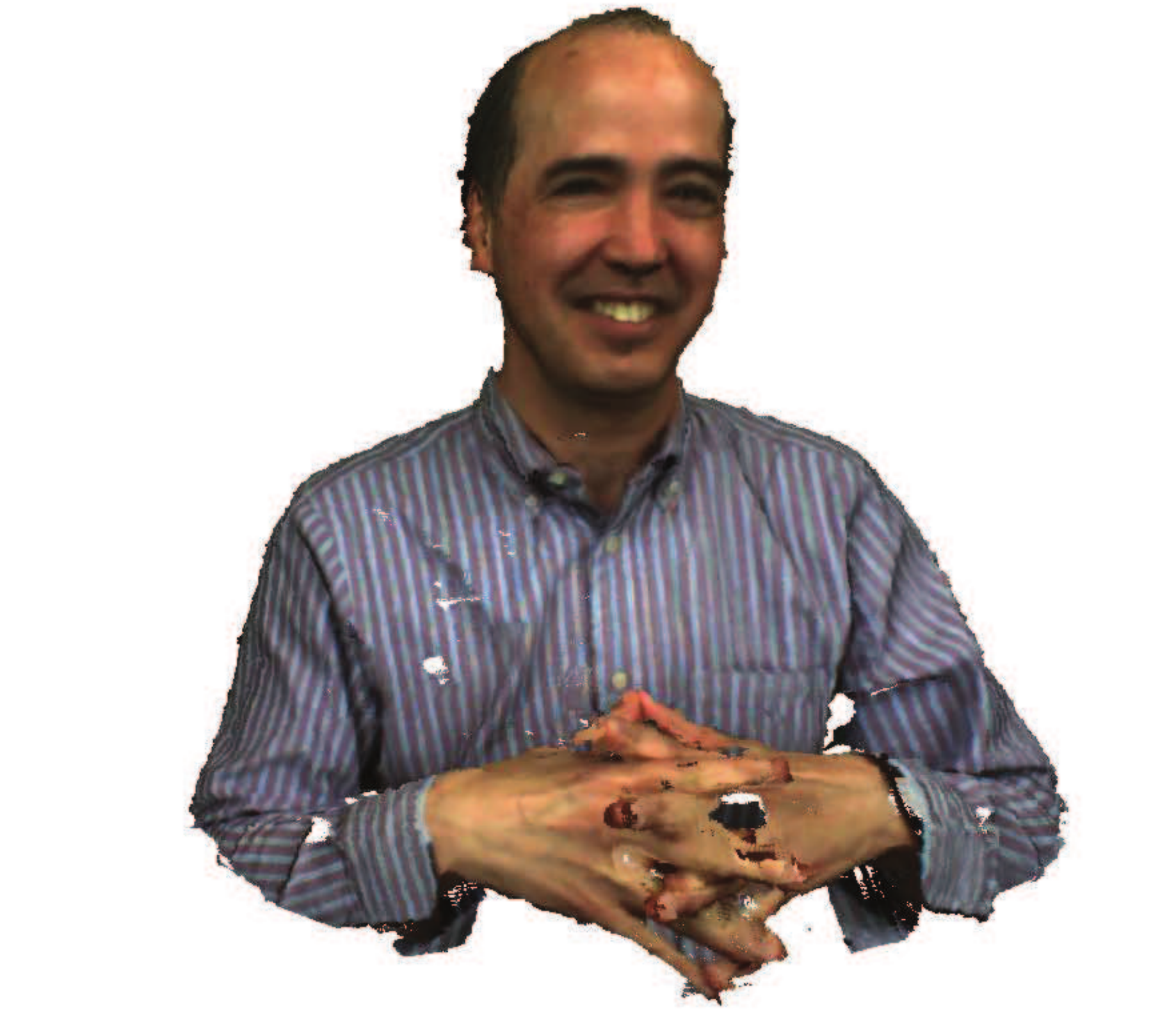}}
\subfigure[]{ \label{fig9:subfig:d}
\includegraphics[width=0.485\columnwidth,height=3.0cm]{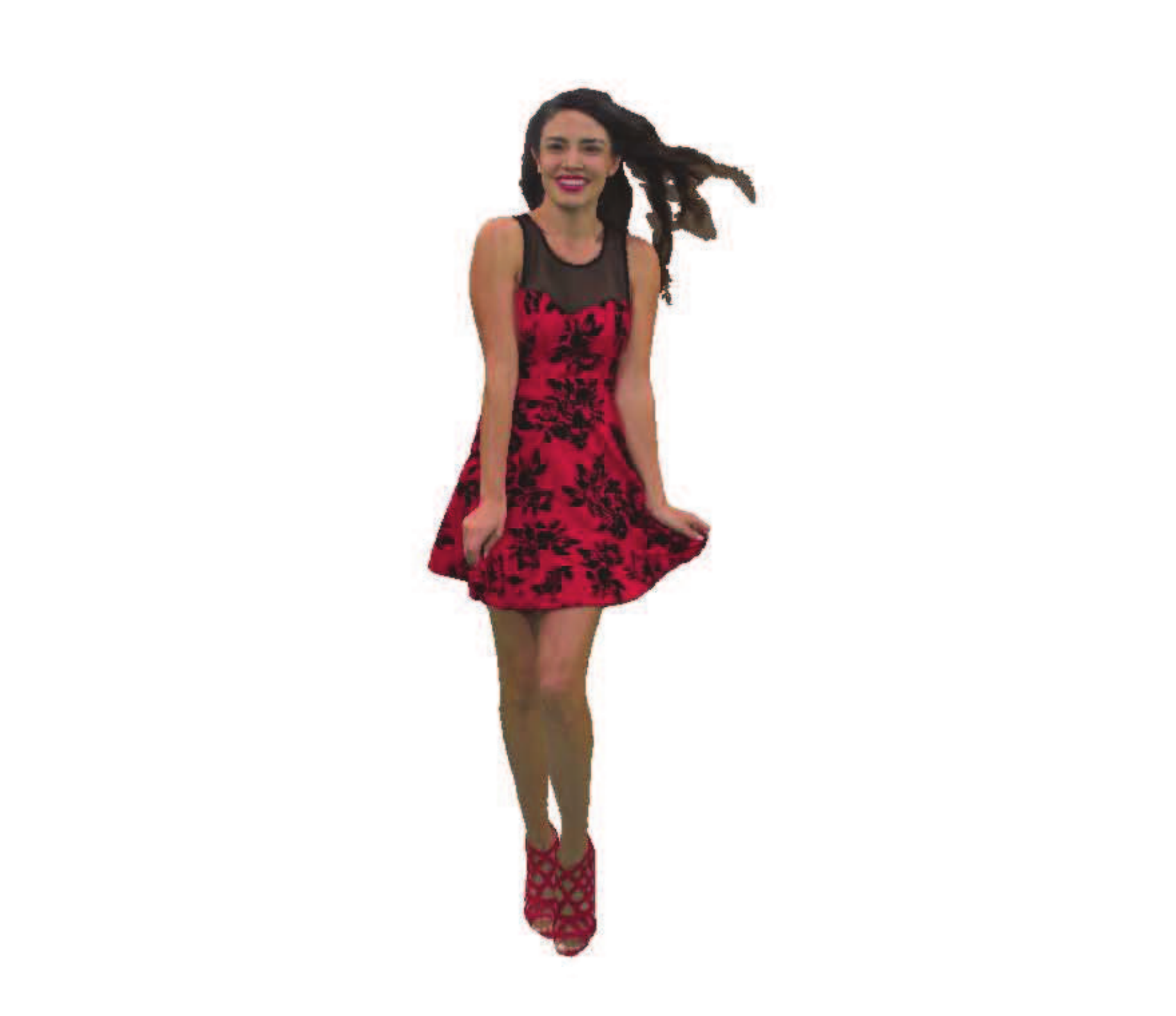}}
\subfigure[]{ \label{fig9:subfig:e}
\includegraphics[width=0.485\columnwidth,height=3.0cm]{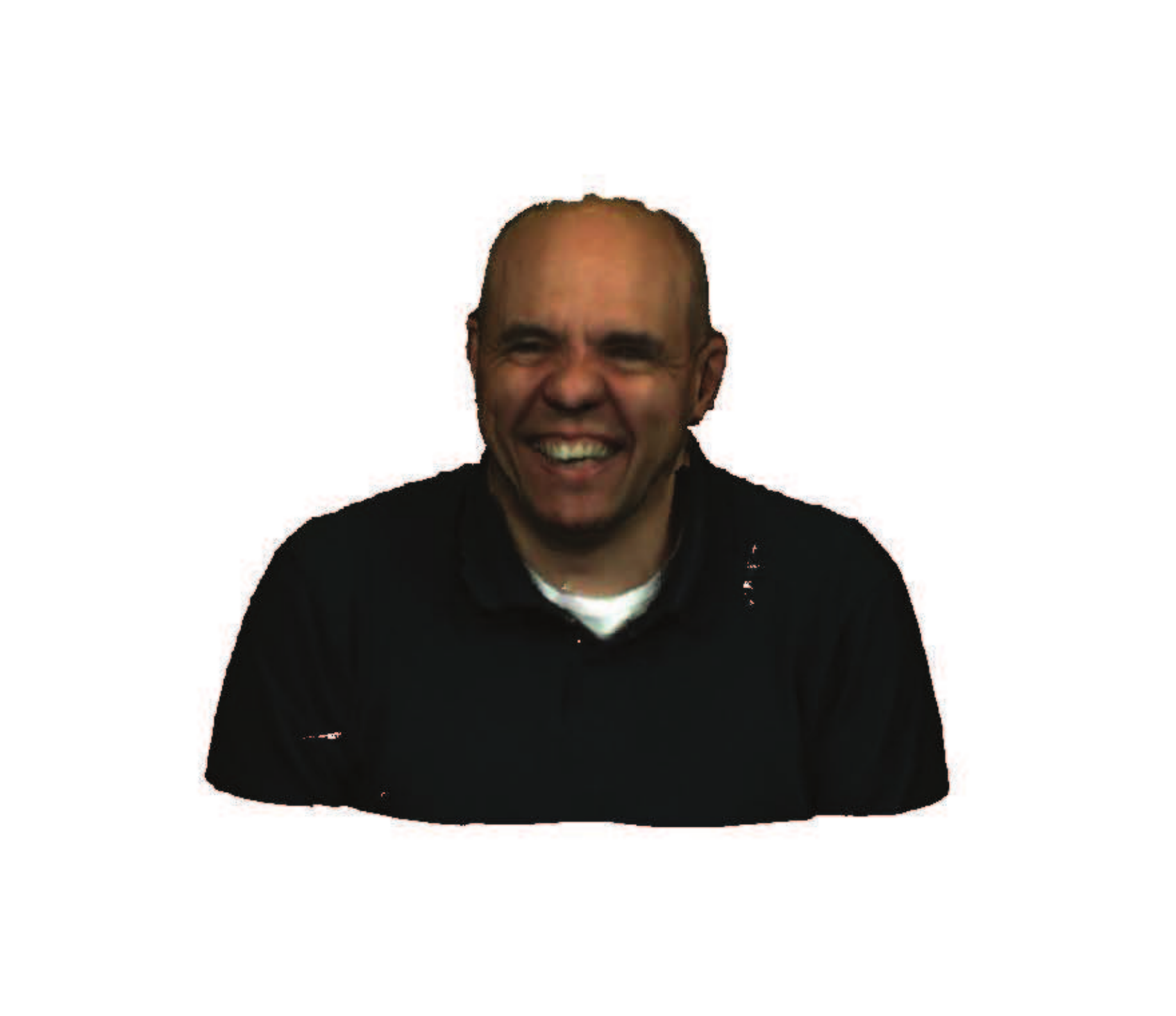}}
\subfigure[]{ \label{fig9:subfig:f}
\includegraphics[width=0.485\columnwidth,height=3.0cm]{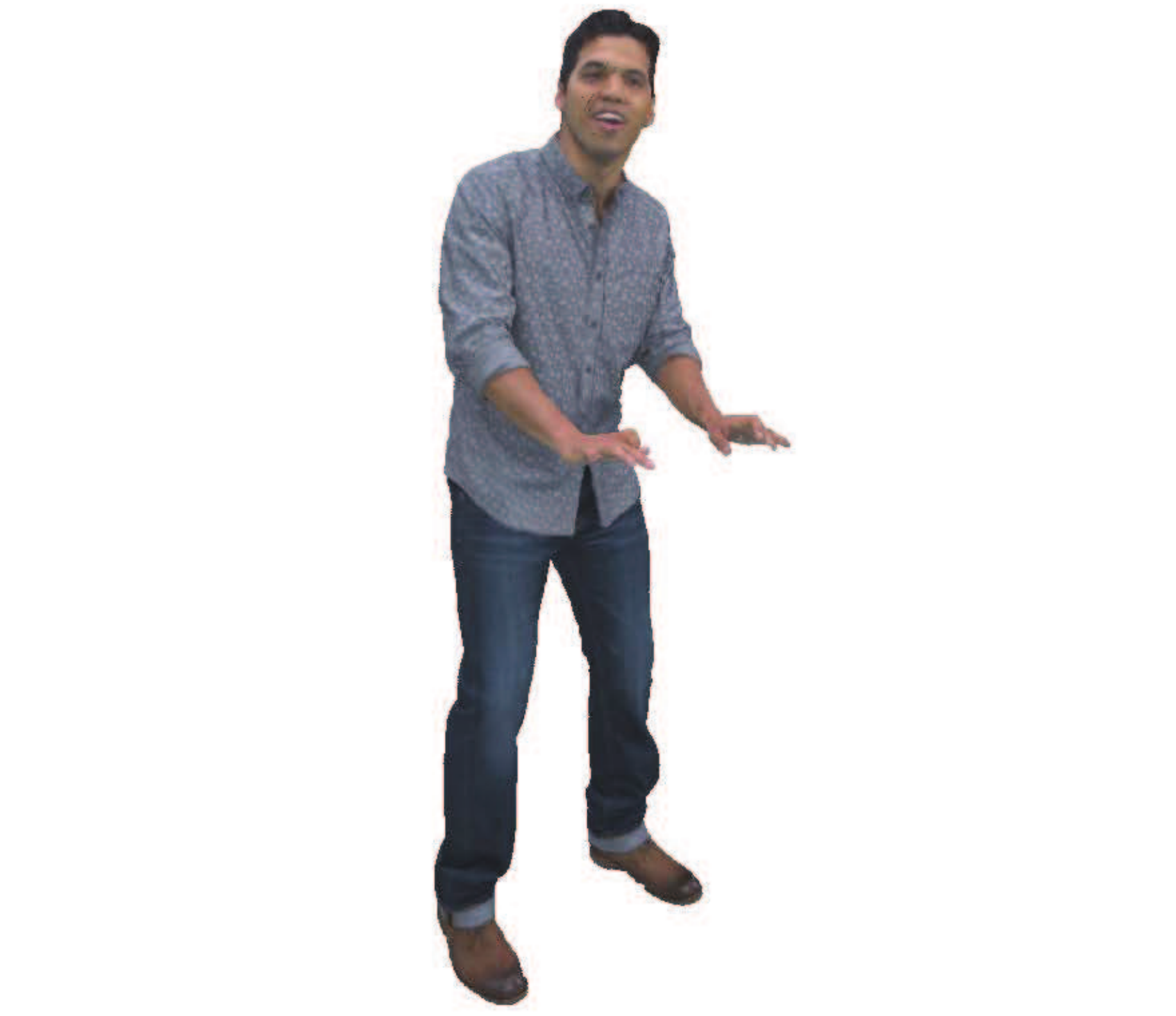}}
\subfigure[]{ \label{fig9:subfig:g}
\includegraphics[width=0.485\columnwidth,height=3.0cm]{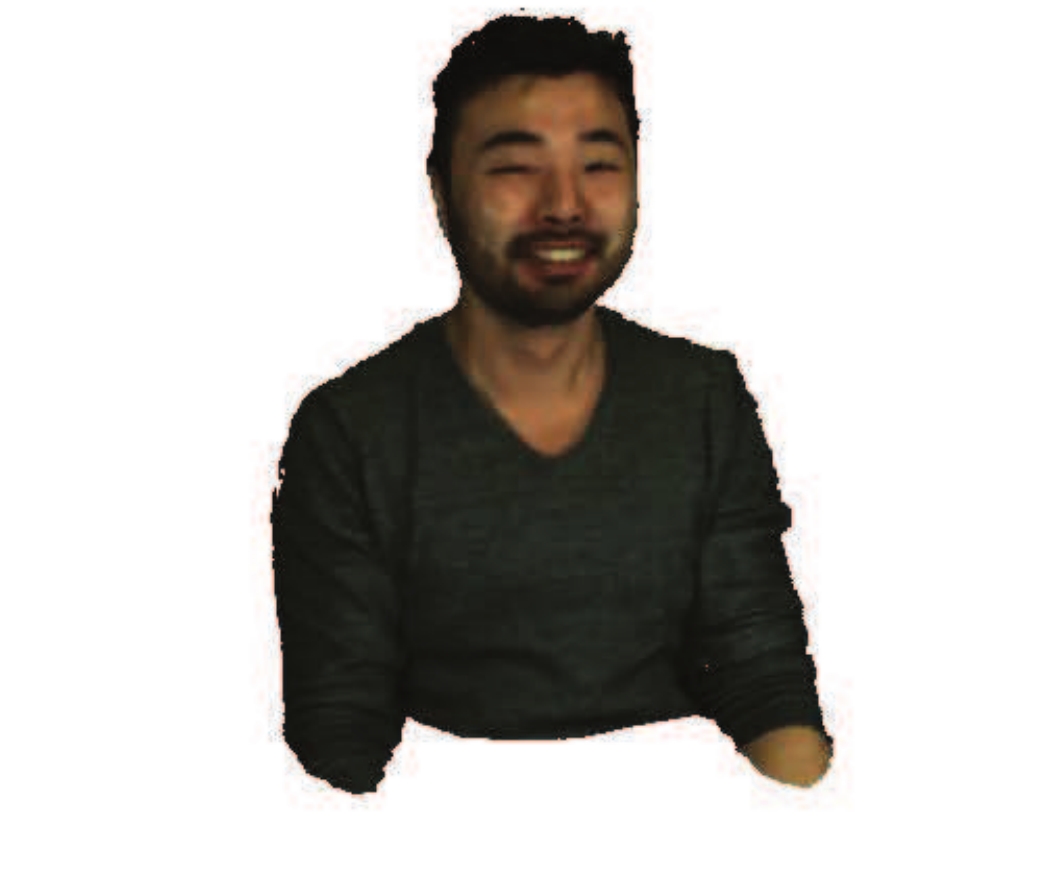}}
\subfigure[]{ \label{fig9:subfig:h}
\includegraphics[width=0.45\columnwidth,height=3.0cm]{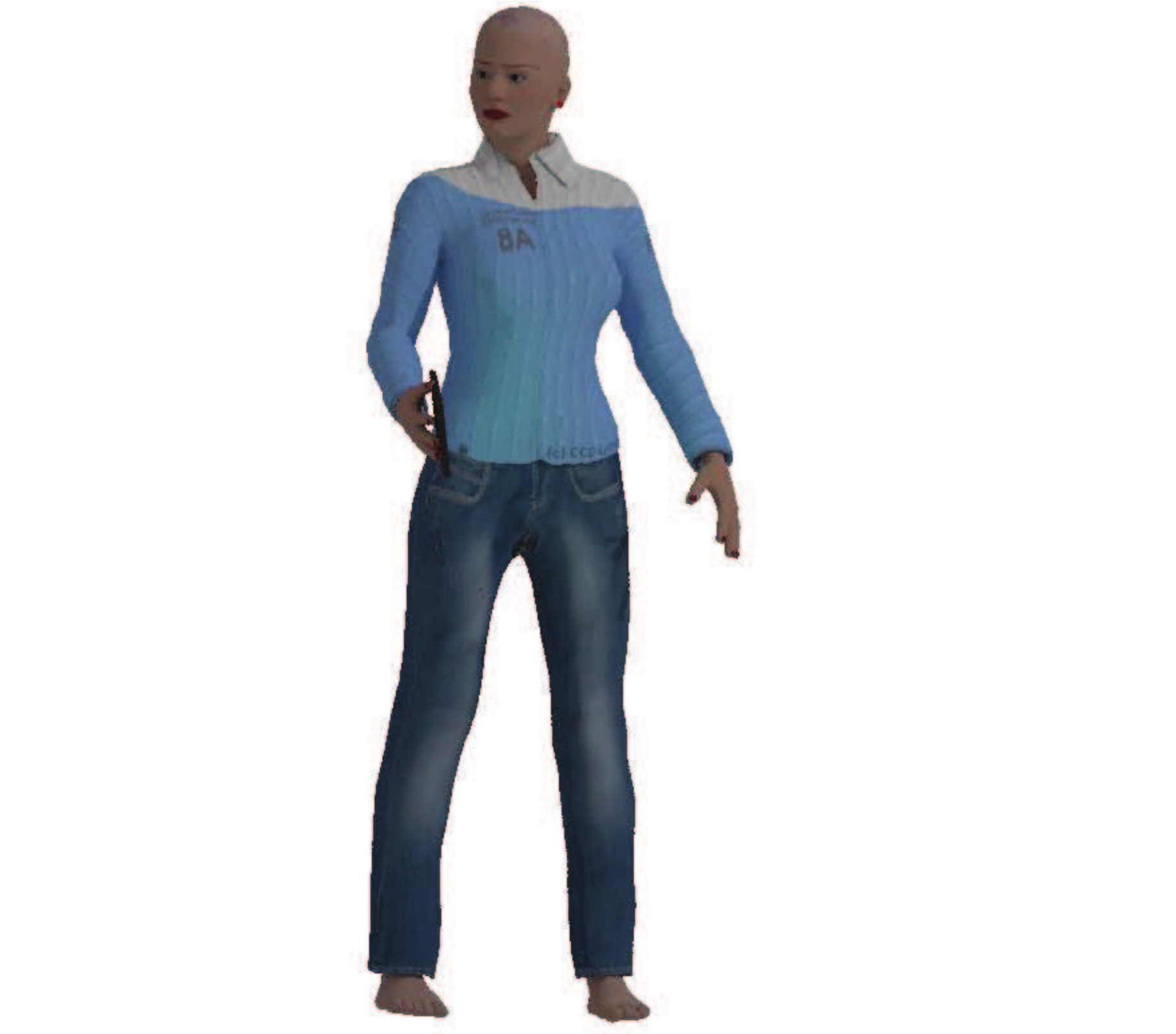}}
\caption{3D point cloud sequences used in the experiments. (a)
\emph{Andrew}, (b) \emph{Longdress}, (c) \emph{Phil}, (d)
\emph{Redandblack}, (e) \emph{Ricardo}, (f) \emph{Loot}, (g)
\emph{David}, (h) \emph{Queen}.} \label{fig9}
\end{figure*}

In this section, we evaluate the accuracy, rate-distortion
performance, and time complexity of the proposed bit allocation
algorithm. We implemented the proposed algorithm in the test model
category 2 version 1.0 (TMC2)~\cite{tmc2v1}, which uses High
Efficiency Video Coding Test Model Version 16.16
(HM16.16)~\cite{hevc} to compress the generated geometry and color
video frames. The barrier parameter $\mu$, decline factor $\eta$,
and level of accuracy $\epsilon$ were set to 0.1, $10^{-6}$, and
$10^{-10}$, respectively. The quality evaluation software
PC\_error~\cite{ctc} was used to calculate the point-to-point
distortion for both color and geometry. The performance of the
proposed algorithm was evaluated on the eight 3D point cloud
sequences~\cite{dataset}~\cite{8I} shown in Fig.~\ref{fig9}.

\begin{table*}[t!]
\newcommand{\tabincell}[2]{\begin{tabular}{@{}#1@{}}#2\end{tabular}}
\centering \caption{Bit Allocation Accuracy for \textbf{ESA} and
\textbf{PBA} ($\omega $=0.25) using the bitrate error ($BE$) and the
QP error ($QPE$). } \label{tab:omega25}
  \begin{tabular}{cccccccccccccc}
      \toprule
      \midrule
\multirow{2}*{\tabincell{l}{Point Cloud}} &\multirow{2}*{\tabincell{l}{Target Bitrate \\\quad($kbpmp$)}} &\multicolumn{3}{c}{\textbf{ESA}} & &\multicolumn{3}{c}{\textbf{PBA}} & &\multicolumn{2}{c}{BE(\%)} &\multirow{2}*{\tabincell{l}{$\left| \Delta BE \right|$}}  &\multirow{2}*{QPE}\\
\cmidrule{3-5}\cmidrule{7-9}\cmidrule{11-12}
      &         &$QP_g$ &$QP_c$ &\tabincell{l}{Bitrate \\($kbpmp$)} & &$QP_g$ &$QP_c$ &\tabincell{l}{Bitrate \\($kbpmp$)}& &\textbf{ESA} &\textbf{PBA} & &\\
\midrule
    \multirow{5}*{\emph{Andrew}}&\cellcolor{mygray}240 &\cellcolor{mygray}27  &\cellcolor{mygray}33  &\cellcolor{mygray}232.7 &\cellcolor{mygray}  &\cellcolor{mygray}27  &\cellcolor{mygray}33  &\cellcolor{mygray}232.7  &\cellcolor{mygray} &\cellcolor{mygray}3.0     &\cellcolor{mygray}3.0     &\cellcolor{mygray}\textcolor[rgb]{0,0,0.6}{\textbf{0.0}}    &\cellcolor{mygray}0 \\
      &450 &22  &29  &444.5  & &22  &29 &444.5 &  &1.2     &1.2     &\textcolor[rgb]{0,0,0.6}{\textbf{0.0}} &0\\
      &\cellcolor{mygray}600 &\cellcolor{mygray}22  &\cellcolor{mygray}27  &\cellcolor{mygray}594.2 &\cellcolor{mygray}  &\cellcolor{mygray}22  &\cellcolor{mygray}27  &\cellcolor{mygray}594.2 &\cellcolor{mygray}  &\cellcolor{mygray}1.0     &\cellcolor{mygray}1.0     &\cellcolor{mygray}\textcolor[rgb]{0,0,0.6}{\textbf{0.0}} &\cellcolor{mygray}0 \\
      &760 &22  &26  &683.4 &  &22  &25  &786.0  & &10.1    &\textbf{3.4}    & \textcolor[rgb]{0,0,0.6}{\textbf{6.7}} &1\\
      &\cellcolor{mygray}1070    &\cellcolor{mygray}22  &\cellcolor{mygray}23  &\cellcolor{mygray}1019.1 &\cellcolor{mygray}  &\cellcolor{mygray}22  &\cellcolor{mygray}23  &\cellcolor{mygray}1019.1 &\cellcolor{mygray}  &\cellcolor{mygray}4.8     &\cellcolor{mygray}4.8    &\cellcolor{mygray}\textcolor[rgb]{0,0,0.6}{\textbf{0.0}}    &\cellcolor{mygray}0\\
\midrule
\multirow{5}*{\emph{Phil}}& 240                 & 26  & 32 & 237.5     &      & 26  & 32 & 237.5     &      &1.1    &1.1  & \textcolor[rgb]{0,0,0.6}{\textbf{0.0}}    & 0   \\
                          &\cellcolor{mygray}{450}                 &\cellcolor{mygray}{22}  &\cellcolor{mygray}28 &\cellcolor{mygray}{423.8}     &\cellcolor{mygray}     &\cellcolor{mygray}22  & \cellcolor{mygray}28 &\cellcolor{mygray}{423.8}     &\cellcolor{mygray}      &\cellcolor{mygray}5.8    & \cellcolor{mygray}5.8  &\cellcolor{mygray}\textcolor[rgb]{0,0,0.6}{\textbf{0.0}}    &\cellcolor{mygray}0   \\
                          & 600                 & 22  & 26 & 552.8     &      & 22  & 25 & 635.6     &      & 7.9    & \textbf{5.9}  & \textcolor[rgb]{0,0,0.6}{\textbf{2.0}}    & 1   \\
                          &\cellcolor{mygray}760                 &\cellcolor{mygray}22  & \cellcolor{mygray}24 &\cellcolor{mygray}729.5     &\cellcolor{mygray}      &\cellcolor{mygray}22  &\cellcolor{mygray}24 &\cellcolor{mygray}729.5     & \cellcolor{mygray}     &\cellcolor{mygray}4.0    &\cellcolor{mygray}4.0  &\cellcolor{mygray}\textcolor[rgb]{0,0,0.6}{\textbf{0.0}}    &\cellcolor{mygray}0   \\
                          & 1070                & 22  & 22 & 949.4     &      & 22  & 22 & 949.4     &      &11.3   &11.3 & \textcolor[rgb]{0,0,0.6}{\textbf{0.0}}    & 0   \\
\midrule
\multirow{5}*{\emph{Longdress}}  &\cellcolor{mygray}240                 & \cellcolor{mygray}30  &\cellcolor{mygray}40 &\cellcolor{mygray}236.6    &\cellcolor{mygray}       &\cellcolor{mygray}30  &\cellcolor{mygray}40 & \cellcolor{mygray}236.6    &\cellcolor{mygray}       &\cellcolor{mygray}1.4    &\cellcolor{mygray}1.4  &\cellcolor{mygray}\textcolor[rgb]{0,0,0.6}{\textbf{0.0}}    &\cellcolor{mygray}0   \\
                                 & 450                 & 24  & 35 & 449.4    &       & 24  & 35 & 449.4    &       &0.1    &0.1  & \textcolor[rgb]{0,0,0.6}{\textbf{0.0}}    & 0   \\
                                 &\cellcolor{mygray} 600                 & \cellcolor{mygray}22  & \cellcolor{mygray}33 & \cellcolor{mygray}585.9    & \cellcolor{mygray}      & \cellcolor{mygray}22  &\cellcolor{mygray}33 &\cellcolor{mygray}585.9    & \cellcolor{mygray}      & \cellcolor{mygray}2.3    & \cellcolor{mygray}2.3  &\cellcolor{mygray}\textcolor[rgb]{0,0,0.6}{\textbf{0.0}}    &\cellcolor{mygray}0   \\
                                 & 760                 & 22  & 31 & 752.4    &       & 22  & 31 & 752.4    &       &1.0    &1.0  & \textcolor[rgb]{0,0,0.6}{\textbf{0.0}}    & 0   \\
                                 & \cellcolor{mygray}1070                &\cellcolor{mygray}22  &\cellcolor{mygray}28 &\cellcolor{mygray}1067.4   &\cellcolor{mygray}       & \cellcolor{mygray}22  &\cellcolor{mygray}28 & \cellcolor{mygray}1067.4   &\cellcolor{mygray}       & \cellcolor{mygray}0.2    &\cellcolor{mygray}0.2  &\cellcolor{mygray}\textcolor[rgb]{0,0,0.6}{\textbf{0.0}}    &\cellcolor{mygray}0   \\
\midrule
\multirow{5}*{\emph{Redandblack}}  & 240                 & 29  & 34 & 238.2   &        & 31  & 34 & 230.2    &       & \textbf{0.8}    & 4.1  & \textcolor[rgb]{0,0,0.6}{\textbf{3.3}}  & 2   \\
                                   &\cellcolor{mygray}450                 &\cellcolor{mygray}23  &\cellcolor{mygray}29 &\cellcolor{mygray}448.0   & \cellcolor{mygray}       &\cellcolor{mygray}25  &\cellcolor{mygray}29 &\cellcolor{mygray}429.6    &\cellcolor{mygray}       &\cellcolor{mygray}\textbf{0.5}    & \cellcolor{mygray}4.5  &\cellcolor{mygray}\textcolor[rgb]{0,0,0.6}{\textbf{4.0}}  &\cellcolor{mygray}2   \\
                                   & 600                 & 22  & 27 & 572.2   &        & 22  & 27 & 572.2    &       &4.6    &4.6  & \textcolor[rgb]{0,0,0.6}{\textbf{0.0}}  & 0   \\
                                   &\cellcolor{mygray}760                 &\cellcolor{mygray}22  &\cellcolor{mygray}25 &\cellcolor{mygray}721.6   &\cellcolor{mygray}        & \cellcolor{mygray}22  &\cellcolor{mygray}25 &\cellcolor{mygray}721.6    & \cellcolor{mygray}      & \cellcolor{mygray}5.1    & \cellcolor{mygray}5.1  &\cellcolor{mygray}\textcolor[rgb]{0,0,0.6}{\textbf{0.0}}  &\cellcolor{mygray}0   \\
                                   & 1070                & 22  & 22 & 1011.4  &        & 22  & 22 & 1011.4   &       &5.5    &5.5  & \textcolor[rgb]{0,0,0.6}{\textbf{0.0}}  & 0   \\
          \midrule
\multirow{5}*{\emph{David}}&\cellcolor{mygray}70                  & \cellcolor{mygray}33  &\cellcolor{mygray}31 &\cellcolor{mygray}69.7    & \cellcolor{mygray}       &\cellcolor{mygray}36  & \cellcolor{mygray}30 &\cellcolor{mygray}72.7     &\cellcolor{mygray}       &\cellcolor{mygray}\textbf{0.4}    &\cellcolor{mygray}3.8  &\cellcolor{mygray}\textcolor[rgb]{0,0,0.6}{\textbf{3.4}}  &\cellcolor{mygray}4   \\
                           & 86                  & 31  & 29 & 86.0    &        & 34  & 29 & 80.9     &       & \textbf{0.0}    & 5.9  & \textcolor[rgb]{0,0,0.6}{\textbf{5.9}}  & 3   \\
                           &\cellcolor{mygray}96                  &\cellcolor{mygray}31  & \cellcolor{mygray}28 &\cellcolor{mygray}93.3    & \cellcolor{mygray}       &\cellcolor{mygray}33  & \cellcolor{mygray}28 &\cellcolor{mygray}90.5     &\cellcolor{mygray}       &\cellcolor{mygray}\textbf{2.8}    &\cellcolor{mygray}5.7  &\cellcolor{mygray}\textcolor[rgb]{0,0,0.6}{\textbf{2.9}}  &\cellcolor{mygray}2   \\
                           & 180                 & 26  & 23 & 176.4   &        & 27  & 23 & 171.8    &       & \textbf{2.0}      & 4.6  & \textcolor[rgb]{0,0,0.6}{\textbf{2.6}}  & 1   \\
                           &\cellcolor{mygray}224                 &\cellcolor{mygray}22  &\cellcolor{mygray}22 &\cellcolor{mygray}221.9   &\cellcolor{mygray}        &\cellcolor{mygray}22  &\cellcolor{mygray}22 & \cellcolor{mygray}221.9    &\cellcolor{mygray}       &\cellcolor{mygray}0.9    &\cellcolor{mygray}0.9  &\cellcolor{mygray}\textcolor[rgb]{0,0,0.6}{\textbf{0.0}}    &\cellcolor{mygray}0   \\
\midrule
\multirow{5}*{\emph{Ricardo}}& 70                  & 36  & 31 & 69.0    &          & 32  & 32 & 69.5    &        & 1.4    & \textbf{0.6}  & \textcolor[rgb]{0,0,0.6}{\textbf{0.8}} & 5   \\
                             & \cellcolor{mygray}86                  &\cellcolor{mygray}30  &\cellcolor{mygray}30 &\cellcolor{mygray}84.6    &\cellcolor{mygray}          &\cellcolor{mygray}30  &\cellcolor{mygray}30 &\cellcolor{mygray}84.6    &\cellcolor{mygray}        &\cellcolor{mygray}1.6    &\cellcolor{mygray}1.6  &\cellcolor{mygray}\textcolor[rgb]{0,0,0.6}{\textbf{0.0}} &\cellcolor{mygray}0   \\
                             & 96                  & 29  & 29 & 94.7    &          & 29  & 29 & 94.7    &        &1.3    &1.3  & \textcolor[rgb]{0,0,0.6}{\textbf{0.0}} & 0   \\
                             &\cellcolor{mygray}180                 &\cellcolor{mygray}24  & \cellcolor{mygray}24 &\cellcolor{mygray}176.5   &\cellcolor{mygray}          & \cellcolor{mygray}23  &\cellcolor{mygray}24 &\cellcolor{mygray}182.4   &\cellcolor{mygray}        &\cellcolor{mygray}1.9    &\cellcolor{mygray}\textbf{1.4}  &\cellcolor{mygray}\textcolor[rgb]{0,0,0.6}{\textbf{0.5}} &\cellcolor{mygray}1   \\
                             & 224                 & 23  & 22 & 219.9   &          & 22  & 22 & 228.2   &        & \textbf{1.8}    & 1.9  & \textcolor[rgb]{0,0,0.6}{\textbf{0.1}} & 1   \\
      \midrule
\multirow{5}*{\emph{Loot}}&\cellcolor{mygray}70                  &\cellcolor{mygray}42  &\cellcolor{mygray}40 &\cellcolor{mygray}69.6     &\cellcolor{mygray}       &\cellcolor{mygray}40  &\cellcolor{mygray}40 & \cellcolor{mygray}70.2    & \cellcolor{mygray}       &\cellcolor{mygray}0.6    & \cellcolor{mygray}\textbf{0.4}  &\cellcolor{mygray}\textcolor[rgb]{0,0,0.6}{\textbf{0.2}} &\cellcolor{mygray}2   \\
                          & 86                  & 36  & 39 & 84.5     &       & 38  & 38 & 93.9    &        & \textbf{1.8}    & 9.2  & \textcolor[rgb]{0,0,0.6}{\textbf{7.4}}  & 3   \\
                          &\cellcolor{mygray}96                  &\cellcolor{mygray}35  &\cellcolor{mygray}38 &\cellcolor{mygray}94.8     &\cellcolor{mygray}       &\cellcolor{mygray}37  &\cellcolor{mygray}37 &\cellcolor{mygray}100.6   &\cellcolor{mygray}        &\cellcolor{mygray}\textbf{1.2}    &\cellcolor{mygray}4.8  &\cellcolor{mygray}\textcolor[rgb]{0,0,0.6}{\textbf{3.6}}  &\cellcolor{mygray}3   \\
                          & 180                 & 35  & 32 & 178.9    &       & 31  & 32 & 188.6   &        & \textbf{0.6}    & 4.8  & \textcolor[rgb]{0,0,0.6}{\textbf{4.2}}  & 4   \\
                          &\cellcolor{mygray}224                 &\cellcolor{mygray}28  &\cellcolor{mygray}31 &\cellcolor{mygray}223.7    &\cellcolor{mygray}       &\cellcolor{mygray}29  &\cellcolor{mygray}31 &\cellcolor{mygray}220.6   &\cellcolor{mygray}        &\cellcolor{mygray}\textbf{0.1}    &\cellcolor{mygray}1.5  &\cellcolor{mygray}\textcolor[rgb]{0,0,0.6}{\textbf{1.4}}  &\cellcolor{mygray}1   \\
\midrule
\multirow{5}*{\emph{Queen}}& 70                  & 34  & 42 & 68.2     &       & 37  & 42 & 65.8    &        & \textbf{2.6}    & 6.0    & \textcolor[rgb]{0,0,0.6}{\textbf{3.4}}  & 3   \\
                           &\cellcolor{mygray}86                  &\cellcolor{mygray}30  &\cellcolor{mygray}41 &\cellcolor{mygray}84.2     &\cellcolor{mygray}       &\cellcolor{mygray}30  &\cellcolor{mygray}42 &\cellcolor{mygray}77.3    &\cellcolor{mygray}        &\cellcolor{mygray}\textbf{2.1}    &\cellcolor{mygray}10.1   &\cellcolor{mygray}\textcolor[rgb]{0,0,0.6}{\textbf{8.0}}  &\cellcolor{mygray}1   \\
                           & 96                  & 27  & 41 & 94.3     &       & 29  & 41 & 89.0    &        & \textbf{1.8}    & 7.3    &\textcolor[rgb]{0,0,0.6}{\textbf{5.5}}  & 2   \\
                           &\cellcolor{mygray}180                 &\cellcolor{mygray}22  &\cellcolor{mygray}36 &\cellcolor{mygray}171.9    &\cellcolor{mygray}       &\cellcolor{mygray}22  &\cellcolor{mygray}35 &\cellcolor{mygray}184.9   &\cellcolor{mygray}        &\cellcolor{mygray}4.5    &\cellcolor{mygray}\textbf{2.7}    &\cellcolor{mygray}\textcolor[rgb]{0,0,0.6}{\textbf{1.8}}  &\cellcolor{mygray}1   \\
                           & 224                 & 22  & 33 & 220.6    &       & 22  & 33 & 220.6   &        &1.5    &1.5    & \textcolor[rgb]{0,0,0.6}{\textbf{0.0}}  & 0   \\
\midrule
 \rowcolor{mygray}
\multicolumn{10}{>{\columncolor{mygray}}c}{\textbf{Average}}
&\textbf{2.6}&\textbf{3.7}
& \textcolor[rgb]{0,0,0.6}{\textbf{1.7}} & \textbf{1.1} \\
\hline
      \bottomrule
  \end{tabular}
\end{table*}

\begin{table*}[t!]
\newcommand{\tabincell}[2]{\begin{tabular}{@{}#1@{}}#2\end{tabular}}
\centering \caption{Bit Allocation Accuracy for \textbf{ESA} and
\textbf{PBA} ($\omega $=0.5) using the bitrate error ($BE$) and the
QP error ($QPE$).} \label{tab:omega50} \scalebox{1.05}{
  \begin{tabular}{cccccccccccccc}
      \toprule
      \midrule
      \multirow{2}*{\tabincell{l}{Point Cloud}} &\multirow{2}*{\tabincell{l}{Target Bitrate \\\quad($kbpmp$)}} &\multicolumn{3}{c}{\textbf{ESA}} & &\multicolumn{3}{c}{\textbf{PBA}} & &\multicolumn{2}{c}{BE(\%)} &\multirow{2}*{\tabincell{l}{$\left| \Delta BE \right|$}}  &\multirow{2}*{QPE}\\
     \cmidrule{3-5}\cmidrule{7-9}\cmidrule{11-12}
      &         &$QP_g$ &$QP_c$ &\tabincell{l}{Bitrate \\($kbpmp$)} & &$QP_g$ &$QP_c$ &\tabincell{l}{Bitrate \\($kbpmp$)}& &\textbf{ESA} &\textbf{PBA} & &\\
\midrule
\multirow{5}*{\emph{Andrew}}& \cellcolor{mygray}240                 & \cellcolor{mygray}27  & \cellcolor{mygray}33 & \cellcolor{mygray}232.7     &\cellcolor{mygray}      & \cellcolor{mygray}27  & \cellcolor{mygray}33 & \cellcolor{mygray}232.7    &\cellcolor{mygray}       & \cellcolor{mygray}3.0      & \cellcolor{mygray}3.0    & \cellcolor{mygray}\textcolor[rgb]{0,0,0.6}{\textbf{0.0}}    & \cellcolor{mygray}0   \\
                            & 450                 & 22  & 29 & 444.5    &       & 22  & 29 & 444.5    &       &1.2    &1.2  & \textcolor[rgb]{0,0,0.6}{\textbf{0.0}}    & 0   \\
                            & \cellcolor{mygray}600                 & \cellcolor{mygray}22  & \cellcolor{mygray}27 & \cellcolor{mygray}594.2    &\cellcolor{mygray}       & \cellcolor{mygray}22  & \cellcolor{mygray}27 & \cellcolor{mygray}594.2     &\cellcolor{mygray}      & \cellcolor{mygray}1.0      & \cellcolor{mygray}1.0    & \cellcolor{mygray}\cellcolor{mygray}\textcolor[rgb]{0,0,0.6}{\textbf{0.0}}    & \cellcolor{mygray}0   \\
                            & 760                 & 22  & 26 & 683.4    &       & 22  & 25 & 786.0     &        & 10.1   & \textbf{3.4}  & \textcolor[rgb]{0,0,0.6}{\textbf{6.7}} & 1   \\
                            & \cellcolor{mygray}1070                & \cellcolor{mygray}22  & \cellcolor{mygray}23 & \cellcolor{mygray}1019.1   &\cellcolor{mygray}       & \cellcolor{mygray}22  & \cellcolor{mygray}23 & \cellcolor{mygray}1019.1    &\cellcolor{mygray}      & \cellcolor{mygray}4.8    & \cellcolor{mygray}4.8  & \cellcolor{mygray}\textcolor[rgb]{0,0,0.6}{\textbf{0.0}}    & \cellcolor{mygray}0   \\
\midrule
\multirow{5}*{\emph{Phil}}& 240                 & 26  & 32 & 237.5   &        & 26  & 32 & 237.5    &       &1.1   &1.1  & \textcolor[rgb]{0,0,0.6}{\textbf{0.0}}    & 0   \\
                          & \cellcolor{mygray}450                 & \cellcolor{mygray}22  & \cellcolor{mygray}28 & \cellcolor{mygray}423.8   &\cellcolor{mygray}       & \cellcolor{mygray}22  & \cellcolor{mygray}28 & \cellcolor{mygray}423.8    &\cellcolor{mygray}       & \cellcolor{mygray}5.8    & \cellcolor{mygray}5.8  & \cellcolor{mygray}\textcolor[rgb]{0,0,0.6}{\textbf{0.0}}    & \cellcolor{mygray}0   \\
                          & 600                 & 22  & 26 & 552.8   &        & 22  & 25 & 635.6    &       & 7.9    & \textbf{5.9}  & \textcolor[rgb]{0,0,0.6}{\textbf{2.0}}    & 1   \\
                          & \cellcolor{mygray}760                 & \cellcolor{mygray}22  & \cellcolor{mygray}24 & \cellcolor{mygray}729.5   &\cellcolor{mygray}        & \cellcolor{mygray}22  & \cellcolor{mygray}24 & \cellcolor{mygray}729.5    &\cellcolor{mygray}       & \cellcolor{mygray}4.0     & \cellcolor{mygray}4.0 & \cellcolor{mygray}\textcolor[rgb]{0,0,0.6}{\textbf{0.0}}    & \cellcolor{mygray}0   \\
                          & 1070                & 22  & 22 & 949.4   &        & 22  & 22 & 949.4    &       &11.3   &11.3 & \textcolor[rgb]{0,0,0.6}{\textbf{0.0}}    & 0   \\
          \midrule
\multirow{5}*{\emph{Longdress}}& \cellcolor{mygray}240                 & \cellcolor{mygray}30  & \cellcolor{mygray}40 & \cellcolor{mygray}236.6    &\cellcolor{mygray}       & \cellcolor{mygray}30  & \cellcolor{mygray}40 & \cellcolor{mygray}236.6     &\cellcolor{mygray}      & \cellcolor{mygray}1.4    & \cellcolor{mygray}1.4  & \cellcolor{mygray}\textcolor[rgb]{0,0,0.6}{\textbf{0.0}}    & \cellcolor{mygray}0   \\
                               & 450                 & 24  & 35 & 449.4    &       & 24  & 35 & 449.4     &      &0.1    &0.1  & \textcolor[rgb]{0,0,0.6}{\textbf{0.0}}    & 0   \\
                               & \cellcolor{mygray}600                 & \cellcolor{mygray}22  & \cellcolor{mygray}33 & \cellcolor{mygray}585.9    &\cellcolor{mygray}      & \cellcolor{mygray}22  & \cellcolor{mygray}33 & \cellcolor{mygray}585.9     &\cellcolor{mygray}      & \cellcolor{mygray}2.3    & \cellcolor{mygray}2.3  & \cellcolor{mygray}\textcolor[rgb]{0,0,0.6}{\textbf{0.0}}    & \cellcolor{mygray}0   \\
                               & 760                 & 22  & 31 & 752.4    &       & 22  & 31 & 752.4     &      &1.0    &1.0  & \textcolor[rgb]{0,0,0.6}{\textbf{0.0}}    & 0   \\
                               & \cellcolor{mygray}1070                & \cellcolor{mygray}22  & \cellcolor{mygray}28 & \cellcolor{mygray}1067.4   &\cellcolor{mygray}       & \cellcolor{mygray}22  & \cellcolor{mygray}28 & \cellcolor{mygray}1067.4    &\cellcolor{mygray}      & \cellcolor{mygray}0.2    & \cellcolor{mygray}0.2  & \cellcolor{mygray}\textcolor[rgb]{0,0,0.6}{\textbf{0.0}}    & \cellcolor{mygray}0   \\
\midrule
\multirow{5}*{\emph{Redandblack}} & 240                 & 29  & 34 & 238.2     &        & 30  & 34 & 234.4    &       & \textbf{0.8}    & 2.3  & \textcolor[rgb]{0,0,0.6}{\textbf{1.5}}  & 1   \\
                                  & \cellcolor{mygray}450                 & \cellcolor{mygray}23  & \cellcolor{mygray}29 & \cellcolor{mygray}448.0     & \cellcolor{mygray}       & \cellcolor{mygray}24  & \cellcolor{mygray}29 & \cellcolor{mygray}440.1    &  \cellcolor{mygray}      & \cellcolor{mygray}\textbf{0.5}    & \cellcolor{mygray}2.2  & \cellcolor{mygray}\textcolor[rgb]{0,0,0.6}{\textbf{1.7}}  & \cellcolor{mygray}1   \\
                                  & 600                 & 22  & 27 & 572.2     &      & 22  & 27 & 572.2      &     &4.6    &4.6  & \textcolor[rgb]{0,0,0.6}{\textbf{0.0}}    & 0   \\
                                  & \cellcolor{mygray}760                 & \cellcolor{mygray}22  & \cellcolor{mygray}25 & \cellcolor{mygray}721.6     &\cellcolor{mygray}      & \cellcolor{mygray}22  & \cellcolor{mygray}25 & \cellcolor{mygray}721.6      & \cellcolor{mygray}   & \cellcolor{mygray}5.1    & \cellcolor{mygray}5.1  & \cellcolor{mygray}\textcolor[rgb]{0,0,0.6}{\textbf{0.0}}    & \cellcolor{mygray}0   \\
                                  & 1070                & 22  & 22 & 1011.4    &      & 22  & 22 & 1011.4     &     &5.5   &5.5  & \textcolor[rgb]{0,0,0.6}{\textbf{0.0}}    & 0   \\
              \midrule
\multirow{5}*{\emph{David}}& \cellcolor{mygray}70                  & \cellcolor{mygray}33  & \cellcolor{mygray}31 & \cellcolor{mygray}69.7      & \cellcolor{mygray}     & \cellcolor{mygray}34  & \cellcolor{mygray}31 & \cellcolor{mygray}68.5     & \cellcolor{mygray}      & \cellcolor{mygray}\textbf{0.4}    & \cellcolor{mygray}2.2  & \cellcolor{mygray}\textcolor[rgb]{0,0,0.6}{\textbf{1.8}}  & \cellcolor{mygray}1   \\
                           & 86                  & 31  & 29 & 86.0      &        & 32  & 29 & 83.6    &        & \textbf{0.0}      & 2.8  & \textcolor[rgb]{0,0,0.6}{\textbf{2.8}}  & 1   \\
                           & \cellcolor{mygray}96                  & \cellcolor{mygray}31  & \cellcolor{mygray}28 & \cellcolor{mygray}93.3      &\cellcolor{mygray}      & \cellcolor{mygray}31  & \cellcolor{mygray}28 & \cellcolor{mygray}93.3      &\cellcolor{mygray}      & \cellcolor{mygray}2.8    & \cellcolor{mygray}2.8  & \cellcolor{mygray}\textcolor[rgb]{0,0,0.6}{\textbf{0.0}}    & \cellcolor{mygray}0   \\
                           & 180                 & 26  & 23 & 176.4     &      & 25  & 23 & 181.5     &      & 2.0      & \textbf{0.9}  & \textcolor[rgb]{0,0,0.6}{\textbf{1.1}} & 1   \\
                           & \cellcolor{mygray}224                 & \cellcolor{mygray}22  & \cellcolor{mygray}22 & \cellcolor{mygray}221.9     &\cellcolor{mygray}      & \cellcolor{mygray}22  & \cellcolor{mygray}22 & \cellcolor{mygray}221.9     &\cellcolor{mygray}      & \cellcolor{mygray}0.9    & \cellcolor{mygray}0.9  & \cellcolor{mygray}\textcolor[rgb]{0,0,0.6}{\textbf{0.0}}    & \cellcolor{mygray}0   \\
             \midrule
\multirow{5}*{\emph{Ricardo}}& 70                  & 32  & 32 & 69.5     &       & 31  & 32 & 70.1     &       & 0.6    &\textbf{0.1}  & \textcolor[rgb]{0,0,0.6}{\textbf{0.5}} & 1   \\
                             & \cellcolor{mygray}86                  & \cellcolor{mygray}30  & \cellcolor{mygray}30 & \cellcolor{mygray}84.6     &\cellcolor{mygray}       & \cellcolor{mygray}29  & \cellcolor{mygray}30 & \cellcolor{mygray}87.6     &\cellcolor{mygray}       & \cellcolor{mygray}\textbf{1.6}    & \cellcolor{mygray}1.9  & \cellcolor{mygray}\textcolor[rgb]{0,0,0.6}{\textbf{0.3}}  & \cellcolor{mygray}1   \\
                             & 96                  & 29  & 29 & 94.7     &       & 28  & 29 & 97.8     &       & \textbf{1.3}    & 1.9  & \textcolor[rgb]{0,0,0.6}{\textbf{0.6}}  & 1   \\
                             & \cellcolor{mygray}180                 & \cellcolor{mygray}24  & \cellcolor{mygray}24 & \cellcolor{mygray}176.5    &\cellcolor{mygray}       & \cellcolor{mygray}22  & \cellcolor{mygray}24 & \cellcolor{mygray}189.8    &\cellcolor{mygray}       & \cellcolor{mygray}\textbf{1.9}    & \cellcolor{mygray}5.5  & \cellcolor{mygray}\textcolor[rgb]{0,0,0.6}{\textbf{3.6}}  & \cellcolor{mygray}2   \\
                             & 224                 & 23  & 22 & 219.9    &       & 22  & 22 & 228.2    &       & \textbf{1.8}    & 1.9  & \textcolor[rgb]{0,0,0.6}{\textbf{0.1}}    & 1   \\
                \midrule
\multirow{5}*{\emph{Loot}}& \cellcolor{mygray}70                  & \cellcolor{mygray}42  & \cellcolor{mygray}40 & \cellcolor{mygray}69.6     &\cellcolor{mygray}       & \cellcolor{mygray}39  & \cellcolor{mygray}40 & \cellcolor{mygray}71.7     &\cellcolor{mygray}       & \cellcolor{mygray}\textbf{0.6}    & \cellcolor{mygray}2.5  & \cellcolor{mygray}\textcolor[rgb]{0,0,0.6}{\textbf{1.9}}  & \cellcolor{mygray}3   \\
                          & 86                  & 36  & 39 & 84.5     &       & 37  & 38 & 90.2     &       & \textbf{1.8}    & 4.9  & \textcolor[rgb]{0,0,0.6}{\textbf{3.1}}  & 2   \\
                          & \cellcolor{mygray}96                  & \cellcolor{mygray}35  & \cellcolor{mygray}38 & \cellcolor{mygray}94.8     &\cellcolor{mygray}       & \cellcolor{mygray}36  & \cellcolor{mygray}38 & \cellcolor{mygray}92.4     &\cellcolor{mygray}       & \cellcolor{mygray}\textbf{1.2}    & \cellcolor{mygray}3.8  & \cellcolor{mygray}\textcolor[rgb]{0,0,0.6}{\textbf{2.6}}  & \cellcolor{mygray}1   \\
                          & 180                 & 29  & 33 & 179.6    &       & 30  & 33 & 175.0    &         & \textbf{0.2}    & 2.8  & \textcolor[rgb]{0,0,0.6}{\textbf{2.6}}  & 1   \\
                          & \cellcolor{mygray}224                 & \cellcolor{mygray}28  & \cellcolor{mygray}31 & \cellcolor{mygray}223.7    &\cellcolor{mygray}       & \cellcolor{mygray}28  & \cellcolor{mygray}31 & \cellcolor{mygray}223.7    &\cellcolor{mygray}       & \cellcolor{mygray}0.1    & \cellcolor{mygray}0.1  & \cellcolor{mygray}\textcolor[rgb]{0,0,0.6}{\textbf{0.0}}    & \cellcolor{mygray}0   \\
              \midrule
\multirow{5}*{\emph{Queen}}& 70                  & 34  & 42 & 68.2     &       & 37  & 42 & 65.8     &       & \textbf{2.6}    & 6.0    & \textcolor[rgb]{0,0,0.6}{\textbf{3.4}}  & 3   \\
                           & \cellcolor{mygray}86                  & \cellcolor{mygray}30  & \cellcolor{mygray}41 & \cellcolor{mygray}84.2     &\cellcolor{mygray}       & \cellcolor{mygray}30  & \cellcolor{mygray}42 & \cellcolor{mygray}77.3     &\cellcolor{mygray}       & \cellcolor{mygray}\textbf{2.1}    & \cellcolor{mygray}10.1 & \cellcolor{mygray}\textcolor[rgb]{0,0,0.6}{\textbf{8.0}}    & \cellcolor{mygray}1   \\
                           & 96                  & 27  & 41 & 94.3     &       & 28  & 41 & 92.0     &       & \textbf{1.8}    & 4.1  & \textcolor[rgb]{0,0,0.6}{\textbf{2.3}}  & 1   \\
                           & \cellcolor{mygray}180                 & \cellcolor{mygray}22  & \cellcolor{mygray}36 & \cellcolor{mygray}171.9    &\cellcolor{mygray}       & \cellcolor{mygray}22  & \cellcolor{mygray}35 & \cellcolor{mygray}184.9    &\cellcolor{mygray}       & \cellcolor{mygray}4.5    & \cellcolor{mygray}\textbf{2.7}  & \cellcolor{mygray}\textcolor[rgb]{0,0,0.6}{\textbf{1.8}} & \cellcolor{mygray}1   \\
                           & 224                 & 22  & 33 & 220.6    &       & 22  & 33 & 220.6    &       &1.5    &1.5  & \textcolor[rgb]{0,0,0.6}{\textbf{0.0}}    & 0   \\
            \midrule
\rowcolor{mygray}
\multicolumn{10}{>{\columncolor{mygray}}c}{\textbf{Average}} &
\textbf{2.5} & \textbf{3.1} & \textcolor[rgb]{0,0,0.6}{\textbf{1.2}}
& \textbf{0.7} \\\hline
      \bottomrule
  \end{tabular}
}
\end{table*}

\begin{table*}[t!]
\newcommand{\tabincell}[2]{\begin{tabular}{@{}#1@{}}#2\end{tabular}}
\centering \caption{Rate-PSNR Performance of \textbf{ESA} and
\textbf{PBA}} \label{tab:rd}
\renewcommand\tabcolsep{3pt}
  \begin{tabular}{cccccccccccccccc}
      \toprule
      \midrule
\multirow{3}*{\tabincell{l}{Point\\Cloud}} &\multirow{3}*{\tabincell{l}{ Target\\ Bitrate \\($kbpmp$)}} &\multicolumn{5}{c}{\textbf{ESA}} & &\multicolumn{5}{c}{\textbf{PBA}}& &\multicolumn{2}{c}{\tabincell{l}{BD-PSNR\\\quad($dB$)}}\\
     \cmidrule{3-7}\cmidrule{9-13}\cmidrule{15-16}
      &         & \multicolumn{2}{c}{\tabincell{l}{Bitrate\\($kbpmp$)}} & &\multicolumn{2}{c}{\tabincell{l}{PSNR\\($dB$)}}& &\multicolumn{2}{c}{\tabincell{l}{Bitrate\\($kbpmp$)}}& &\multicolumn{2}{c}{\tabincell{l}{PSNR\\($dB$)}}& &\multirow{2}*{$\omega=0.25$}& \multirow{2}*{$\omega=0.5$} \\
      \cmidrule{3-4}\cmidrule{6-7}\cmidrule{9-10}\cmidrule{12-13}
      & &$\omega=0.25$ &$\omega=0.5$& &$\omega=0.25$ &$\omega=0.5$ & &$\omega=0.25$ &$\omega=0.5$& &$\omega=0.25$ &$\omega=0.5$& &&\\
 \midrule
\multirow{5}*{\emph{Andrew}}& \cellcolor{mygray}240      & \cellcolor{mygray}233     & \cellcolor{mygray}233 &\cellcolor{mygray} &\cellcolor{mygray}32.8  &\cellcolor{mygray}34.5 &\cellcolor{mygray} & \cellcolor{mygray}233     & \cellcolor{mygray}233 &\cellcolor{mygray} &\cellcolor{mygray}32.8  &\cellcolor{mygray}34.5 &\cellcolor{mygray} &\cellcolor{mygray}        &\cellcolor{mygray}       \\
                            & 450      & 445     & 445 & &33.8  &35.6 & & 445     & 445 & &33.8  &35.6 & &        &       \\
                            & \cellcolor{mygray}600      & \cellcolor{mygray}594     & \cellcolor{mygray}594 &\cellcolor{mygray} &\cellcolor{mygray}34.2  &\cellcolor{mygray}36.0 &\cellcolor{mygray} & \cellcolor{mygray}594     & \cellcolor{mygray}594 &\cellcolor{mygray} &\cellcolor{mygray}34.2  &\cellcolor{mygray}36.0 &\cellcolor{mygray} & \cellcolor{mygray}\textbf{0.00}      & \cellcolor{mygray}\textbf{0.00}     \\
                            & 760      & 683     & 683 & &34.4  &36.1 & & 786     & 786 & &34.5  &36.3 & &        &       \\
                            & \cellcolor{mygray}1070     & \cellcolor{mygray}1019    & \cellcolor{mygray}1019 &\cellcolor{mygray} &\cellcolor{mygray}34.8  &\cellcolor{mygray}36.5 &\cellcolor{mygray} & \cellcolor{mygray}1019    & \cellcolor{mygray}1019&\cellcolor{mygray} &\cellcolor{mygray}34.8    &\cellcolor{mygray}36.5 &\cellcolor{mygray} &\cellcolor{mygray}        &\cellcolor{mygray}       \\
\midrule
 \multirow{5}*{\emph{Phil}}
                & 240                 & 237     & 237 & & 34.5 & 36.3 &  & 237     & 237 & & 34.5 & 36.3 &  &        &       \\
                & \cellcolor{mygray}450                 & \cellcolor{mygray}424     & \cellcolor{mygray}424 &\cellcolor{mygray} & \cellcolor{mygray}35.6 & \cellcolor{mygray}37.3 &\cellcolor{mygray}  & \cellcolor{mygray}424     & \cellcolor{mygray}424 &\cellcolor{mygray} & \cellcolor{mygray}35.6 & \cellcolor{mygray}37.3 &\cellcolor{mygray} &\cellcolor{mygray}        &\cellcolor{mygray}       \\
                & 600                 & 553     & 553 & & 36.0 & 37.7 &  & 636     & 636 & & 36.1 & 37.9 & & \textbf{0.00}   & \textbf{0.00}  \\
                & \cellcolor{mygray}760                 & \cellcolor{mygray}730     & \cellcolor{mygray}730 &\cellcolor{mygray} & \cellcolor{mygray}36.3 & \cellcolor{mygray}38.1 &\cellcolor{mygray}  & \cellcolor{mygray}730     & \cellcolor{mygray}730 &\cellcolor{mygray} & \cellcolor{mygray}36.3 & \cellcolor{mygray}38.1 &\cellcolor{mygray} &\cellcolor{mygray}        &\cellcolor{mygray}       \\
                & 1070                & 949     & 949 & & 36.6 & 38.3 &  & 949     & 949 & & 36.6 & 38.3 & &        &       \\
\midrule
\multirow{5}*{\emph{Longdress}}
                & \cellcolor{mygray}240                 & \cellcolor{mygray}237     & \cellcolor{mygray}237 &\cellcolor{mygray} & \cellcolor{mygray}28.8   & \cellcolor{mygray}30.6 &\cellcolor{mygray} & \cellcolor{mygray}237     & \cellcolor{mygray}237 &\cellcolor{mygray} & \cellcolor{mygray}28.8   & \cellcolor{mygray}30.6 &\cellcolor{mygray} &\cellcolor{mygray}        &\cellcolor{mygray}       \\
                & 450                 & 449     & 449 & & 30.5 & 32.3 & & 449     & 449 & & 30.5 & 32.3 & &        &       \\
                 & \cellcolor{mygray}600                 & \cellcolor{mygray}586     & \cellcolor{mygray}586 &\cellcolor{mygray} & \cellcolor{mygray}31.2 & \cellcolor{mygray}33.0 &\cellcolor{mygray} & \cellcolor{mygray}586     & \cellcolor{mygray}586 &\cellcolor{mygray} & \cellcolor{mygray}31.2 & \cellcolor{mygray}33.0 &\cellcolor{mygray} & \cellcolor{mygray}\textbf{0.00}  & \cellcolor{mygray}\textbf{0.00} \\
                 & 760                 & 752     & 752 & & 31.7 & 33.4 & & 752     & 752 & & 31.7 & 33.4 & &        &       \\
                 & \cellcolor{mygray}1070                & \cellcolor{mygray}1067    & \cellcolor{mygray}1067 &\cellcolor{mygray} & \cellcolor{mygray}32.2 & \cellcolor{mygray}34.0 &\cellcolor{mygray} & \cellcolor{mygray}1067    & \cellcolor{mygray}1067 &\cellcolor{mygray} & \cellcolor{mygray}32.2 & \cellcolor{mygray}34.0 &\cellcolor{mygray} &\cellcolor{mygray}        &\cellcolor{mygray}       \\
\midrule
\multirow{5}*{\emph{Redandblack}}
                & 240                 & 238     & 238 &  & 36.0 & 37.8 & & 230     & 234 & & 35.9 & 37.7 & &        &       \\
                 & \cellcolor{mygray}450                 & \cellcolor{mygray}448     & \cellcolor{mygray}448 &\cellcolor{mygray} & \cellcolor{mygray}37.7 & \cellcolor{mygray}39.4 &\cellcolor{mygray}  & \cellcolor{mygray}430     & \cellcolor{mygray}440 &\cellcolor{mygray} & \cellcolor{mygray}37.6   & \cellcolor{mygray}39.4 &\cellcolor{mygray} &\cellcolor{mygray}        &\cellcolor{mygray}       \\
                 & 600                 & 572     & 572 & & 38.3 & 40.1 &  & 572     & 572 & & 38.3 & 40.1 & & \textbf{0.01}      & \textbf{0.00}     \\
                 & \cellcolor{mygray}760                 & \cellcolor{mygray}722     & \cellcolor{mygray}722 &\cellcolor{mygray} & \cellcolor{mygray}38.7 & \cellcolor{mygray}40.5 &\cellcolor{mygray} & \cellcolor{mygray}722     & \cellcolor{mygray}722  &\cellcolor{mygray} & \cellcolor{mygray}38.7 & \cellcolor{mygray}40.5 &\cellcolor{mygray} &\cellcolor{mygray}        &\cellcolor{mygray}       \\
                 & 1070                & 1011    & 1011 & & 39.3 & 41.0 &  & 1011    & 1011 & & 39.3 & 41.0  &  &        &       \\
\midrule
\multirow{5}*{\emph{David}}
                & \cellcolor{mygray}70                  & \cellcolor{mygray}70      & \cellcolor{mygray}70 &\cellcolor{mygray}  & \cellcolor{mygray}42.4 & \cellcolor{mygray}44.1 &\cellcolor{mygray}  & \cellcolor{mygray}73      & \cellcolor{mygray}68 &\cellcolor{mygray}  & \cellcolor{mygray}42.7 & \cellcolor{mygray}44.1 &\cellcolor{mygray} &\cellcolor{mygray}        &\cellcolor{mygray}       \\
                 & 86                  & 86      & 86 &   & 43.0 & 44.8 & & 81      & 84  & & 42.9 & 44.7 & &        &       \\
                 & \cellcolor{mygray}96                  & \cellcolor{mygray}93      & \cellcolor{mygray}93 &\cellcolor{mygray}  & \cellcolor{mygray}43.4   & \cellcolor{mygray}45.1 &\cellcolor{mygray} & \cellcolor{mygray}91      & \cellcolor{mygray}93 &\cellcolor{mygray}  & \cellcolor{mygray}43.3 & \cellcolor{mygray}45.1 &\cellcolor{mygray} & \cellcolor{mygray}\textbf{0.04}  & \cellcolor{mygray}\textbf{0.03}\\
                 & 180                 & 176     & 176 & & 44.9 & 46.7 & & 172     & 182 & & 44.9 & 46.8 & &        &       \\
                 & \cellcolor{mygray}224                 & \cellcolor{mygray}222     & \cellcolor{mygray}222 &\cellcolor{mygray} & \cellcolor{mygray}45.3   & \cellcolor{mygray}47.1 &\cellcolor{mygray}  & \cellcolor{mygray}222     & \cellcolor{mygray}222 &\cellcolor{mygray} & \cellcolor{mygray}45.3   & \cellcolor{mygray}47.1 &\cellcolor{mygray}  &\cellcolor{mygray}        &\cellcolor{mygray}       \\
\midrule
\multirow{5}*{\emph{Ricardo}}
                & 70                  & 69      & 70 &  & 41.8 & 43.5 & & 70      & 70 &  & 41.8 & 43.5 & &        &       \\
                 & \cellcolor{mygray}86                  & \cellcolor{mygray}85      & \cellcolor{mygray}85 &\cellcolor{mygray}  & \cellcolor{mygray}42.3 & \cellcolor{mygray}44.0 &\cellcolor{mygray} & \cellcolor{mygray}85      & \cellcolor{mygray}88 &\cellcolor{mygray}  & \cellcolor{mygray}42.3 & \cellcolor{mygray}44.1 &\cellcolor{mygray} &\cellcolor{mygray}        &\cellcolor{mygray}       \\
                 & 96                  & 95      & 95 &  & 42.6 & 44.4 & & 95      & 98 &  & 42.6 & 44.3 & & \textbf{0.01}     & \textbf{-0.07}  \\
                 & \cellcolor{mygray}180                 & \cellcolor{mygray}177     & \cellcolor{mygray}177 &\cellcolor{mygray} & \cellcolor{mygray}43.9 & \cellcolor{mygray}45.7 &\cellcolor{mygray}  & \cellcolor{mygray}182     & \cellcolor{mygray}190 &\cellcolor{mygray} & \cellcolor{mygray}44.0  & \cellcolor{mygray}45.8 &\cellcolor{mygray} &\cellcolor{mygray}        &\cellcolor{mygray}       \\
                 & 224                 & 220     & 220 & & 44.3 & 46.1 & & 228     & 228 & & 44.3 & 46.1 & &        &       \\
\midrule
\multirow{5}*{\emph{Loot}}
                & \cellcolor{mygray}70                  & \cellcolor{mygray}70      & \cellcolor{mygray}70 &\cellcolor{mygray}  & \cellcolor{mygray}34.0 & \cellcolor{mygray}35.8 &\cellcolor{mygray} & \cellcolor{mygray}70      & \cellcolor{mygray}72 &\cellcolor{mygray}  & \cellcolor{mygray}34.1 & \cellcolor{mygray}35.9 &\cellcolor{mygray} &\cellcolor{mygray}        &\cellcolor{mygray}       \\
                 & 86                  & 84      & 84 &  & 34.7 & 36.4 & & 94      & 90 &  & 34.9 & 36.7 & &        &       \\
                 & \cellcolor{mygray}96                  & \cellcolor{mygray}95      & \cellcolor{mygray}95 &\cellcolor{mygray}  & \cellcolor{mygray}35.1 & \cellcolor{mygray}36.8 &\cellcolor{mygray}  & \cellcolor{mygray}101     & \cellcolor{mygray}92 &\cellcolor{mygray}  & \cellcolor{mygray}35.3 & \cellcolor{mygray}36.8 &\cellcolor{mygray} & \cellcolor{mygray}\textbf{0.00}   & \cellcolor{mygray}\textbf{0.05} \\
                 & 180                 & 179     & 180 & & 37.2 & 38.9 & & 189     & 175 & & 37.4   & 38.9 &  &        &       \\
                 & \cellcolor{mygray}224                 & \cellcolor{mygray}224     & \cellcolor{mygray}224 &\cellcolor{mygray} & \cellcolor{mygray}37.9 & \cellcolor{mygray}39.7 &\cellcolor{mygray} & \cellcolor{mygray}221     & \cellcolor{mygray}224 &\cellcolor{mygray} & \cellcolor{mygray}37.9 & \cellcolor{mygray}39.7 &\cellcolor{mygray} &\cellcolor{mygray}        &\cellcolor{mygray}       \\
\midrule
\multirow{5}*{\emph{Queen}}
                & 70                  & 68      & 68 &  & 30.3 & 32.0 & & 66      & 66  &  & 29.5 & 31.2 & &        &       \\
                 & \cellcolor{mygray}86                  & \cellcolor{mygray}84      & \cellcolor{mygray}84  &\cellcolor{mygray} & \cellcolor{mygray}31.4   & \cellcolor{mygray}33.2 &\cellcolor{mygray} & \cellcolor{mygray}77      & \cellcolor{mygray}77  &\cellcolor{mygray}  & \cellcolor{mygray}31.1 & \cellcolor{mygray}32.8 &\cellcolor{mygray} &\cellcolor{mygray}        &\cellcolor{mygray}       \\
                 & 96                  & 94      & 94  & & 32.0 & 33.7 & & 89      & 92  &  & 31.7 & 33.5 & & \textbf{0.02}   & \textbf{-0.01} \\
                 & \cellcolor{mygray}180                 & \cellcolor{mygray}172     & \cellcolor{mygray}172 &\cellcolor{mygray} & \cellcolor{mygray}33.9 & \cellcolor{mygray}35.7 &\cellcolor{mygray} & \cellcolor{mygray}185     & \cellcolor{mygray}185 &\cellcolor{mygray}  & \cellcolor{mygray}34.1 & \cellcolor{mygray}35.9 &\cellcolor{mygray} &\cellcolor{mygray}        &\cellcolor{mygray}       \\
                 & 224                 & 221     & 221 & & 34.6 & 36.4 & & 221     & 221  & & 34.6 & 36.4 & &        &       \\
\midrule
\rowcolor{mygray}
\multicolumn{14}{>{\columncolor{mygray}}c}{\textbf{Average}}
&\textbf{0.01} &\textbf{0.00}
\\\hline
      \bottomrule
  \end{tabular}
\end{table*}

\begin{figure*}[t!]
\centering \subfigure[]{ \label{fig10:subfig:a}
\includegraphics[width=0.485\columnwidth]{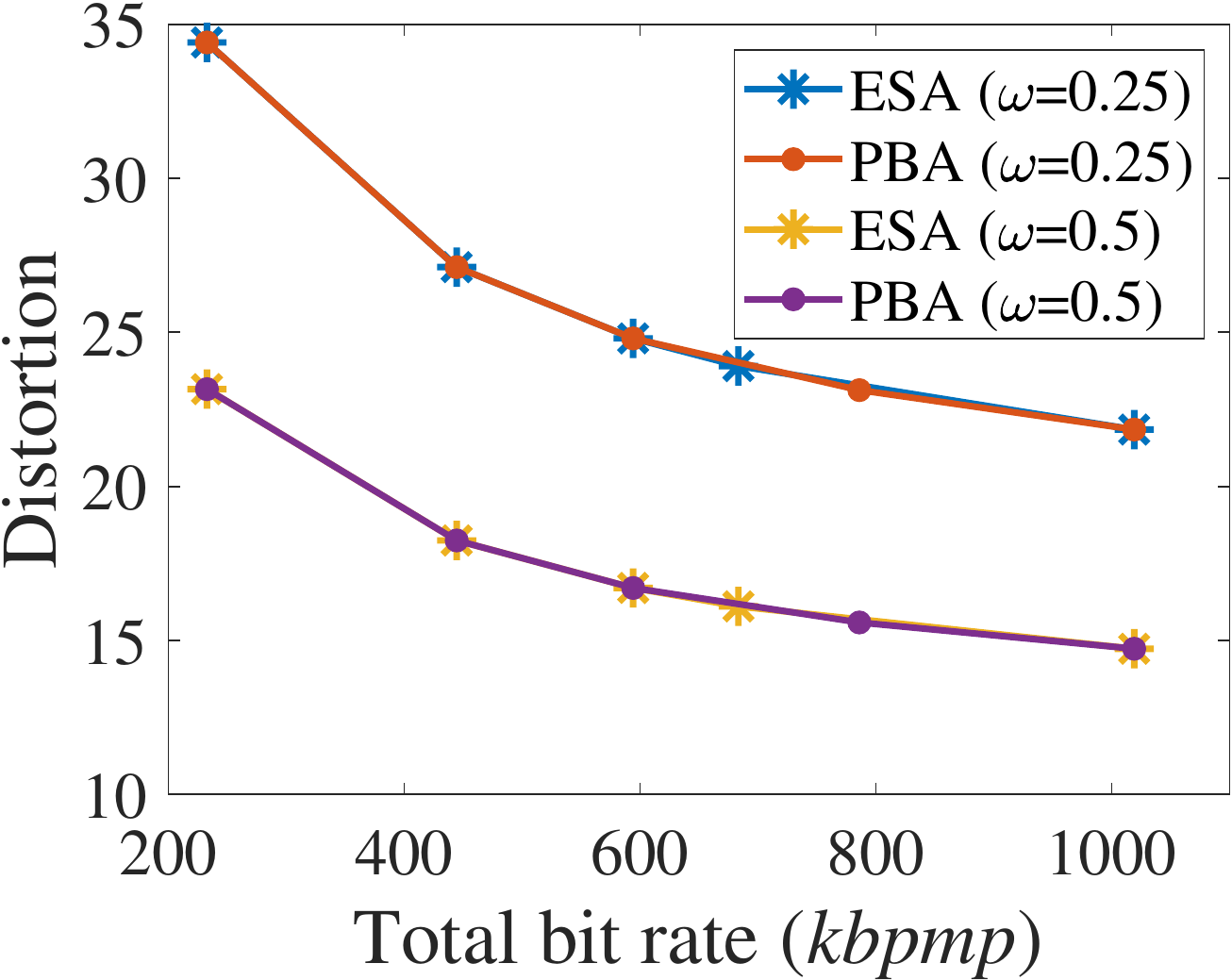}} 
\subfigure[]{ \label{fig10:subfig:b}
\includegraphics[width=0.485\columnwidth]{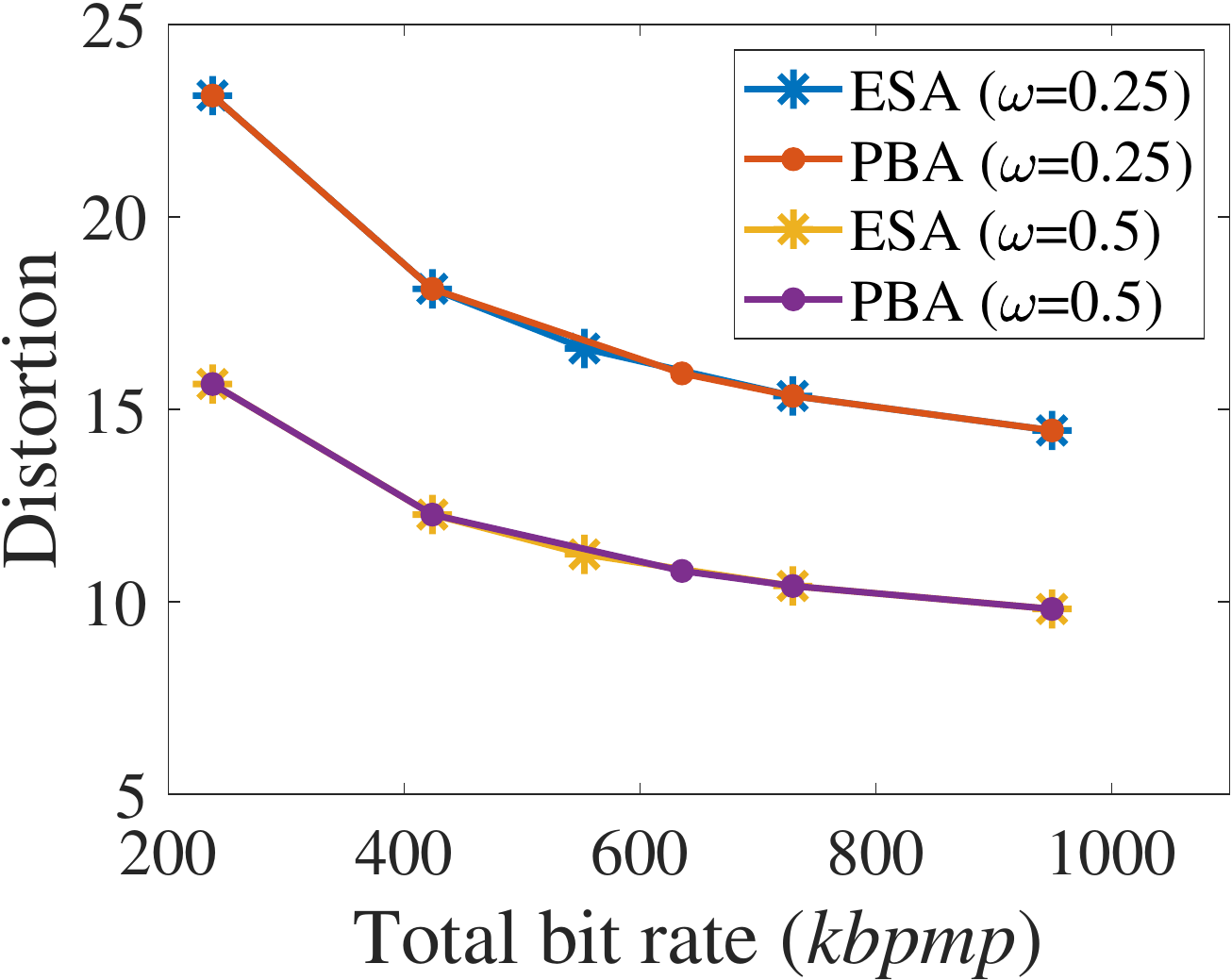}}
\subfigure[]{ \label{fig10:subfig:c}
\includegraphics[width=0.485\columnwidth]{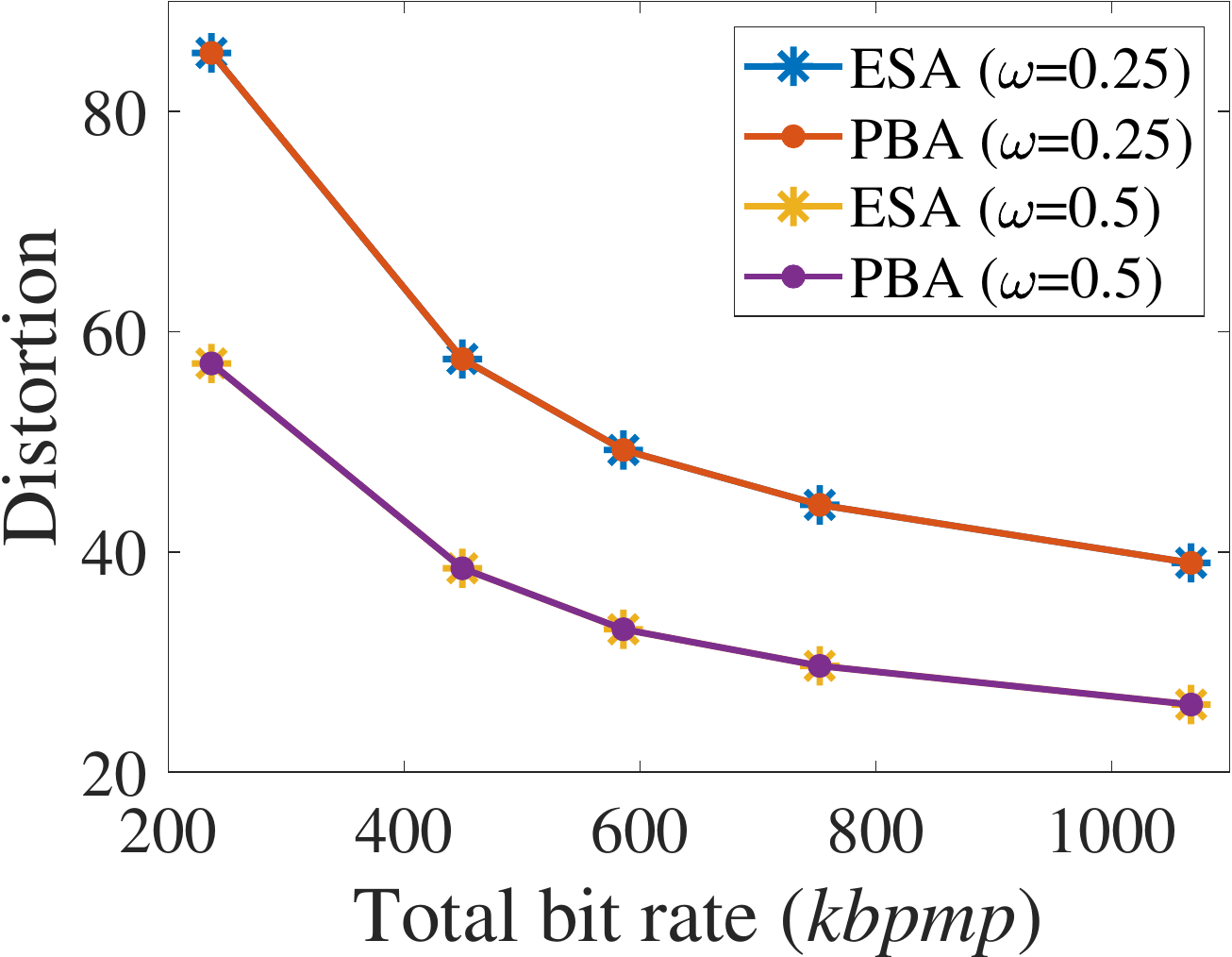}}
\subfigure[]{ \label{fig10:subfig:d}
\includegraphics[width=0.485\columnwidth]{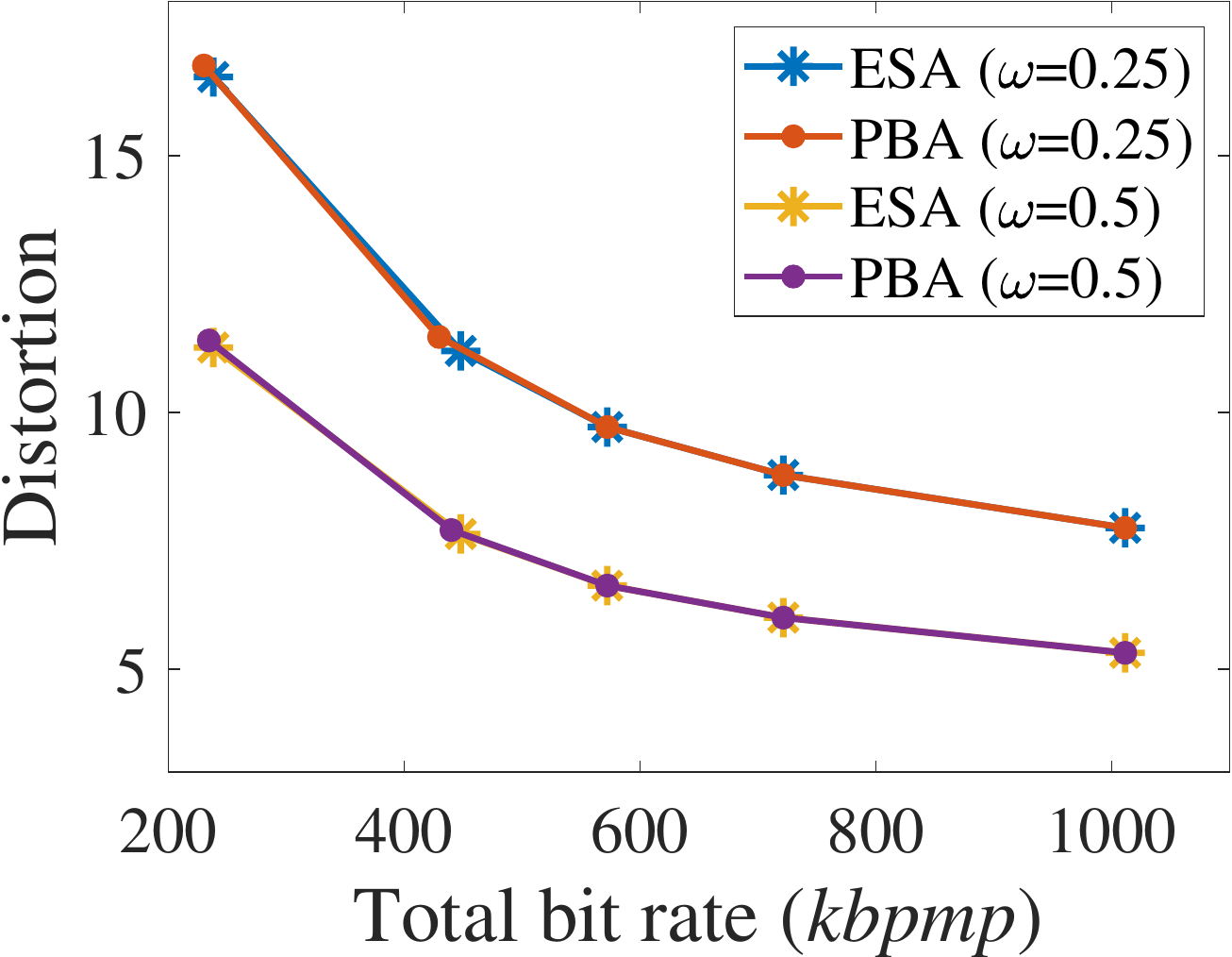}}
\subfigure[]{ \label{fig10:subfig:e}
\includegraphics[width=0.485\columnwidth]{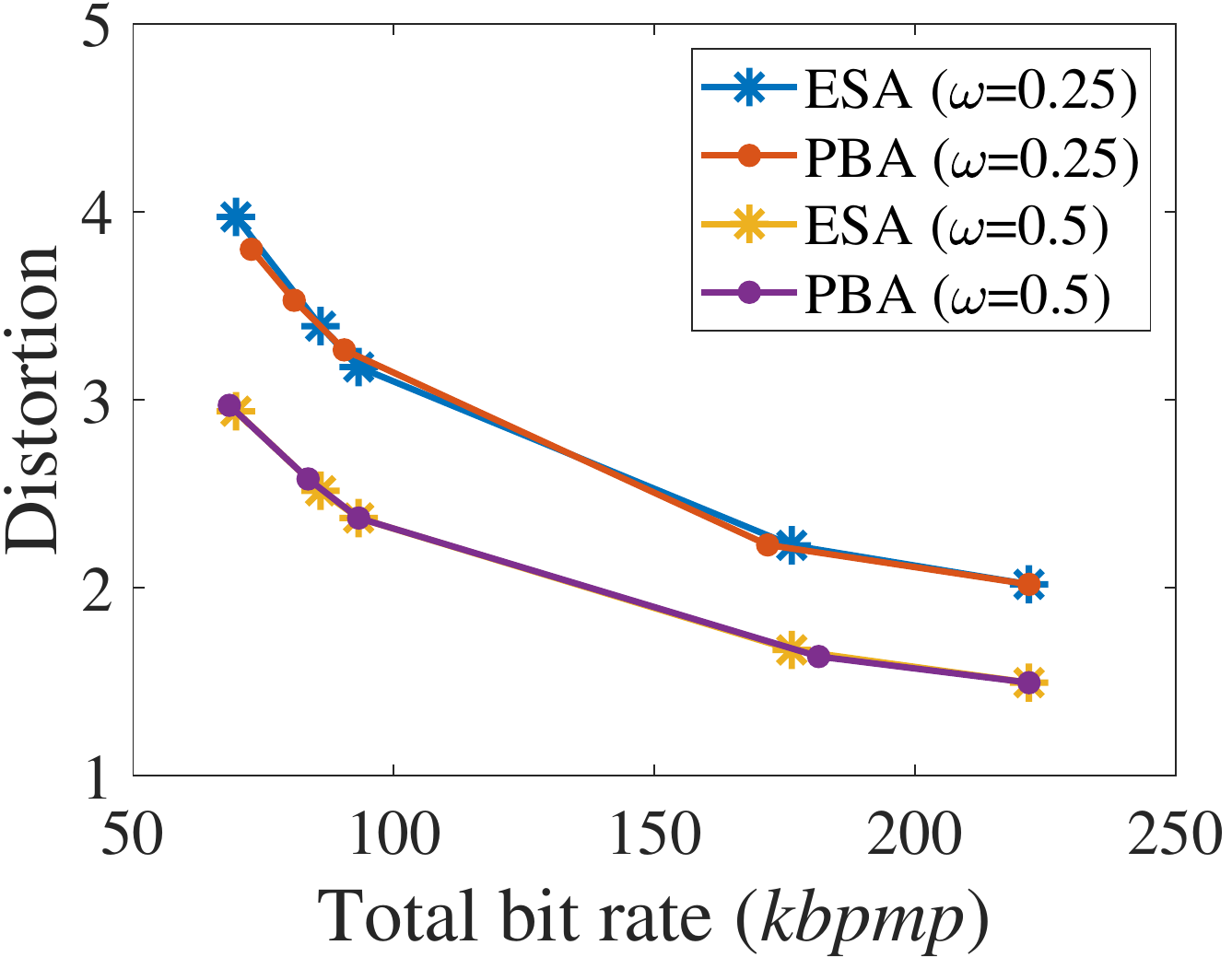}}
\subfigure[]{ \label{fig10:subfig:f}
\includegraphics[width=0.485\columnwidth]{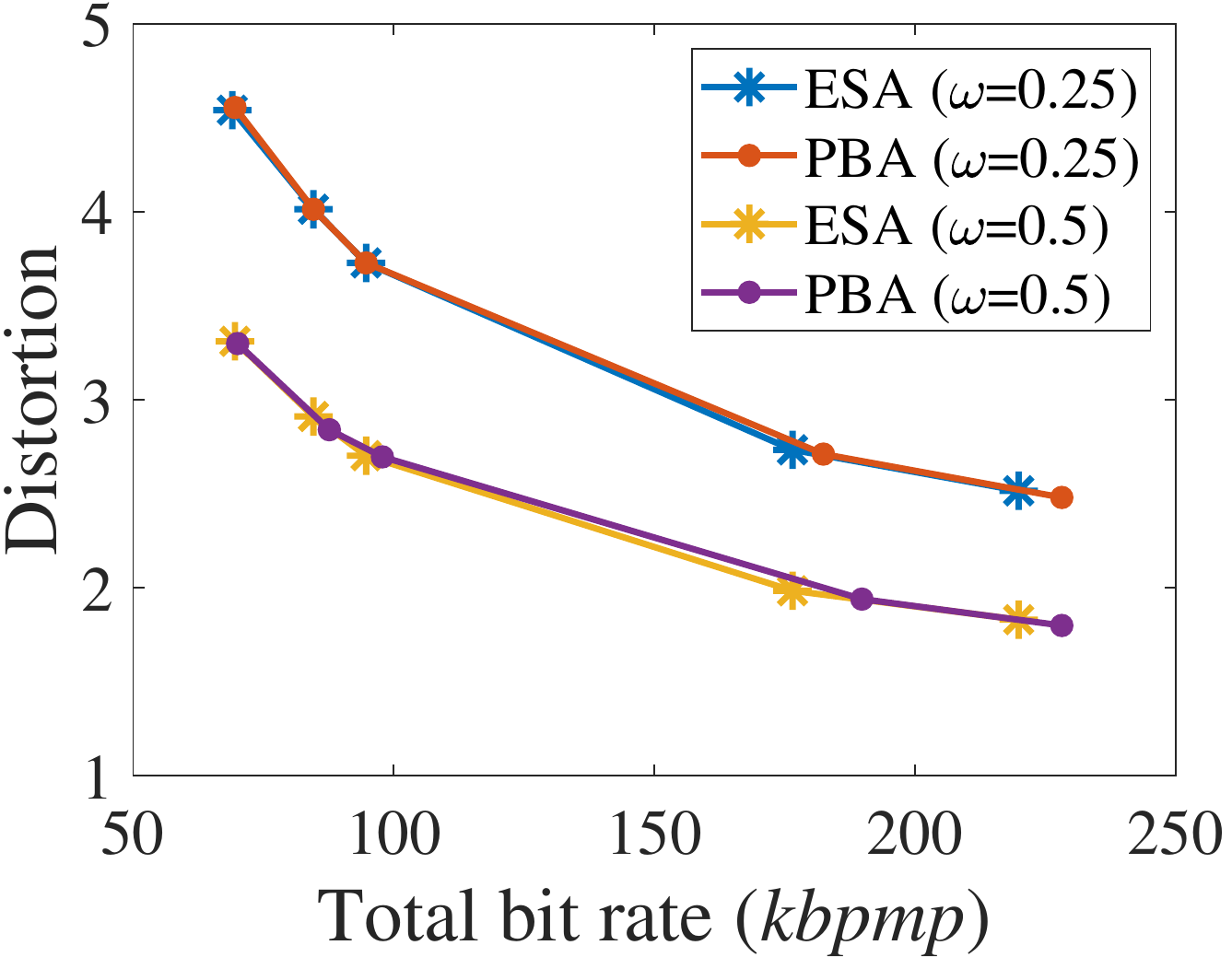}}
\subfigure[]{ \label{fig10:subfig:g}
\includegraphics[width=0.485\columnwidth, height=3.35cm]{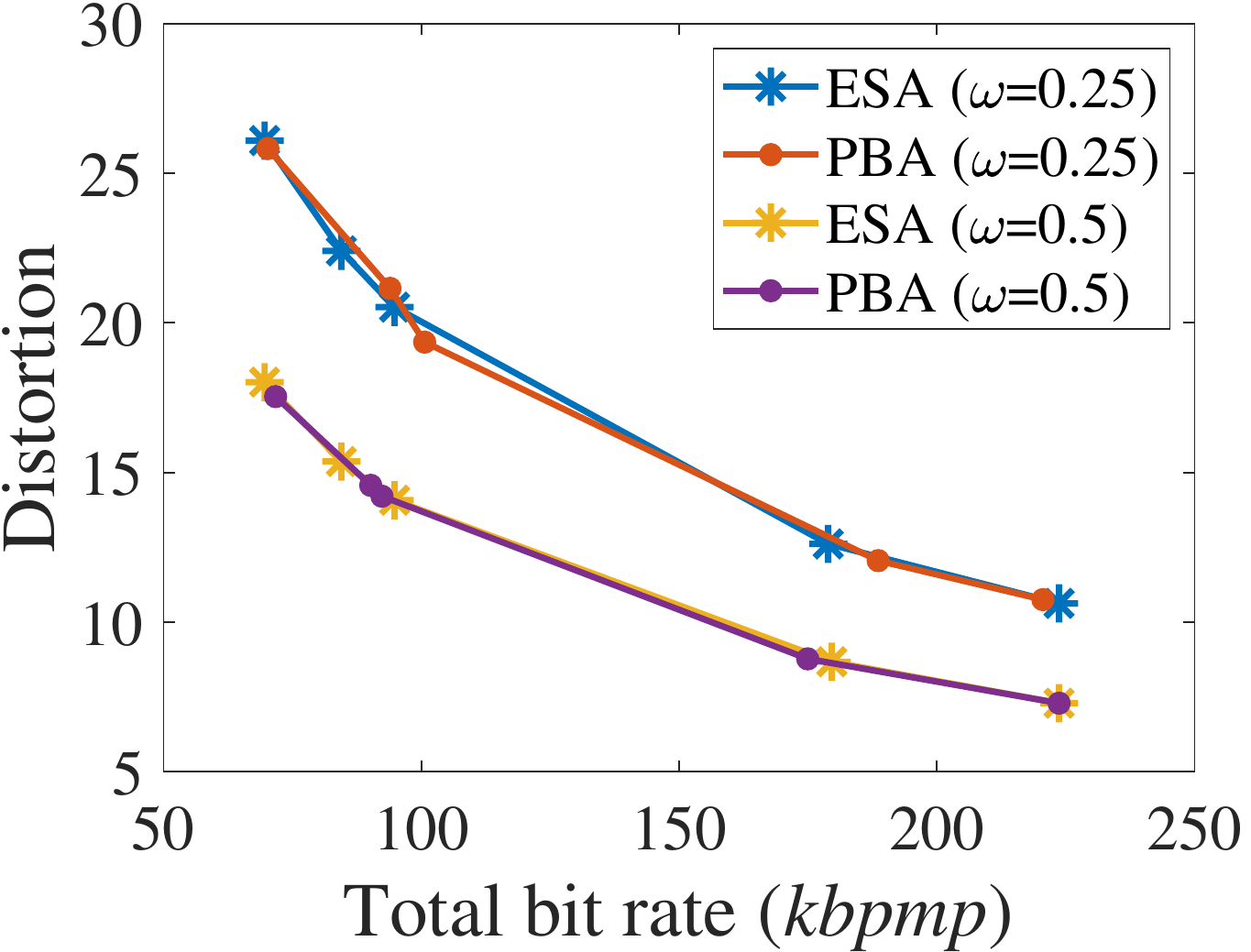}}
\subfigure[]{ \label{fig10:subfig:h}
\includegraphics[width=0.485\columnwidth, height=3.35cm]{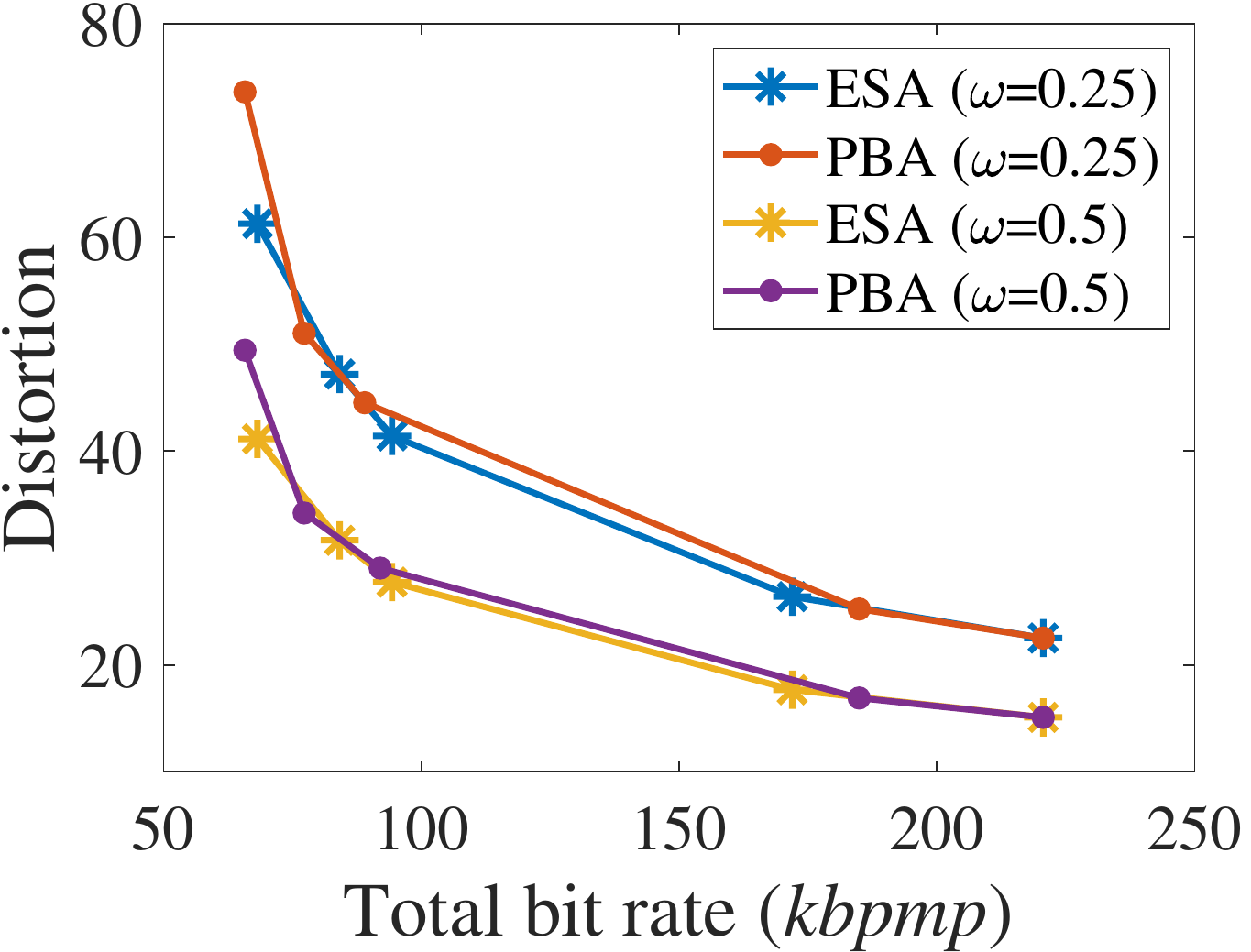}}
\caption{R-D performance of the proposed algorithm (\textbf{PBA})
and exhaustive search (\textbf{ESA}). (a) \emph{Andrew},
(b)\emph{Phil} , (c) \emph{Longdress}, (d) \emph{Redandblack}, (e)
\emph{David}, (f) \emph{Ricardo}, (g) \emph{Loot}, (h)
\emph{Queen}.}
 \label{fig10}
\end{figure*}

Because the color of the 3D point cloud sequences in
Figs.~\ref{fig9:subfig:a}-\ref{fig9:subfig:d} is more diverse than
that of the sequences in
Figs.~\ref{fig9:subfig:e}-\ref{fig9:subfig:h}, we divided the point
clouds in Fig.~\ref{fig9} into two groups, a complex 3D point cloud
group (\emph{Andrew}, \emph{Longdress}, \emph{Phil} and
\emph{Redandblack}) and a simple 3D point cloud group
(\emph{Ricardo}, \emph{Loot}, \emph{David}, and \emph{Queen}). The
performance of exhaustive search was used as the benchmark. In
exhaustive search, a 3D point cloud was first encoded by all the
possible geometry and color quantization step pairs (ranging from 8
to 80), which correspond to $QP$ values 22, 23, 24, \dots, 42. Then
the subset of admissible pairs (that is, those for which the bitrate
is smaller than the target bitrate) was identified. Finally, the
pair that gave the smallest distortion was selected from this
subset.

In the proposed method, to derive the rate and distortion models,
the point clouds were empirically pre-encoded with three
geometry-color QP pairs (33, 25), (34, 35), and (24, 33). The
distortion model parameters $a$, $b$ and $c$ were computed by
solving~\eqref{eq:deqs}, and the rate model parameters $\gamma_g $,
$\theta_g$, $\gamma_c$, and $\theta_c$ were obtained by
solving~\eqref{eq:reqs}. Then, given the target bitrate $R_T$, the
optimal $Q_{g,opt}$ and $Q_{c,opt}$ were obtained by
solving~\eqref{eq:rdo2} using the interior point method.

\subsection{Bit Allocation Accuracy of Proposed Algorithm} 
To evaluate the accuracy of the proposed bit allocation algorithm
for the tested point clouds, we set different target bitrates
according to the geometry and color characteristics of each sequence
to cover different compression levels. For the complex 3D point
clouds, the common target bitrates are 240 kbpmp, 450 kbpmp, 600
kbpmp, 760 kbpmp, and 1070 kbpmp. For the simple 3D point cloud
group, the target bitrates are 70 kbpmp, 86 kbpmp, 96 kbpmp, 180
kbpmp, and 224 kbpmp. In practical applications, the target bitrate
for geometry and color can be obtained by subtracting the bitrate of
the occupancy map and the auxiliary information, which can be
obtained by pre-encoding. To evaluate the bit allocation accuracy,
we used the bitrate error ($BE$), defined as
\begin{equation}
\label{eq:BE}
\begin{aligned}
BE &=\frac{\left|B_{actual}-B_{target} \right|}{B_{target}} \times
100\%,
\end{aligned}
\end{equation}
where $B_{actual}$ is the actual bitrate and $B_{target}$ represents
the target bitrate. The lower the BE, the more accurate the
algorithm. Because the proposed bit allocation algorithm allocated
the bits for geometry and color components by selecting the QPs, QP
error ($QPE$) was also used to measure the performance as follows:
\begin{equation}
\label{eq:QPE}
\begin{aligned}
QPE &=\left|QP_{g,PBA}-QP_{g,ESA} \right| +
\left|QP_{c,PBA}-QP_{c,ESA} \right|,
\end{aligned}
\end{equation}
where $QP_{g,PBA}$ and $QP_{c,PBA}$ denote the geometry and color
QPs obtained from the proposed algorithm, and $QP_{g,ESA}$ and
$QP_{c,ESA}$ represent the geometry and color QPs obtained from
exhaustive search.

Table~\ref{tab:omega25} and Table~\ref{tab:omega50} show the $BE$
and $QPE$ of the proposed bit allocation algorithm (\textbf{PBA})
and exhaustive search (\textbf{ESA}) for different values of
$\omega$ (0.25 and 0.5). Note that \textbf{ESA} produces an optimal
solution but has a much higher computational cost. As shown in
Table~\ref{tab:omega25} and Table~\ref{tab:omega50}, the $BE$ of
\textbf{ESA} was as small as zero, while its average was 2.6\% and
2.5\% when $\omega$ was set to 0.25 and 0.5, respectively. For
\textbf{PBA}, $BE$ was as low as 0.1\%, while its average was about
3.7\% and 3.1\% when $\omega$ was set to 0.25 and 0.5, respectively.
The average absolute difference in BE between \textbf{ESA} and
\textbf{PBA} was only 1.7\% and 1.2\% for $\omega =0.25$ and 0.5,
respectively. On the other hand, the average $QPE$ was only 1.1 when
$\omega$ was set to 0.25 and only 0.7 when $\omega$ was set to 0.5.
In 50\% of the cases, our algorithm (\textbf{PBA}) found the same
solution as exhaustive search (\textbf{ESA}), so the BE of the two
algorithms was the same. In the remaining cases, our algorithm found
a suboptimal solution (see the discussion at the end of Section
III), which had a higher BE in 36.25\% of the cases and a lower one
in 13.75\% of the cases.

\subsection{Rate-Distortion Performance} 
In addition to bit allocation accuracy, the rate-distortion
performance should also be taken into account. After determining the
coding parameters with \textbf{ESA} and \textbf{PBA}, we compressed
the point clouds and computed their respective distortion and peak
signal-to-noise ratio (PSNR) using the PC\_error reference
software~\cite{ctc}.

Fig.~\ref{fig10} shows the distortion $D=\omega D_g+(1-\omega) D_c$
as a function of the bitrate (in $kbpmp$) for \textbf{PBA} and
\textbf{ESA}. We can see that the rate-distortion performance of the
proposed algorithm was very close to that of exhaustive search.

In addition to the distortion (measured in MSE), we used the BD-PSNR
~\cite{bdpsnr} to compare the performance of the two algorithms. In
general, it is necessary to normalize with respect to the peak value
when converting MSE into $PSNR$. However, the peak values of
geometry and color are completely different. To calculate the $PSNR$
of the reconstructed 3D point cloud reasonably, the geometry and
color values were both normalized to [0,1]. Hence, the $PSNR$ of the
reconstructed 3D point cloud was calculated as:
\begin{equation}
\label{eq:psnr}
\begin{aligned}
PSNR &=10\log_{10}\left[\frac{1}{NMSE(NMSE_g,NMSE_c)}\right],
\end{aligned}
\end{equation}
where $NMSE_g$ and $NMSE_c$ are the normalized geometry and color (Y
channel) distortion (i.e., $D_g$ and $D_c$), respectively and
$NMSE(NMSE_g,NMSE_c)=\omega NMSE_g+(1-\omega) NMSE_c$.

Table~\ref{tab:rd} shows that there is almost no BD-PSNR degradation
of \textbf{PBA} when $\omega=0.5$. Interestingly, an average 0.01 dB
BD-PSNR gain was achieved when $\omega=0.25$. The main reason is
that \textbf{ESA} is optimal for the distortion $D=\omega
D_g+(1-\omega) D_c$ but not necessarily optimal for the
PSNR~\eqref{eq:psnr}.

\subsection{Complexity Comparison} 
We run the experiments on a PC with a 3.40 GHz Intel Core i7
Processor and 8.00 GB RAM and used the encoding time to evaluate the
time complexity. The ratio between the encoding time of \textbf{PBA}
and that of \textbf{ESA} was used to define the complexity quotient
(CQ) as
\begin{equation}
\label{eq:CQ}
\begin{aligned}
CQ &=\frac{T_{PBA}}{T_{ESA}} \times 100\%,
\end{aligned}
\end{equation}
where $T_{PBA}$ and $T_{ESA}$ denote the encoding time of
\textbf{PBA} and \textbf{ESA}, respectively. The time complexity of
\textbf{ESA} and \textbf{PBA} mainly depends on the pre-encoding
times. \textbf{ESA} needs to pre-encode the 3D point cloud for all
possible combinations of QPs for the geometry and color components.
Because both the geometry and color QP search range was [22, 42]
with a search step size of 1, a 3D point cloud needs to be encoded
$21 \times 21 = 441$ times to find the optimal $Q_g$ and $Q_c$ with
\textbf{ESA}. In contrast, only three pre-encodings were required by
\textbf{PBA} to compute the model parameters. As the time complexity
of the interior point method is very small compared to the
pre-encoding procedure (for example the time spent to obtain the
optimal $Q_g$ and $Q_c$ by the interior point method was only 1.47 s
for the $Andrew$ point cloud, while a single pre-encoding required
120.96 s), the time complexity of \textbf{PBA} was only about 0.68\%
of that of \textbf{ESA}, as shown in Table~\ref{tab:timecomplity}.
\begin{table}[t!]
\centering \caption{Complexity Comparison for \textbf{ESA} and
\textbf{PBA}} \label{tab:timecomplity}
  \begin{tabular}{c|c|c|c}
    \toprule
    \midrule
    \multirow{2}*{Point Cloud} &\multicolumn{2}{c|}{Encoding Time
    ($s$)} &\multirow{2}*{CQ(\%)}\\
    \cline {2-3}
    & \textbf{ESA} & \textbf{PBA} &\\\hline
    \emph{Andrew}  &53342.84    &364.08  &0.68\\
    \emph{Phil}   &56716.51    &366.50  &0.65\\
    \emph{Longdress} &38476.52    &247.17  &0.64\\
    \emph{Redandblack} &33958.65    &234.23  &0.69\\
    \emph{David}   &50537.59    &335.58  &0.66\\
    \emph{Ricardo}    &38867.25    &251.00  &0.65\\
    \emph{Loot}   &35673.09    &256.62  &0.72\\
    \emph{Queen} &41749.56    &298.17  &0.71\\\hline
    \multicolumn{3}{c|}{Average} &\textbf{0.68} \\
    \midrule
    \bottomrule
  \end{tabular}
\end{table}
\section{Conclusion} 
This paper presented a model-based joint bit allocation algorithm
for the V-PCC encoder. To reduce the time complexity of exhaustive
search as well as preserve its rate-distortion performance, we first
derived rate and distortion models for point clouds through
theoretical analysis and statistical validation. Based on the
derived rate and distortion models, the optimal bit allocation
problem was formulated as a convex constrained optimization problem
and solved by an interior point method. Model parameters were
calculated by pre-encoding a 3D point cloud only three times.
Experimental results showed that the bit allocation accuracy and the
rate-distortion performance of the PBA were very close to those of
exhaustive search at only 0.68\% of its computational cost. As
future work, we plan to use our rate and distortion models to
develop a rate control algorithm for 3D point clouds.

\appendices
\section{}\label{appendix:a}
From the law of large numbers~\cite{idele1979law}, the average value
of the coding error of the point cloud approximates to its
expectation value. Thus,~\eqref{eq:Dc1} can be written as
\begin{equation}
\label{eq:Dc2}
\begin{aligned}
e^{B,A}_{c} &\approx E\{(\textbf{\emph{C}}_{v}-\textbf{\emph{C}}_{v^*})^2\},
\end{aligned}
\end{equation}
where $E\{.\}$ denotes the expectation operator,
$\textbf{\emph{C}}_{v}$ is the random variable corresponding to the
color of the point cloud $A$, and $\textbf{\emph{C}}_{v^*}$
is the random variable corresponding to the color of the point cloud $B$. In the V-PCC encoder, the color information of the points is reassigned to the reconstructed geometry information~\cite{tmc2v1}. Let $\textbf{\emph{C}}_{v^g}$ denote the reassigned color that is not compressed. Then~\eqref{eq:Dc2} can be rewritten as:
\begin{equation}
\label{eq:Dc3}
\begin{aligned}
e^{B,A}_{c} &\approx E\{(\textbf{\emph{C}}_{v}-\textbf{\emph{C}}_{v^g}+\textbf{\emph{C}}_{v^g}-\textbf{\emph{C}}_{v^*})^2\}\\
&=E\{(\textbf{\emph{C}}_{v}-\textbf{\emph{C}}_{v^g})^2\}+E\{(\textbf{\emph{C}}_{v^g}-\textbf{\emph{C}}_{v^*})^2\}\\
&\quad
+2E\{(\textbf{\emph{C}}_{v}-\textbf{\emph{C}}_{v^g})(\textbf{\emph{C}}_{v^g}-\textbf{\emph{C}}_{v^*})\},
\end{aligned}
\end{equation}
where $E\{(\textbf{\emph{C}}_{v}-\textbf{\emph{C}}_{v^g})^2\}$
represents the color distortion induced only by $Q_g$, and
$E\{(\textbf{\emph{C}}_{v^g}-\textbf{\emph{C}}_{v^*})^2\}$
represents the color distortion induced only by $Q_c$. Because
$(\textbf{\emph{C}}_{v}-\textbf{\emph{C}}_{v^g})$ and
$(\textbf{\emph{C}}_{v^g}-\textbf{\emph{C}}_{v^*})$ can be thought
of as two independent random variables,
$E\{(\textbf{\emph{C}}_{v}-\textbf{\emph{C}}_{v^g})(\textbf{\emph{C}}_{v^g}-\textbf{\emph{C}}_{v^*})\}$
approximates to zero~\cite{yuan2011model}~\cite{6544278}. Accordingly,~\eqref{eq:Dc3} can be further written as
\begin{equation}
\label{eq:Dc4}
\begin{aligned}
e^{B,A}_{c} &=f^{B,A}_g(Q_g)+f^{B,A}_c(Q_c),
\end{aligned}
\end{equation}
where $f^{B,A}_g(Q_g)=E\{(\textbf{\emph{C}}_{v}-\textbf{\emph{C}}_{v^g})^2\}$ and $f^{B,A}_c(Q_c)=E\{(\textbf{\emph{C}}_{v^g}-\textbf{\emph{C}}_{v^*})^2\}$.

\ifCLASSOPTIONcaptionsoff
  \newpage
\fi
\bibliographystyle{IEEEtran}
\bibliography{ref}
\end{document}